\documentclass[11pt]{article}
\usepackage[utf8]{inputenc}
\usepackage{cite}
\usepackage{amsmath,amssymb,amsbsy,amstext, amsthm, simplewick, amsfonts}
\usepackage{hyperref}
\usepackage{graphicx}
\usepackage{upgreek}
\usepackage{bm}
\usepackage{framed}
\usepackage{bbm}
\usepackage{textcomp}
\usepackage{tikz}
\usepackage{pifont}
\usetikzlibrary{matrix,shapes,fit,tikzmark,calc}
\usepackage{adjustbox}
\usepackage{makecell}
\usepackage{bbold}
\usepackage{tcolorbox}
\usepackage{physics}
\usepackage{empheq}
\usepackage[normalem]{ulem}

\numberwithin{equation}{section}

\def \me {\mathbb{e}}

\def\be {\begin{equation}}
\def\ee {\end{equation}}

\def\d{\delta}

\def\e{\epsilon}

\def\ba{\begin{eqnarray}}
\def\ea{\end{eqnarray}}

\def \bfx {\textbf{x}}

\def\vpi {\varphi}
\def \bfk {\textbf{k}}

\def \bfs {\textbf{s}}
\def \del {\partial}

\usepackage[framemethod=default]{mdframed}
\newmdenv[skipabove=7pt,
skipbelow=7pt,
rightline=false,
leftline=false,
topline=false,
bottomline=false,
backgroundcolor=gray!10,
linecolor=gray,
innerleftmargin=5pt,
innerrightmargin=5pt,
innertopmargin=5pt,
innerbottommargin=5pt,
leftmargin=0cm,
rightmargin=0cm,
linewidth=4pt]{eBox}
\newmdenv[skipabove=7pt,
skipbelow=7pt,
rightline=false,
leftline=false,
topline=false,
bottomline=false,
backgroundcolor=gray!10,
linecolor=gray,
innerleftmargin=5pt,
innerrightmargin=5pt,
innertopmargin=-5pt,
innerbottommargin=5pt,
leftmargin=0cm,
rightmargin=0cm,
linewidth=4pt]{eBox2}

\usepackage{array}

\definecolor{blue3}{RGB}{31, 119, 180}
\definecolor{red3}{RGB}{	214, 39, 40}
\definecolor{orange3}{RGB}{255, 127, 14}
\definecolor{green3}{RGB}{44, 160, 44}

\usepackage{colortbl}
\definecolor{lightgreen}{cmyk}{0.2, 0, 0.2, 0.2}
\definecolor{lightgray}{cmyk}{0.1,0.2,0,0.1}
\definecolor{lightgray2}{cmyk}{0.1,0.1,0,0.1}

\makeatletter
\newlength{\apb@width}
\newcommand{\autoparbox}[2][c]{\settowidth{\apb@width}{#2}\parbox[#1]{\apb@width}{#2}}

\makeatother


\def\d{{\rm d}}

\def\disc{\text{Disc}}

\def\eps{\boldsymbol\epsilon}
\def \bfp {\textbf{p}}
\def \BCOT {B_{\text{COT}}}
\def \BMLT {B_{\text{MLT}}}
\def \Bcontact {B_{\text{contact}}}
\def \psires {\psi_{\text{Res}}}

\def \RHS {\Xi}

\def\beq{\begin{equation}}
\def\eeq{\end{equation}}

\allowdisplaybreaks[1]
\begin{document}



\begin{titlepage}
\setcounter{page}{1} \baselineskip=15.5pt 
\thispagestyle{empty}

\begin{center}
{\fontsize{28}{28} \centering \bf From Locality and Unitarity \\[0.3cm] to Cosmological Correlators \vspace{0.1cm}
\;}\\
\end{center}

\vskip 18pt
\begin{center}
\noindent
{\fontsize{12}{18}\selectfont Sadra Jazayeri\footnote{\tt
			jazayeri@iap.fr}$^{,\star}$, Enrico Pajer\footnote{\tt
			ep551@damtp.cam.ac.uk}$^{,\dagger}$ and David Stefanyszyn\footnote{\tt dps56@cam.ac.uk}$^{,\dagger}$}
\end{center}

\begin{center}
  \vskip 8pt
  $\star$\textit{ Institut d'Astrophysique de Paris, GReCO, UMR 7095 du CNRS et de Sorbonne Universit\'{e},\\ 98bis
boulevard Arago, 75014 Paris, France} \\ 
$\dagger$\textit{ Department of Applied Mathematics and Theoretical Physics, University of Cambridge, Wilberforce Road, Cambridge, CB3 0WA, UK}
\end{center}

\vspace{0.4cm}

\noindent In the standard approach to deriving inflationary predictions, we evolve a vacuum state in time according to the rules of a given model. Since the only observables are the future values of correlators 
and not their time evolution, this brings about a large degeneracy: a vast number of different models are mapped to the same minute number of observables. Furthermore, due to the lack of time-translation invariance, even tree-level calculations require an increasing number of nested integrals that quickly become intractable. Here we ask how much of the final observables can be ``bootstrapped" directly from locality, unitarity and symmetries. 

\noindent To this end, we introduce two new ``boostless'' bootstrap tools to efficiently compute tree-level cosmological correlators/wavefunctions without any assumption about de Sitter boosts. The first is a \textit{Manifestly Local Test} (MLT) that any $n$-point (wave)function of massless scalars or gravitons must satisfy if it is to arise from a manifestly local theory. When combined with a sub-set of the recently proposed Bootstrap Rules, this allows us to compute explicitly all bispectra to all orders in derivatives for a single scalar. Since we don't invoke soft theorems, this can also be extended to multi-field inflation. The second is a \textit{partial energy recursion relation} that allows us to compute exchange correlators. Combining a bespoke complex shift of the partial energies with Cauchy's integral theorem and the Cosmological Optical Theorem, we fix exchange correlators up to a boundary term. The latter can be determined up to contact interactions using unitarity and manifest locality. As an illustration, we use these tools to bootstrap scalar inflationary trispectra due to graviton exchange and inflaton self-interactions.


\end{titlepage}


\newpage
\setcounter{tocdepth}{2}
\tableofcontents

\newpage




\section{Introduction}


The space of all consistent field theories is vast, and even the number of different models that have been proposed for particle physics and cosmology is enormous and continuously growing. In contrast, the fundamental principles on which these models are based and that define what we deem to be consistent are actually very few: quantum mechanics and its main pillar of unitarity together with locality and symmetries. The journey from physical theories to predictions can then take one of two paths. On the one hand, we can propose theories that obey the above fundamental principles and use well-established computational methods to derive observables. This is a very convenient approach when we have \textit{few concrete theories} that we trust and we want to explore a wide range of diverse phenomena that they might lead to. In this case, we can compactly define the theories through an action and use it to compute a multitude of predictions. For example, this is a very natural approach when deriving precise predictions from General Relativity and the standard model of particle physics. \\

\noindent On the other hand, there are cases in which, either due to the lack of data or to the richness of imagination of theorists, we are interested in a \textit{large number of possible theories}. Furthermore, it might also be the case that we probe these theories through one and the same observable. An example is much of particle physics beyond the standard model: the number of extensions to the standard model for collider phenomenology is large and ultimately many of them are constrained through the scattering amplitudes they predict. Another example is the many models of the very early universe that have been proposed, which are ultimately confronted with observations in the statistics of correlators of primordial perturbations. In these cases, the previous approach is inefficient because we have to compute the same observables over and over again for new models. It may instead be simpler to \textit{apply the fundamental principles directly to the observables} that we want to compute and ask which predictions are compatible with those principles. This bypasses altogether the explicit construction of concrete models. While both approaches have their merit, it is this second approach, which sometimes goes under the name of \textit{the bootstrap programme}, that we advocate in this work in the context of early universe cosmology. \\

\noindent The natural observables of cosmology are the correlation functions of perturbations of the density and velocity of the constituents of the universe. In models of single-clock inflation, these observables are fixed by the quantum expectation value of the product of field fluctuations at the end of inflation. These so-called primordial correlators naturally live at the future conformal boundary of quasi-de Sitter spacetime. In the past few years, our understanding of how fundamental principles constrain primordial correlators has dramatically improved. Much information has emerged from improvements in perturbative calculations \cite{Arkani-Hamed:2015bza,Arkani-Hamed:2017fdk,Arkani-Hamed:2018bjr,Benincasa:2018ssx,Benincasa:2019vqr,Sleight:2019mgd,Sleight:2019hfp,Sleight:2020obc,Baumgart:2019clc,Gorbenko:2019rza,Cohen:2020php,Baumgart:2020oby}. At the same time, it has become clear that very powerful results can be bootstrapped directly from fundamental principles taking advantage of the restrictive symmetries of de Sitter spacetime \cite{Maldacena:2011nz,Creminelli:2011mw,Kehagias:2012pd,Mata:2012bx,Bzowski:2013sza,Ghosh:2014kba,Kundu:2014gxa,Kundu:2015xta,Pajer:2016ieg,CosmoBootstrap1,Bzowski:2019kwd,CosmoBootstrap2,Isono:2020qew,CosmoBootstrap3}. For these bootstrap methods to make contact with the rich world of inflationary phenomenology and eventually observations, it is necessary to abandon the requirement of full de Sitter isometries since de Sitter boosts are incompatible with large primordial non-Gaussianities \cite{Green:2020ebl}. Fortunately, several lessons from the de Sitter studies can be exported to the more general ``boostless" case, where we allow for interactions coming from the coupling to the inflationary background, which break boosts and can give rise to phenomenologically large primordial non-Gaussianity \cite{EFTofI}. Indeed, building upon several results derived in \cite{Arkani-Hamed:2015bza,Arkani-Hamed:2017fdk,Benincasa:2018ssx,Benincasa:2019vqr,CosmoBootstrap1}, a set of (boostless) Bootstrap Rules (Rules $1$-$6$) was proposed in \cite{BBBB} to bootstrap a large number of bispectra involving (massless) scalars and gravitons. The Bootstrap Rules allow one to quickly derive many classical results in the literature, e.g. those in \cite{Maldacena:2002vr,Chen:2006nt}, but also make the role of fundamental principles explicit. In this paper we will improve the Bootstrap Rules by better understanding the implications of locality. In particular, we do not fix the leading total energy pole in terms of a local amplitude (Rule $3$) and we do not use soft theorems (Rule $6$)\footnote{We leave the possibility of imposing soft theorems as additional constraints.}. Rather, we introduce a more powerful Manifestly Local Test (MLT), that selects theories where the interactions of on-shell degrees of freedom are manifestly local. Our primary objects of interest will be the \textit{wavefunction coefficients} appearing in the wavefunction of the universe, which are related to cosmological correlators via simple algebraic relations (see e.g. \cite{COT,WFCtoCorrelators1,WFCtoCorrelators2}). We choose to work with the wavefunction as opposed to correlators for several reasons. At the technical level, the constraints from unitarity and the choice of vacuum derived in \cite{COT,Cespedes:2020xqq,sCOTt,Gordon} can be expressed very straightforwardly in terms of the wavefunction coefficients, but become very cumbersome in terms of the correlators, since in general one needs to include correlators of products of both the fields and their conjugate momenta. At a more abstract level, in a conjectural holographic approach to quantum field theory and quantum gravity in de Sitter, it is the wavefunction that is naturally computed by a conformal field theory, in analogy with the AdS/CFT correspondence and along the lines of the proposal in \cite{Maldacena:2002vr} (see also \cite{Witten:2001kn,Strominger:2001pn} for a different perspective). When combined with Rules $1,2,4$ and $5$ of \cite{BBBB}, we will show that the Manifestly Local Test enables us to derive all possible shapes of scalar $3$-point functions arising from manifestly local interactions. Our MLT enables us to capture some important constraints that were missed in \cite{BBBB}. We will construct both real and imaginary parts in this paper. \\

\noindent Another important recent achievement has been the formulation of the consequences of unitary time evolution for cosmological correlators. The main insight is that, assuming a Bunch-Davies initial state, unitarity manifests itself in a relation between wavefunction coefficients and their analytical continuation to negative energies \cite{COT}. For more general initial states this can be shown to correspond to a set of conserved quantities \cite{Cespedes:2020xqq}. This relation can be thought of as a Cosmological Optical Theorem (COT) providing a non-linear relation for the discontinuities of correlators. The COT is also remarkably general: it is valid for fields of any mass and spin on a very general class of FLRW spacetimes, including de Sitter, inflation and $\Lambda$CDM \cite{Gordon}, as well as to any loop order in perturbation theory \cite{sCOTt} (where it takes the form of Cosmological Cutting Rules). In this paper we show how these relations can be used to bootstrap exchange $n$-point functions arising from boost-breaking interactions. In particular, we use a set of \textit{partial energy shifts} which deform the partial energies of exchange wavefunction coefficients by a complex parameter. When combined with the COT, these shifts lead to \textit{partial energy recursion relations} that combine lower-order diagrams into higher-order exchange diagrams. Similar energy shifts were first introduced in \cite{Arkani-Hamed:2017fdk} to fix the residues of simple poles. Our particular choice of partial energy shifts combined with the COT enables us to go beyond simple poles, which is crucial for cosmology\footnote{Indeed, energy shifts and corresponding recursion relations for wavefunction coefficients were recently introduced in \cite{Arkani-Hamed:2017fdk} within the context of a toy model for cosmology, namely a conformally coupled field with non-derivative self-interactions. It was further assumed that no boundary term arises when the shifting parameter approaches complex infinity.  In such cases, the correlators can only possess simple poles with residues that are encoded in the amplitude limit.  As such, these correlators are entirely dictated by the knowledge of the flat space limit of the diagram and its constituent subdiagrams.  In this work, we want to study phenomenologically relevant correlators an so we consider a massless field with arbitrary derivative interactions. Our method can be used to bootstrap exchange diagrams with poles of any degree, the residues of which cannot be read off from the corresponding flat space amplitude except for the most singular one.  Also,  our energy shifts will generically induce boundary terms that need to be added to the residue sector in order to have a consistent correlator. We will see that unitarity and locality combine to pinpoint both the residue and the boundary sectors up to unspecified local contact terms.}. While this approach is inspired by analogous methods used in amplitudes \cite{Benincasa:2013,Elvang:2013cua,TASI}, e.g. BCFW recursion relations \cite{BCFW}, the features of cosmological spacetime, and in particular the lack of time translation and boost invariance, make the details of the method quite different from the flat spacetime counterpart. \\

\noindent Our results in this paper apply to de Sitter geometries and inflation, at tree-level. Indeed, our MLT and the ansatz we employ for our partial-energy shifts follow from de Sitter mode functions. It would be interesting to extend our results to other accelerating FLRW spacetimes, with help from the recent generalisation of the COT to backgrounds away from de Sitter space \cite{Gordon}. The ansatz that we work with, for both the three-point and four-point functions, are also applicable to tree-level only. We expect the MLT constraint that we derive in Section \ref{sec:MLT} to be valid away from the tree-level approximation, however technical difficulties in going beyond tree-level lie within writing down the appropriate ansatz. Indeed, a rational ansatz is very important in Section \ref{PartialEnergyShifts} such that we can use Cauchy's integral theorem. We leave the interesting generalisation of our results to loop level for future work. \\

\noindent The bootstrap approach for cosmology is just at its infancy and there is a lot more to learn. For example, here we primarily study external scalar fields, but from our experience with amplitudes, one expects even more interesting results and constraints to emerge in the presence of massless spinning particles. We hope our progress here will eventually contribute to that larger goal.


\subsection{Summary of the results}

For the convenience of the reader, we summarize our main results below.
\begin{itemize}
    \item In Section \ref{sec:MLT}, we derive the following simple condition that all wavefunction coefficients $\psi_n$ involving massless scalars and gravitons (and any field with the same $\Delta^+=3$ mode functions given in \eqref{ModeFunctionsMassless}) must satisfy for the theory to be \textit{manifestly local}, namely involve only local interactions of the dynamical fields\footnote{This in particular excludes some gravitational interactions during inflation where, after integrating out the non-dynamical lapse and the shift, interactions with inverse Laplacians appear, which violate manifest locality \cite{Maldacena:2002vr}.}:
    \begin{align} 
    \frac{\partial }{\partial k_{c}} \psi_{n}(k_1,...,k_n;\{p \}; \{\bfk\})\Big|_{k_{c}=0}=0\,,\quad \quad \forall c=1,\dots,n\,, 
    \end{align}
    where $\bfk_a$ with $a=1,\dots, n$ are the $n$ external momenta with ``energies" $k_a=|\bfk_a|$, $\{p \}$ denotes possible internal energies if $\psi_{n}$ arises due to an exchange process and $\{\bfk\}$ denotes possible contractions of momenta and polarisation vectors. Here the derivative is taken with all other variables kept fixed. This condition is satisfied, for example, for interactions to any order in derivatives of any number of (massless) inflatons. However, this condition may be violated in the presence of massless spinning particles, where the solution of the gauge constraints may induce non-manifestly local interactions involving inverse Laplacians, as is the case for General Relativity \cite{Maldacena:2002vr}. We name this constraint the \textit{Manifestly Local Test (MLT)}. In Section \ref{MLTforMassive} we discuss analogous conditions for fields of arbitrary masses. For conformally coupled scalars the condition is much weaker than for their massless counterparts.
    \item We show in Section \ref{BootstrapThreePoint} that the MLT is a surprisingly powerful computational tool to derive wavefunction coefficients within the recently proposed \textit{boostless bootstrap} approach \cite{BBBB}. As the name suggests, this approach makes no assumption about invariance under de Sitter boosts and can therefore be applied to most models in the literature. Using a set of Bootstrap Rules that enforce the correct singularities, symmetries and the Bunch-Davies vacuum, one can easily derive a simple \textit{bootstrap Ansatz} for general bispectra. Without any reference to local amplitudes and soft theorems (which conversely were invoked in \cite{BBBB}), we show here that the Manifestly Local Test determines precisely the scalar bispectra generated by the Effective Field of Inflation to all orders in derivatives and excludes those gravitational interactions that are not manifestly local (and also slow-roll suppressed). We prove that the number $N_{\text{total}}$ of possible manifestly local scalar bispectra matches the number $N_{\text{amplitudes}}$ of independent cubic scalar amplitudes plus one:
    \begin{align}
    N_{\text{total}}(p) = N_{\text{amplitudes}}(p) + 1= 1+\sum_{q=0}^{\lfloor \frac{p+3}{2}\rfloor}\, \lfloor \dfrac{p+3-2q}{3}\rfloor\,,
    \end{align}
    where $\lfloor \dots \rfloor$ is the floor function and $p$ is the maximum number of derivatives. The additional bispectrum corresponds to the only allowed manifestly local field redefinition. 
    \item In Section \ref{PartialEnergyShifts} we derive for the first time \textit{partial-energy recursion relations} as a tool to bootstrap exchange diagrams from lower-point vertices and show this explicitly for exchange $4$-point wavefunction coefficients. To this end, we perform complex shifts of the partial energies of a given diagram, which are the sums of the energies flowing into any given sub-diagram, and are the only allowed singularities at tree-level together with the total energy. The residues of all the poles in the complex shift are fully fixed by the Cosmological Optical Theorem \cite{COT} and its perturbative manifestation in the recently derived Cosmological Cutting Rules \cite{Gordon,sCOTt}. This is in contrast to the factorization limits discussed for example in \cite{CosmoBootstrap3}, which fix only the leading order singularities. For example, for the \textit{exchange} scalar $4$-point wavefunction coefficient $\psi_4$ (related to the trispectrum) our bootstrap result comprises of three terms derived in three subsequent steps,
    \begin{align}
    \psi_4&=\psires+\BCOT + \BMLT \,.
    \end{align}
    The first term $\psires$ is derived in Step I (Section \ref{sec:step1}) via Cauchy's integral theorem and is found to be
    \begin{align}
      \psires &=\sum_{0<n\leq m}\dfrac{A_n(E_L,E_R,k_1k_2,k_3k_4,s)}{E_L^n}+\sum_{0<n\leq m}\dfrac{A_n(E_R,E_L, k_3k_4,k_1k_2,s)}{E_R^n}\,, \\
      A_n  &=\dfrac{1}{(m-n)!}\partial^{m-n}_z \Big\{ (z+E_L)^m\, \RHS(E_L+z,E_R-z,k_1k_2,k_3k_4,s) \Big\} _{z=-E_L}\,,
    \end{align}
    where $m$ is the number of derivatives in the cubic interaction (also the order of the $k_T^{(3)}$ pole), $s=|\bfk_1+\bfk_2|$, $E_L=k_1+k_2+s$, $E_R=k_3+k_4+s$ and $\RHS$ is the right-hand side of the COT given by \cite{COT}
    \begin{align}
         \RHS(E_L,E_R,k_1k_2,k_3k_4,s)= P(s) &\left[ \psi_3(k_1,k_2,s)-\psi_3(k_1,k_2,-s) \right] \nonumber \\
         \qquad \times &\left [ \psi_3(k_3,k_4,s)-\psi_3(k_3,k_4,-s) \right]\,,
    \end{align}
    with $P(s)$ the power spectrum of the exchanged field. The remaining terms are boundary terms not fixed by Cauchy's integral theorem. The second term $\BCOT$ is derived in Step II (Section \ref{sec:step2}) and ensures that $\psi_4$ obeys the COT. It is given by 
    \begin{align}
        \BCOT=\frac{1}{12}\,\dfrac{\partial^3 }{\partial s^3} \, \RHS(E_L,E_R,k_1k_2,k_3k_4,s) \Big|_{s=0} s^3\,,
    \end{align}
    where the derivative is computed keeping fixed $\{E_L,E_R,k_1k_2, k_3 k_4\}$. The third and last term $\BMLT$ obeys the contact COT and is computed in Step III (Section \ref{BoundaryTerm}) by applying our manifestly local test to $\psi_{4}$. For external fields with massless mode functions, it can be found by solving 
    \begin{align} 
    \dfrac{\partial }{\partial k_c}\BMLT \Big|_{k_c=0}= -\dfrac{\partial }{\partial k_c}\psi_{\text{Res}}\Big|_{k_c=0} \quad \quad \forall c=1,\dots,n\,.
\end{align}

    \item In Section \ref{Results} we show how a combination of the MLT and the partial-energy recursion relations can be used to explicitly compute the scalar trispectrum in a few \textit{examples}: first we demonstrate our techniques in Minkowski spacetime for the polynomial interaction $\phi^3$, where the algebra is minimal. Then, we derive the scalar trispectrum from graviton exchange, which, together with a contact diagram, contributes to leading order to the trispectrum for a minimally coupled canonical inflaton \cite{Seery:2008ax}. Finally, we compute the potentially much larger trispectrum from scalar exchange for the leading cubic interactions in the Effective Field Theory of inflation, namely $\dot \phi^3$ and $\dot \phi (\nabla \phi)^2$. We also consider $\phi \dot\phi^2$ as a simple example. Our derivation is computationally faster than the bulk in-in computation and in all cases we find agreements with the results in the literature, where available. 
\end{itemize}


\paragraph{Notation and conventions} \label{Conventions}
We work with the mostly positive metric signature $(-+++)$. The 3d Fourier transformation is defined as
\begin{align}
f(\bfx)&=\int \dfrac{d^3\bfk}{(2\pi)^3}{f}(\bfk)\exp(i\bfk\cdot\bfx)\equiv\int_{\bfk}{f}(\bfk)\exp(i\bfk\cdot\bfx) \,,\\
{f}(\bfk)&=\int d^3\bfx f(\bfx)\exp(-i\bfk\cdot\bfx)\equiv \int_{\bfx}  f(\bfx)\exp(-i\bfk\cdot\bfx)\,.
\end{align}
We use bold letters to refer to vectors, e.g. $\bf x$ for spatial co-ordinates and $\bf k$ for spatial momenta, and we write the magnitude of a vector as $k\equiv |\bfk|$. We will sometimes refer to these objects as ``energies". We will use $i,j,k,\ldots = 1,2,3$ to label the components of $SO(3)$ vectors, and $a,b,c=1,\dots,n$ to label the $n$ external fields. For wavefunction coefficients and cosmological correlators we use $\psi_{n}$ and $B_n$ respectively:
\begin{align}
\psi_n(\bf{k_1},\dots ,\bf{k_n})& \equiv \psi_n'({\bf{k_1},\dots ,\bf{k_n}})(2\pi)^3 \delta^3 \left(\sum \bf{k_a} \right)\,, \\ \nonumber
\langle {\cal{O}} {(\bf{k_1})\dots {\cal{O}}(\bf{k_n})} \rangle &\equiv \langle {\cal{O}} {(\bf{k_1})\dots {\cal{O} }({\bf{k_n}}) }\rangle' (2\pi)^3\delta^3 \left(\sum \bf{k_a} \right)\\
&\equiv B_n ({\bf{k_1},...,\bf{k_n}})\,  (2\pi)^3\delta^3\left(\sum \bf{k_a} \right)\,,
\end{align}
and we will drop the primes on $\psi_n$ when no confusion arises. We will also use a prime to denote a derivative with respect to the conformal time e.g. $\phi'=\del_\eta \phi$. When computing exchange $4$-point functions we use the following variables
\begin{align}
    s&\equiv |\bfk_1+\bfk_2|\,,& t&\equiv |\bfk_1+\bfk_3|\,,& u&\equiv |\bfk_1+\bfk_4|\,, &\label{sum}
s^2+t^2+u^2&=\sum_{a=1}^4\,k_a^2\,.
\end{align}
We define the ``total energy'' $k_{T}$ of an $n$-point function as
\be\label{kT}
k_T^{(n)} \equiv \sum_{a=1}^{n}  k_a\,.
\ee
We will often drop the superscript ``$  (n) $'' in $  k_{T} $ when it is clear from the context. We define the $s$-channel ``partial energies" in a $4$-point exchange diagram as
\begin{align}
E_L=k_1+k_2+s\,,\qquad E_R=k_3+k_4+s\,.
\end{align}
We denote the $n$ external momenta of a tree-level Feynman diagram by $\bfk_a$ ($a=1,..,n$) while referring to its $I$ internal lines by $\bfp_b$ ($b=1,...,I$), apart from for $4$-point exchanges where we use the more familiar notation in \eqref{sum}. We will often use $S$ as the energy of an internal line that is ``cut". The internal momenta are fixed with the knowledge of $\bfk_a$'s due to the conservation of spatial momentum at each vertex. The symbols $a_\bfk$ and $a_{\bfk}^\dagger$ refer to annihilation and creation operators, respectively. 
The flat space amplitude will be written as 
\be
{\cal S}-\mathbf{1}\equiv i\, A_n(p_1^\mu\,,\dots\, p_n^\mu) \,(2\pi)^4\,\delta^4\left( \sum\,p_a^{\mu} \right)\,,
\ee
with all the four-momenta defined to be ingoing. We write symmetric polynomials in terms of elementary symmetric polynomials (ESP). For three variables, $k_{1}, k_{2}$ and $k_{3}$, the elementary symmetric polynomials are
\begin{align}
k_{T}^{(3)} &=e_1 = k_{1}+k_{2}+k_{3}, & e_{2} &= k_{1}k_{2}+k_{1}k_{3}+k_{2}k_{3}, & e_{3} &= k_{1}k_{2}k_{3}.
\end{align}


\section{A review of the Cosmological Optical Theorem} \label{Review}

\noindent We begin by reviewing aspects of a scalar quantum field in de Sitter (dS) space including the formalism of the wavefunction of the universe and the recently derived Cosmological Optical Theorem (COT) \cite{COT} which will be used throughout this work.

\paragraph{A quantum scalar field in de Sitter} We take the background geometry to be that of dS space, which we write using the conformal time co-ordinate $\eta \in (-\infty,0]$ as
\begin{align}
ds^2 = a^2(\eta) (-\d \eta^2 + d {\bf x^2}), \qquad a(\eta) =  -\frac{1}{\eta H},
\end{align}
where $H$ is the constant Hubble parameter which we will often set to unity. Our methods apply to general quantum field theories, but we will primarily work with a single scalar $\sigma(\eta, \bf x)$ which we assume does not perturb the background dS geometry. We will therefore refer to this field as  a ``spectator field". In some examples we will also include a coupling to the graviton, and a more general analysis for spinning fields will appear elsewhere \cite{JPSS}. The free action for this scalar is 
\begin{align}
S_{\sigma, \text{free}} = \int d \eta d^3 {\bf x} a^2(\eta) \left[\frac{1}{2}\sigma'^2 - \frac{1}{2}c_{\sigma}^2 \partial_{i}\sigma \partial^{i}\sigma - \frac{1}{2}a^{2}(\eta)m_{\sigma}^2 \sigma^2 \right],
\end{align}
where we have allowed for an arbitrary, constant speed of sound $c_{\sigma}$ which we will also often set to unity. Working in momentum space, we write the quantum free field operator as 
\begin{align}
\hat{\sigma}(\eta, {\bf k}) = \sigma^{-}(\eta, k) a_\bfk + \sigma^{+}(\eta,k) a_{-\bfk}^\dagger\,,
\end{align}
where the mode functions $\sigma^{\pm} (\eta, k)$ correspond to solutions of the free classical equation of motion and are given by
\begin{align} \label{ModeFunctionsGeneral}
\sigma^{+}(k,\eta) = i \frac{\sqrt{\pi}H}{2}e^{-i\frac{\pi}{2}(\nu +1/2)}\left(\frac{-\eta}{c_{\sigma}}\right)^{\frac{3}{2}}H_{\nu}^{(2)}(-c_{\sigma}k\eta), \qquad \sigma^{-}(k,\eta) = (\sigma^{+}(k,\eta))^{*},
\end{align}
where $\nu = \sqrt{\frac{9}{4} - \frac{m_{\sigma}^2}{H^2}}$, and $H_{\nu}^{(2)}(z)$ is the Hankel function of the second kind and order $\nu$. 
In analogy with the AdS/CFT literature, this is often expressed in terms of the two scaling dimensions $\Delta^+=3-\Delta^-$, defined as
\begin{align}\label{delta}
    \Delta^\pm\equiv 
     \frac{3}{2}\pm \sqrt{\frac{9}{4} - \frac{m_{\sigma}^2}{H^2}} =\frac{3}{2}\pm \nu \,.
\end{align}
We will primarily illustrate our methods in the massless ($m_{\phi} = 0$) and conformally coupled ($m_{\varphi} = \sqrt{2} H$) limits where we use $\phi$ and $\varphi$ respectively to denote these fields. In these limits the mode functions take the simpler forms
\begin{align}
\phi^{\pm}(\eta, k) &= \frac{H}{\sqrt{2 c_{\phi}^3 k^3}} (1 \mp i c_{\phi} k \eta)e^{\pm i c_{\phi} k \eta} &  \text{(massless, $\nu=3/2$, $\Delta^+=3$)}&\,, \label{ModeFunctionsMassless} \\
\varphi^{\pm}(\eta, k) &= \mp \frac{iH}{\sqrt{2 c_{\varphi}^3 k^3}} \eta e^{\pm i c_{\varphi} k \eta} &\text{(conformally coupled, $\nu=1/2$, $\Delta^+=2$)}& \label{ModeFunctionsCC}. 
\end{align}
Note that the mode functions for a massless graviton are the same as for a massless scalar, up to the necessary polarisation factor. As explained above, we allow interactions to break de Sitter boost symmetry but we keep the remaining symmetries of the dS group intact (translations, rotations and dilations). A general interaction vertex with $n$ fields therefore takes the schematic form
\begin{align}
S_{\sigma, \text{int}} =  \int d \eta d^3 {\bf x} \, a(\eta)^{4 - N_{\text{deriv}}} \partial^{N_{\text{deriv}}} \sigma^{n},
\end{align}
where $\partial$ stands for either temporal derivatives $\partial_{\eta}$ or spatial derivatives $\partial_{i}$, and $N_{\text{deriv}}$ is the total number of derivatives. The powers of the scale factor are fixed by scale invariance and all spatial derivatives are contracted with the $SO(3)$ invariant objects $\delta_{ij}$ and $\epsilon_{ijk}$. Our system is a very good approximation to inflationary models where any deviations from exact scale invariance are slow-roll suppressed. Note that, in the context of a conformally coupled scalar, we do not assume that the interactions respect the conformal symmetry: it is only the quadratic operators in $\varphi$ that are conformally coupled such that the mode functions become tractable.

\begin{figure} 
\begin{center}
\includegraphics[width=10cm]{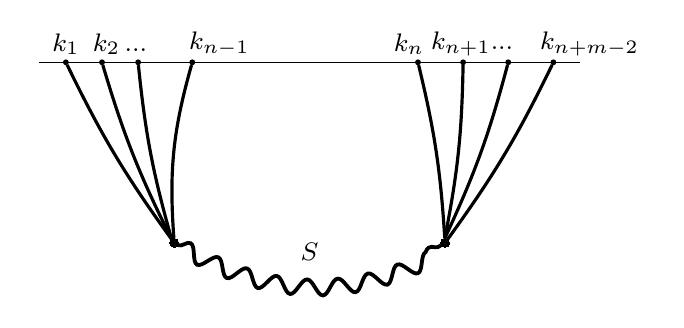}
\caption{A single exchange diagram with external scalars exchanging a generic field.}\label{SingleExchange}
\end{center}
\end{figure}

\paragraph{Wavefunction of the universe} Our primary object of interest is the wavefunction of the universe $\Psi$ evaluated at the late-time boundary of dS space, which we denote as $\eta_{0}$. For simplicity, consider $\sigma$ as the only field in the theory. The wavefunction then has an expansion in $\sigma({\bf k}) \equiv \sigma(\eta_{0}, {\bf k})$ given by
\begin{align}
\Psi[\eta_{0},\sigma({\bf k})] = \text{exp}\left[-\sum_{n=2}^{\infty} \frac{1}{n!} \int_{{\bfk}_{1}, \ldots, {\bf{k}}_{n}} \psi_{n}({\bf k_{1}} \ldots {\bf k_{n}})\sigma({\bf k}_{1}) \ldots \sigma({\bf k}_{n}) \right],
\end{align}
with the dynamics of the theory encoded in the \textit{wavefunction coefficients} $\psi_{n}({\bf k_{1}} \ldots {\bf k_{n}})$. Invariance of the theory under spatial translations ensures that the $\psi_{n}({\bf k_{1}} \ldots {\bf k_{n}})$ always contain a momentum conserving delta function and so we can write
\begin{align}
\psi_{n}({\bf k_{1}}, \ldots, {\bf k_{n}}) = \psi'_{n}({\bf k_{1}}, \ldots, {\bf k_{n}}) (2 \pi)^3 \delta^{3}({\bf k_{1}}+ \ldots+ {\bf k_{n}}).
\end{align}
We will often drop the prime even when we do not explicitly include the delta function. In the saddle-point approximation, which is exact for the tree-level processes of interest here, we have
\begin{align}
\Psi[\eta_{0},\sigma({\bf k})] \approx e^{i S_{\text{cl}}[\sigma({\bf k})]}.
\end{align}
As reviewed in detail in \cite{COT,WFCtoCorrelators2}, in bulk perturbation theory $S_{\text{cl}}[\sigma({\bf k})]$ is computed in a diagrammatic fashion using the bulk-to-boundary propagator $K_{\sigma}(\eta,k)$ and bulk-to-bulk propagator $G_{\sigma}(\eta, \eta', k)$. When it's clear from the context, we'll often drop the field label and simply write $K$ and $G$. Both of these propagators are represented in Figure \ref{SingleExchange} and, denoting the free equation of motion as $\mathcal{O}(\eta,k) \sigma$ = 0, they satisfy
\begin{align}
&\mathcal{O}(\eta,k) K(\eta, k) = 0, \\
&\mathcal{O}(\eta,k) G(\eta, \eta', k) = -\delta(\eta - \eta'), 
\end{align}
with boundary conditions
\begin{align}
\lim_{\eta \rightarrow \eta_{0}} K(\eta, k) &= 1, & \lim_{\eta \rightarrow -\infty(1 - i \epsilon)} K(\eta, k) &= 0 \\
\lim_{\eta, \eta' \rightarrow \eta_{0}} G(\eta, \eta', k) &= 0, & \lim_{\eta, \eta' \rightarrow -\infty(1 - i \epsilon)} G(\eta, \eta', k) &= 0.
\end{align}
Both propagators can be written in terms of the positive and negative frequency mode functions and are given by
\begin{align}
    K(k,\eta)&=\frac{\sigma^+_k(\eta)}{\sigma^+_k(\eta_0)}\,,\\
G ( p,\eta, \eta') &= i\left[ \theta(\eta-\eta')\left(\sigma^+_p(\eta')\sigma^-_p(\eta)-\frac{\sigma^-_p(\eta_0)}{\sigma^+_p(\eta_0)}\sigma^+_p(\eta)\sigma^+_p(\eta')\right)+(\eta \leftrightarrow \eta')\right]\\
&=iP(p)\left[\theta(\eta-\eta')\frac{\sigma^+_p(\eta')}{\sigma^+_p(\eta_0)}\left(\frac{\sigma^-_p(\eta)}{\sigma^-_p(\eta_0)}-\frac{\sigma^+_p(\eta)}{\sigma^+_p(\eta_0)}\right)+(\eta \leftrightarrow \eta')\right]
\end{align}
with $P(p)$ the power spectrum of $\sigma$. When $p$ is real, $G$ can also be written as
\begin{align} \label{BulkToBulk}
G (p, \eta, \eta') &=
 2 P(p)  \left[  \theta( \eta - \eta') K ( p, \eta'  )\text{Im} \, K (p, \eta  )  + \left( \eta \leftrightarrow \eta' \right)    \right]\; \\
 &=iP(p)  \left[  \theta(\eta-\eta') K^{\ast}(p,\eta ) K(p,\eta' )+\theta(\eta'-\eta) K^{\ast}(p,\eta')K(p,\eta) \right. \nonumber \\
 & \qquad \left. -K(p,\eta )K(p,\eta' )\right]. \label{BulkToBulkThreeTerms}
\end{align}
Notice that with the overall $i$ in the definition of $G$, our Feynman rules require a factor of $-i$ for every diagram, a factor of $G$ for every internal line and no factor of $i$ for the vertices (e.g. the vertex corresponding to $\lambda \phi^n/n!$ is simply $\lambda$). In perturbation theory, the real part of the wavefunction coefficients $\text{Re}(\psi_{n})$ can be used to compute correlation functions via simple algebraic relations \cite{COT,WFCtoCorrelators1,WFCtoCorrelators2}, at least for parity-even scalar and graviton interactions. The bulk computations of these $n$-point functions, however, requires computing involved time integrals with the integrands products of $a(\eta)$, $K(\eta,k)$, $G(\eta, \eta', k)$ and their derivatives (see Appendix \ref{AppendixA} for more details).  In this paper we take a different approach: using the \textit{Cosmological Optical Theorem} (COT) \cite{COT}, the singularity structure of correlators/wavefunction coefficients and our manifestly local test, we will show how to bootstrap the wavefunction coefficients without the need to perform any time integrals. Our interests will lie primarily in the real part of these coefficients as this is related to cosmological correlators. We will also briefly discuss how to bootstrap the imaginary part of contact diagrams which, for parity even theories, is related to correlators involving (an odd number of) momentum conjugates of the fields. These imaginary parts are of less phenomenological interest because the corresponding correlators are associated to observables that decay exponentially with (cosmological) time are practically unobservable.

\paragraph{Kinematics}
Before reviewing the Cosmological Optical Theorem, let's first state the kinematic variables we will use\footnote{We are closely following the definitions adopted in \cite{Gordon}.}. After imposing rotational and spatial translation symmetries, one requires $3n-6$ independent variables\footnote{For large $n$ it is possible that one cannot find explicitly a set of independent rotation invariant contractions. This is not a problem for our approach because one can simply work with the larger set of dependent variables. We thank James Bonifacio for pointing this out to us.} to fully specify $\psi_n$. We will always work with \textit{energy} variables where we employ \textit{all} $n$ external energies $\{k\}=k_1,\dots k_n$, which appear in the bulk-to-boundary propagators, and all $I$ internal energies $\{p\}=p_{1},\dots,p_I$ that appear in the bulk-to-bulk propagators of a given diagram. The remaining independent variables are chosen from contractions of external momenta with $\delta_{ij}$, $\e_{ijk}$ or polarisation vectors. We refrain from using angles. In compliance with the little group scaling, the dependence on the polarisation vectors must be in a factorised form. We only encounter two examples that involve spinning fields, in Sections \eqref{grav1} and \eqref{grav2}. Until then we focus on scalar $n$-point functions. Our choice of variables therefore depends on the Feynman diagram of interest. \\

\noindent For \textit{contact} $n$-point functions we need $2n-6$ independent inner products and we write 
\begin{equation}
    \text{contact}: \qquad \psi_n(\{k\};\{\bfk\})\equiv \psi_n(k_1,...,k_n;\bfk_a \cdot \bfk_b, \bfk_a \cdot (\bfk_b\times \bfk_c),\bfk_a \cdot \e(\bfk_b))\,.
\end{equation}
For our discussion any set of $2n-6$ choices will do the job. Let's see this in two examples. For a contact $3$-point function we require three independent variables and these are already provided by the three external energies and so there is no need to introduce any inner products. Indeed, each inner product can be written as a linear combination of squared energies due to momentum conservation. For a contact $4$-point function we need six independent variables, namely the four external energies and two inner product such as $\bfk_1.\bfk_2$ and $\bfk_3.\bfk_4$, or $\bfk_1.\bfk_2$ and $\bfk_1.\bfk_3$.\\

\noindent For \textit{exchange diagrams}, we use all the external $\{k\}$ and internal $\{p\}$ energy variables and add to these the appropriate number of additional contractions. We therefore write
\begin{equation}
    \text{exchange}:\qquad \psi_n(\{k\}; \{p\}; \{\bfk\})\equiv \psi_n(\{k\}; p_1,...,p_I;  \{\bfk\} )\,,
\end{equation}
where $\{k\}$ stands for external energies and $\{\bfk\}$ stand for rotation invariant contractions, as above. 
For example, for say the $s$-channel $4$-point exchange diagram we actually only need five variables, which we choose to be $(k_1,k_2,k_3,k_4,s)$. The remaining channels, $t$ and $u$, also only require five energy variables. The sum of the three channels naively therefore contains seven variables but due to the relation in \eqref{sum}, there is a one-variable redundancy and so the number of independent variables is six, as expected. It is worth noting that, as we have seen in this $4$-point example, not all the internal energies are independent. However, in this paper our treatment is channel-by-channel and so we will not encounter sums of internal energies due to different channels. We can therefore take each internal energy to be an independent variable. In the rest of this paper we will employ these variables, and one can easily find expressions in a new set of variables by employing the chain rule, where appropriate. Furthermore, we will often employ these variables away from the physical configuration. Indeed, we will sometimes need to analytically continue the energy variables in $\psi_n$ to the lower complex half-plane, namely $\{k \}, \{p \}\in \mathbb{C}^-$. In the majority of cases, however, we keep all energies real but allow them to be negative. Throughout we will keep the inner products real. \\

\noindent In summary, our notation for our choice of variables is
\begin{equation}
    \psi_n=\psi_n(\,n\text{ external}\,;\,I\text{ internal} \,; \,\text{rotation-invariant contractions} \,)\,.
\end{equation}

\begin{figure}
    \centering
    \includegraphics[width=10cm]{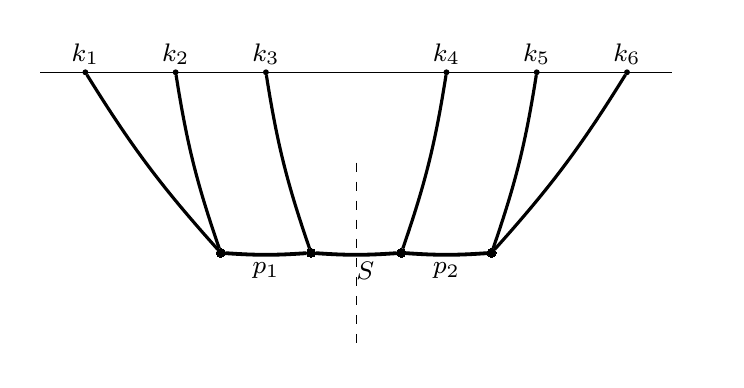}
    \caption{Triple exchange $6$-point diagram for scalars cut into the product of two $4$-point exchange diagrams.\label{6pt}}
\end{figure}

\paragraph{The Cosmological Optical Theorem (COT)} Under a limited number of assumptions, it was shown in \cite{COT} that perturbative unitarity implies a set of powerful constraints on both contact and exchange contributions to the wavefunction coefficients in the form of a Cosmological Optical Theorem (COT) (see also \cite{Cespedes:2020xqq} for a complementary derivation of the COT and \cite{Aharony:2016dwx,Meltzer:2019nbs,Meltzer:2020qbr} for analogous statements in anti-de Sitter (AdS) space). The derivation assumed de Sitter mode functions and Bunch-Davies initial conditions\footnote{The original proof of the COT focused on scalar fields and restricted to contact and single exchange diagrams. However, similar expressions hold for fields of any mass and any spin, both on external and internal lines, as well as for more general tree diagrams with an arbitrary number of vertices \cite{Gordon}.}. Two immediate consequences of these assumptions are  $(i)$ the \textit{Hermitian analyticity} of the bulk-to-boundary propagator:
\begin{align}
    K^*(-k^*,\eta)=K(k,\eta)\,,\qquad k \in \mathbb{C}^-\,,
\end{align}
and $(ii)$ the factorisation property of the bulk-to-bulk propagator:
\begin{align}
    \text{Im}\,G(k,\eta,\eta')=P(k)\,\text{Im}\,K(k,\eta)\,\text{Im}\,K(k,\eta')\,,\qquad k\in \mathbb{R}^+\,.
\end{align}
Exploiting these two features in the bulk formalism, one can derive the Cosmological Optical Theorem for contact  $n$-point functions which must be satisfied by any contact $n$-point function arising from unitary evolution in the bulk spacetime: 
\begin{align}
\label{cotcon}
    \disc \left[ i \psi_n(k_1,...,k_n;\{ \bfk\}) \right] =0\,,
\end{align}    
where we have introduced the discontinuity function $\disc$, which acts on a general wavefunction coefficient as 
\begin{align}
 &  \underset{ k_1 \dots k_j  }{\disc } f(k_1,\dots,k_n ;\{\bfk \}) \nonumber \\ 
 &\equiv f(k_1,\dots,k_n ;\{\bfk \}) - f^\ast(k_1,\dots,k_j,-k^\ast_{j+1},\dots,-k^\ast_n ;-\{\bfk \}) \,. \label{defdisc}
\end{align} 
Note that all spatial momenta (internal or external) in the second term get a minus sign, $ {\bfk} \to -\bfk$, while only the energies that do \textit{not} that appear in the argument of $\disc$ are analytically continued. In other words, the argument of $\disc$ indicates the spectator energies that are untouched by the $\disc$. For example, \eqref{cotcon} becomes
\begin{align}\label{finalCOT}
    \disc \left[ i \psi_{n}(k_1,...,k_n; \{\bfk\} ) \right] = i \left[ \psi_n(k_1,...,k_n;  \{\bfk\}) + \psi^*_n(-k^*_1,...,-k^*_n; - \{\bfk\} ) \right] \,.
\end{align} 
For general tree-level diagrams this generalises to ``single-cut" rules (see Figure \ref{SingleExchange}) \cite{COT,Gordon}. Denoting by $\bf{S}$ the momentum of the internal line to be cut and by $\bfp_m$ all other internal momenta, we have
\begin{align}
\label{cotex}
 \disc_S \left[ i \psi_{n+m-2}(\{k\};\{p\},S; \{\bfk\}) \right]= -i P(S) & \disc_S \left[ i \psi_{n}(k_1,...,k_{n-1},S;\{p\};  \{\bfk\}) \right] \nonumber \\
 \quad \times &\disc_S \left[ i \psi_{m}(k_{n},...,k_{n+m-2},S;\{p\};  \{\bfk\}) \right]\,,
 \end{align}
where in the physical domain of momenta we have
\begin{align}
S=|\textbf{S}|=|\sum_{a=1}^{m-1} \bfk_a|=|\sum_{b=m}^{m+n-2} \bfk_b|\,.
\end{align}
These expressions can be simplified for massless and conformally coupled fields, in the absence of IR-divergences. In such cases there is no branch-cut, and so we can freely assume that all the energies are real: $-k_a^*=-k_a$ (in general the negative real axis is always approached from the lower-half complex plane). In this paper it will be important to go beyond single exchange diagrams and, as was elaborated on in \cite{Gordon}, the single-cut rules carry over to such cases. In Section \ref{BoundaryTerm} we will need the Cosmological Optical Theorem (COT) that relates a triple exchange 6-point function of a massless field to the product of its constituent 4-point exchange sub-diagrams, see Figure \ref{6pt}. The resulting COT reads 
\begin{align}
    \psi_6(k_1,\dots k_6&;p_1,S,p_2)+\psi_6^\ast(-k_1,\dots -k_6;-p_1,S,-p_2)= \nonumber \\
    \quad P(S) &\left[ \psi_4(k_1,\dots,k_3,S;p_1)+\psi_4^\ast(-k_1,\dots,-k_3,S;-p_1)\right] \nonumber\\
    &\left[ \psi_4(k_4,\dots,k_6,S;p_2)+\psi_4^\ast(-k_4,\dots,-k_6,S;-p_2)\right],
\end{align}
where for simplicity we have dropped a possible dependence on inner products. We refer the reader to \cite{COT, Gordon} for more details. 

\section{A Manifestly Local Test (MLT) for $n$-point functions}\label{sec:MLT}

In this section we introduce a manifestly local test (MLT) that must be satisfied by $n$-point functions arising from manifestly local theories. We remind the reader that manifestly local interactions do not contain any inverse Laplacians (see \cite{BBBB} for a recent discussion in the context of cosmology and for additional necessary conditions for locality). For example, $\sigma'^2 \nabla^{2}\sigma$ is a manifestly local interaction, whereas $\sigma'^2 \nabla^{-2} \sigma$ is not. As we shall see, the MLT is a necessary condition for manifest locality, but it becomes sufficient when combined with a sub-set of the Bootstrap Rules of \cite{BBBB} which we review in Section \ref{BootstrapThreePoint}.\\

\noindent In the following two subsections we present the MLT for massless mode functions given in \eqref{ModeFunctionsMassless}. Our test applies equally well to contact and exchange diagrams and to fields of any spin. We first present an argument for the MLT based on the allowed singularity structure of consistent wavefunction coefficients in Section \ref{MLTfromCOT}. We then provide an alternative derivation of the test from a purely bulk perspective in Section \ref{MLTfromBulk}. Finally, we extend the MLT to massive mode functions in Section \ref{MLTforMassive}.


\subsection{Manifest locality from singularities} \label{MLTfromCOT}

The singularity structure of tree-level wavefunction coefficients is now well understood (see e.g. \cite{Arkani-Hamed:2015bza,Arkani-Hamed:2017fdk,Maldacena:2011nz,CosmoBootstrap3,COT,Raju:2012zr}). The only kinematical singularities that can arise are the following:
\begin{itemize}
    \item Total energy singularities in $1/k_T^p$, which arise when the sum of all the (analytically continued) energies flowing to the boundary are taken to zero. The residue of the highest $k_T$ singularity is fixed by a corresponding amplitude, as first noticed in \cite{Maldacena:2011nz,Raju:2012zr} and proven in \cite{COT}. The order $p$ of the singularity is linearly related to the mass dimension of the involved interactions according to a simple formula \cite{BBBB}. Here we will show for the first time that also the residue of \textit{subleading} $k_T$ poles is fixed by a corresponding amplitude, as dictated by the manifestly local test.
\item Partial energy singularities in the sum of (analytically continued) energies that enter a connected subdiagram. These singularities are absent for contact diagrams, so the simplest example is $E_L=k_1+k_2+|\bfk_1+\bfk_2| = k_{1}+k_{2}+s$ in the $s$-channel $4$-point exchange diagram. This can be seen explicitly in the bulk representation of these quantities, as reviewed in appendix \ref{AppendixA}. The residue of these leading-order partial-energy singularities is fixed by factorization, as used extensively in e.g. \cite{CosmoBootstrap3}. In Section \ref{PartialEnergyShifts} we will leverage the fact that all residues of leading \textit{and subleading} partial-energy singularities are fixed by the Cosmological Optical Theorem (see also \cite{COT} for a preliminary discussion) to bootstrap $4$-point exchange diagrams.
\end{itemize}

\noindent One may wonder about the possibility of singularities as one of the external or internal energies vanishes. If the vertices of the theory are manifestly local, then it is clear that no singularities in the external energies can appear. Indeed, no such singularities arise from manifestly local vertices, and any poles arising from time integration correspond to a sum of partial energies, e.g. $k_{T}, E_{L}$ or $E_{R}$. However, what happens when internal energies go to zero? The bulk-to-bulk propagator \eqref{BulkToBulk} contains a factor of the power spectrum which is indeed singular in this limit. For example, $P(S) \sim 1/S^3$ for massless mode functions. Consider one of the possible bulk-bulk propagators appearing in an exchange diagram. We have
\begin{align} \label{BulkToBulkSimplify}
G_{\phi}(\eta,\eta', S) = 2 i P_{\phi}(S)[\text{Im}(K_{\phi}(\eta,S))K_{\phi}(\eta',S) \theta(\eta - \eta')+ \left( \eta \leftrightarrow \eta' \right)  ].
\end{align}
Note that here we include the $\phi$ subscript to indicate that this discussion applies to massless mode functions. For any finite time $\eta$ we have
\begin{align}
\lim_{S\to 0} K_{\phi}(\eta,S)  &= 1+\frac{1}{2}(c_{\phi} S\eta)^2+\frac{i}{3}(c_{\phi} S \eta)^3 + \mathcal{O}(S^4),\\
\lim_{S\to0}\text{Im}{K_{\phi}(\eta,S)} &= \frac{1}{3}(c_{\phi} S \eta)^3 + \mathcal{O}(S^5), 
\end{align}
and so any potential $S=0$ singularities due to the power spectrum are cancelled by the factors of $\text{Im}{K_{\phi}(\eta,S)}$ and $\text{Im}{K_{\phi}(\eta',S)}$. Note that the final term in \eqref{BulkToBulkThreeTerms}, which enforces the vanishing of $G$ on the boundary, is crucial since it is this term that ensures that each term in \eqref{BulkToBulkSimplify} contains a factor of $\text{Im}(K)$. One may worry about the behaviour of the bulk integrals at $\eta = -\infty$. However, this limit could only affect our argument if there were exponential factors containing $e^{i S \eta}$ which would yield additional inverse powers of $S$. However, no such exponents occur: all exponents contain a \textit{sum} of energies which is finite in the limit $S \rightarrow 0$ as long as the other energies are kept fixed. We therefore conclude that in manifestly local theories, exchange diagrams are not singular as an internal energy is taken to zero\footnote{We thank Austin Joyce for discussions about the absence of such singularities.}.\\

\noindent Let's now consider the consequences of this result for the $n$-point functions that can contribute to an exchange diagram. Consider the COT given in \eqref{cotex} which is valid for a single-cut. The left-hand side of this equation is regular as $S \rightarrow 0$, as we have just argued, and so the right-hand side must be too. Given that $P(S) \sim 1/S^3$, this tells us that the product
\begin{align} \label{COTLimit}
&\left[ \psi_{n}(k_1,...,k_{n-1},S;\{p\}; \{\bfk\})+\psi_{n}^*(-k_1^*,...,-k_{n-1}^*,S;-\{p\}; -\{\bfk\}) \right] \nonumber \\
 &\times\left[\psi_{m}(k_{n},...,k_{n+m-2}, S;\{p\}; \{\bfk\})+\psi_{m}^*(-k^*_{n},...,-k^*_{n+m-2},S;-\{p\}; -\{\bfk\}) \right],
\end{align}
must cancel this $1/S^3$ contribution from the power spectrum. Now, given that \eqref{cotex} holds for all diagrams individually, we can take $m=n$ and consider an exchange diagram with the same sub-diagram on each side of the cut. For IR-finite $\psi_{n}$ that satisfy the contact COT, we can keep the $k_{a}$ real and then 
\begin{align}
&\psi_{n}(k_1,...,k_{n-1},S;\{p\}; \{\bfk\})+\psi_{n}^*(-k_1^*,...,-k_{n-1}^*,S;-\{p\}; -\{\bfk\}) \nonumber \\ = & \psi_{n}(k_1,...,k_{n-1},S;\{p\}; \{\bfk\})-\psi_{n}(k_1,...,k_{n-1},-S;\{p\}; \{\bfk\}),
\end{align}
by scale invariance. This is odd in $S$, and so must its Taylor expansion around $S=0$. Given that the square of this quantity has to cancel the $S^{-3}$ pole in the power spectrum, and that at tree-level there are only integer powers of momenta, we must require
\begin{align} \label{MLT}
\dfrac{\partial }{\partial S}\psi_{n}(k_1,...,k_{n-1},S;\{p\}; \{\bfk\})\Big|_{S=0}=0\,,
\end{align}
to ensure that the right-hand side of the COT is regular at $S=0$. To emphasise that the energy $S$ is now external, after the cut, we write this constraint as 
\begin{align} \label{MLTGeneral}
\boxed{\dfrac{\partial }{\partial k_{c}}\psi_{n}(k_1,...,k_n;\{p\};  \{\bfk\})\Big|_{k_{c}=0}=0\,, \quad (\text{$\Delta^+=3$, e.g. massless scalar or graviton})}\,,
\end{align}
where $c=1,\ldots n$ and we have relabelled $S$ to $k_{n}$. Any of the external fields in this $n$-point function could be used as an internal line in an exchange diagram and so \eqref{MLT} should hold for each external energy (as long as the corresponding field's mode functions are those in \eqref{ModeFunctionsMassless}). Note that here we are taking the derivative with respect to one of the energies while keeping all other variables fixed\footnote{We are therefore working away from the physical configuration.}, and we have used the fact that $n$-point functions, arising from manifestly local theories, are finite in the soft limit of an external energy. Wavefunction coefficients of fields with a $\Delta^+=3$ mode function (see \eqref{ModeFunctionsMassless}), as for example massless scalars and gravitons, arising from unitarity, manifestly local theories must satisfy \eqref{MLTGeneral}, which we call the \textit{Manifestly Local Test} (MLT).\\

\noindent It would be interesting to go beyond manifest locality because non-manifestly local interactions appear in the presence of massless spinning fields on non-trivial backgrounds, for example for a scalar field coupled to gravity (after solving the gravitational constraints \cite{Maldacena:2002vr}). We leave such explorations for future work and for the rest of this paper concentrate on manifestly local theories. 


\subsection{Manifest locality from the bulk representation} \label{MLTfromBulk}

As outlined in Appendix \ref{AppendixA}, the computation of tree-level diagrams in the bulk reduces to nested time integrals of the following form
\begin{align}
\psi_{n}(\{k\};\{p\};\{\bfk\}) \sim \int \left(\prod_A^V d \eta_A F_A\right) \left( \prod_a^n \partial_\eta^{\#} K_{\phi}(k_{a}) \right) \left(\prod_m^I \partial_\eta^\# G(p_m)\right) \,,
\end{align}
where the $F_A$'s collect the momentum dependence of the spatial derivatives in the $V$ vertices and we allowed for arbitrary time-derivative interactions denoted by $\partial_\eta^{\#}$ acting on $n$ external bulk-to-boundary propagators $K$ and $I$ internal bulk-to-bulk propagators $G$. Now consider the form of derivatives of $ K_{\phi}(k,\eta)$,
\begin{align}
\frac{d^N}{d\eta^N} K_{\phi}(\eta,k)=( i k)^N (1-N-ik\eta)e^{ik\eta}\,.
\end{align}
By direct calculation they satisfy 
\begin{align} \label{BulkMLT}
\dfrac{\partial}{\partial k} \left( \frac{d^N}{d\eta^N} K_{\phi}(\eta,k) \right) \Big|_{k=0}=0\,,
\end{align}
for \textit{any} non-negative integer $N\geq 0$. This property is inherited by $\psi_{n}$ when this $\partial_{k_a}$ derivative is taken keeping all other internal and external energies and rotation-invariant contractions fixed. This is ensured by the two following observations: (\textit{i}) the time integral does not affect this property since \eqref{BulkMLT} is valid for all $\eta$; (\textit{ii}) the vertices $F_A$ either depend on contractions of the momenta, which are kept fixed when we take the derivative in the MLT in \eqref{MLTGeneral} (they are independent variables), or they involve squared energies and so $\partial_k k^2$ vanishes at $k=0$. It therefore follows that \eqref{BulkMLT} implies the MLT \eqref{MLTGeneral}. So the MLT in fact follows from a neat property of the bulk-boundary propagators of massless mode functions ($\Delta^+=3$) in dS space. Notice that whether a $\psi_n$ satisfies the MLT or not is independent of which rotation-invariant contractions we choose as variables, since all these are kept fixed. Also, note that we have not assumed anything about the ultimate form of the $\psi_{n}$. Indeed, this argument holds for IR-finite and IR-divergent $n$-point functions alike.\\

\noindent The above argument is also easily adapted to bosonic spinning fields where we also allow for contractions with polarisation vectors in $F$. The mode functions for spinning fields with the same $\Delta^+$ as given in \eqref{delta} coincide with their scalar counterparts, the only difference being polarisation factors required to make up the indices and symmetries of the spinning field. If we keep the polarization factors fixed in taking the derivative in \eqref{MLTGeneral}, the behaviour of the bulk-boundary propagators is unaltered and so the MLT holds too for spinning fields with $\Delta^+=3$. Importantly, this includes massless gravitons.\\

\noindent In Section \ref{BootstrapThreePoint} we will show that the MLT allows us to bootstrap \textit{all} contact $3$-point functions arising from scalar self-interactions and those arising from coupling the scalar to an on-shell massless graviton. Then in Section \ref{PartialEnergyShifts} we will show that the MLT provides a key ingredient when bootstrapping complete $4$-point exchange diagrams.


\subsection{The Manifestly Local Test for massive fields} \label{MLTforMassive}

In the previous two subsections we have derived a manifestly local test for $n$-point functions involving fields with massless mode functions. In this subsection we extend those arguments to massive fields with particular attention paid to the case of conformally coupled fields. First we follow the derivation based on singularities and the Cosmological Optical Theorem, then we discuss the bulk perspective.\\

\noindent Equation \eqref{BulkToBulkSimplify} holds for any mass so again we are interested in the behaviour of $P_{\sigma}(S) \text{Im}K_{\sigma}(S,\eta)$ in the limit $S \rightarrow 0$. Given the general mode functions \eqref{ModeFunctionsGeneral}, one can show that  
\begin{align}
    \lim_{S\to 0^+}P_\sigma(S,\eta_0)\,\text{Im}K_{\sigma}(S,\eta)&=\dfrac{1}{2i}\lim_{S\to 0^+}\,\left(\sigma^+(S,\eta)\sigma^-(S,\eta_0)-\sigma^-(S,\eta)\sigma^+(S,\eta_0)\right)\, \nonumber\\ 
    &=\dfrac{1}{4\nu}(\eta\,\eta_0)^{3/2-\nu}\,\left((-\eta_0)^{2\nu}-(-\eta)^{2\nu}\right)\,,
\end{align}
and therefore the bulk-to-bulk propagator remains finite when the internal energy is taken soft. It follows that any $n$-point function arising due to the exchange of a massive field is also finite in this limit. Now turning to the right-hand side of the COT \eqref{cotex}, the power spectrum of the massive field behaves as
\begin{align} \label{PSMassive}
   \lim_{S\to 0^+} P_\sigma(S,\eta_0)=\dfrac{\eta_0^3\,\cot(\pi\nu)}{2\nu}-\dfrac{2^{-2-2\nu}\pi \eta_0^3}{\Gamma^2(\nu+1)\sin^2(\pi\nu)}(-S\eta_0)^{2\nu}-\dfrac{2^{2\nu-2}\,\eta_0^3\,\Gamma^2(\nu)}{\pi}(-S\eta_0)^{-2\nu}\,.\,
\end{align}
For a generic light field ($m<\frac{3}{2}H$), the last term in \eqref{PSMassive} dominates in the soft limit, and the power spectrum exhibits a non-analytic singularity around $S=0$, with the two exceptional cases of conformally coupled ($\nu=1/2$) and massless ($\nu=3/2$) mode functions, where the power spectrum behaves as $1/S$ and $1/S^3$ respectively. Given that the right-hand side of the COT must be regular in this limit, for $m<3 H/2$ we must have
\begin{align}
\label{MLTnu}
   \lim_{S\to 0^+} \left[\psi_{n}(k_1,...,k_{n-1},S;\{p\}; \{\bfk\})+\psi_{n}^*(-k_1^*,...,-k_{n-1}^*,S;-\{p\}; -\{\bfk\})\right]={\cal O}(S^\nu)\,,
\end{align}
where $\psi_n$ is a generic $n$-point function involving at least one massive field with energy $S$. 
For heavy fields ($m>3/2 H$), the power spectrum is regular at $S=0$, and so we have
\begin{align}
\label{MLTheavy}
   \lim_{S\to 0^+} \left[\psi_{n}(k_1,...,k_{n-1},S;\{p\}; \{\bfk\})+\psi_{n}^*(-k_1^*,...,-k_{n-1}^*,S;\{p\}; -\{\bfk\})\right]={\cal O}(S^0) \,.
\end{align}
Recall that for massless exchange, we replaced the complex energies with real ones and used scale invariance to arrive at the MLT \eqref{MLT}. However, when a massive field is exchanged, $\psi_n$ has a branch cut at negative energies and so it is important to keep the energies complex (or alternatively one could insert $k-i\epsilon$ as the modified energy in $\psi_n(-k_1^*,..,-k_n^*,S)$ in order to ensure the condition $\text{Im}(-k_a)<0$ is satisfied). We therefore take \eqref{MLTnu} and \eqref{MLTheavy} as the Manifestly Local Tests for $n$-point functions with external massive fields of generic mass. Given a candidate $n$-point function due to the exchange of a massive field, one can use these equations to check if such a wavefunction coefficient is describing the bulk dynamics of a manifestly local theory. These conditions might provide a useful tool to bootstrap massive exchange $n$-point functions, which are at the heart of the cosmological collider physics programme \cite{Arkani-Hamed:2015bza,Lee:2016vti}. \\

\noindent Among light fields, the case of a conformally coupled field ($m = \sqrt{2}H$) is particularly interesting as many of the expressions simplify. In this case, IR-finite $n$-point functions are analytic in all the energies and scale invariance dictates that $\psi_n^\vpi \sim k^{3-n}$ (in contrast to $k^3$ for the massless fields). Therefore, for real valued energies, equation \eqref{MLTnu} reduces to
\begin{align}
\label{MLTcc}
    \lim_{S\to 0^+}\psi_n(k_1,...,k_n,S;\{p\}; \{\bfk\} )- \psi_n(k_1,...,k_n,-S;\{p\}; \{\bfk\})={\cal O}(S^{1/2})\,,
\end{align}
 where we have used the contact COT \eqref{cotcon}. This condition is automatically satisfied for manifestly local theories. Indeed, $\psi_{n}$ is analytic in all external energies and so the left-hand side of \eqref{MLTcc} scales as $\mathcal{O}(S)$, or softer. The MLT for a conformally coupled field is therefore the simple requirement that $\psi_{n}$ is finite as the energy of such an external field is taken soft. This can also easily be derived from the bulk representation since, given \eqref{ModeFunctionsCC}, we have
\begin{align}
    \lim_{S\to 0}K_{\vpi}(S,\eta)=\dfrac{\eta}{\eta_0}\left(1+i\,S\,(\eta-\eta_0)+\mathcal{O}(S^2)\right)\,,
\end{align}
which does not have vanishing coefficients in the first three terms of the expansion. In conclusion, when a conformally coupled field with momentum $\bfk_a$ appears in $\psi_n$, we have the following weaker version of the MLT 
\begin{align} \label{MLTConformal}
\dfrac{\partial }{\partial k_a}\psi_{n}(k_1,...,k_{n};\{p\};  \{\bfk\})\Big|_{k_a=0}=\text{finite}\,, \quad (\text{$\Delta^+=2$, e.g. conformally coupled scalar}).
\end{align}
The existence of the ${\cal O}(S)$ term in $K_\vpi$, which is absent for a massless field, explains why the MLT for conformally coupled field is different to that of a massless field. In Section \ref{Results}, we will show that the differing forms of the two MLTs can explain an intriguing difference between the $4$-point functions for massless and conformally coupled fields due to graviton exchange. 


\section{Bootstrapping 3-point functions using the MLT} \label{BootstrapThreePoint}

In this section we illustrate the power of the MLT by using it in combination with  a sub-set of the Bootstrap Rules of \cite{BBBB} to bootstrap $3$-point functions of spectator scalars and gravitons, which capture inflationary bispectra to leading order in slow-roll. We first consider a massless scalar and concentrate on the $3$-point function due to the scalar's self-interactions, $\psi_{\phi \phi \phi}$. We then consider general manifestly local interactions between a scalar and the graviton $\gamma$ and bootstrap $\psi_{\phi \phi \gamma}$. Finally, we do the same for a conformally coupled scalar and bootstrap $\psi_{\varphi \varphi \gamma}$. In all these cases we are able to bootstrap both the IR-finite, which is $  \eta $-independent, and the IR-divergent part, which depends on regulated position of the boundary. We do not consider the cubic self-interactions for a conformally coupled scalars since the COT ensures that the real part of $\psi_{\varphi \varphi \varphi}$ is zero \cite{COT}.

\paragraph{Comparison with the previous literature} Before proceeding, let's compare our derivation with the boostless bootstrap for the bispectrum presented in \cite{BBBB}. Here we will use some of the Bootstrap Rules proposed there, with the minor difference that here we use the language of the wavefunction rather than that of correlators. We focus on the real part of the wavefunction and briefly comment on how to bootstrap the imaginary part of contact diagrams in Section 
\ref{image}. The bootstrap rules allow us to write a simple bootstrap Ansatz with some numerical coefficients, whose number increases as we consider higher and higher dimension operators, just like in the Lagrangian description of EFT's. Our derivation presents some advantages and improvements over that in \cite{BBBB}:
\begin{itemize}
    \item We will not impose any soft limits (Rule 6) or the amplitude limit (Rule 3). Instead, we fix all the free parameters in the bootstrap Ansatz with the manifestly local test (MLT). This means that our derivation can be straightforwardly extended to multifield inflation, where there are no soft theorems in the most general case (but see \cite{Hui:2018cag}). It should be stressed that here we calculate the $n$-point function of a spectator scalar, as opposed to $\zeta$, which differ by slow-roll suppressed terms.
    \item We will be able to fix all free coefficients with the MLT in all manifestly local cases. This is contrast with the results of \cite{BBBB} where for the zero-derivative scalar bispectrum only five of the six parameters in the Ansatz could be fixed with soft limits (the sixth could be fixed with boost invariance, which we never invoke here). Furthermore, for the scalar bispectrum to higher derivatives in \cite{BBBB}, the constraint of a manifestly local amplitude was given explicitly only for the amplitude corresponding to the leading $k_T$ pole. This missed some constraints starting at four derivatives. We will show that the MLT is able to enforce these constraints also on the amplitudes appearing at subleading poles.  
    \item For the scalar-scalar-graviton bispectrum, in \cite{BBBB} the ad hoc assumption had to be made that the bispectrum is symmetric in all momenta, which does not follow from Bose symmetry because the fields are distinct. In contrast, here we will see that this property is a consequence of the MLT.
\end{itemize}
Finally, our approach here has one shortcoming as compared to the that in \cite{BBBB}: since we only discuss manifestly local theories, we miss those gravitational interactions that arise in the off-shell description from integrating out constrained fields (the lapse and the shift in the ADM formalism). We hope to return to this problem in the future.


\subsection{Self-interactions of a massless scalar}
For a single massless scalar, IR-finite $3$-point functions only depend on the three external energies $k_{1},k_{2},k_{3}$. These energies appear in symmetric combinations by Bose symmetry and so without loss of generality we apply the manifestly local test (MLT) to $k_{3}$ only, and require
\begin{align} \label{MLT3Point}
\dfrac{\partial }{\partial k_{3}}\psi_3^{(p)}(k_1,k_2,k_{3})\Big|_{k_{3}=0}=0\,,
\end{align}
where the superscript $(p)$ denotes the degree of the leading $k_{T}$ pole which is equal to the largest number of derivatives in the EFT expansion \cite{BBBB} (unless this vanishes by symmetry, as e.g. in DBI \cite{GJS}). We have dropped the $\phi \phi \phi$ subscript since it will be clear throughout this section that we are considering a scalar self-interaction. We will now use the MLT to bootstrap $\textit{all}$ $3$-point functions for \textit{any} $p$. Initially our discussion will concentrate on the real part of the wavefunction but then we will turn to the imaginary part. \\

\noindent The Boostless Bootstrap Rules of \cite{BBBB} enforce the following properties:
\begin{itemize}
    \item Homogeneity, isotropy and scale invariance (but no assumption about dS boosts): this enforces
    \begin{align}
\psi_n &= \sum_\text{contractions} \left[ \eps^{h_1}(\bfk_1)\eps^{h_2}(\bfk_2)\eps^{h_3}(\bfk_3)\bfk_1^{\alpha_1}\bfk_2^{\alpha_2}\bfk_3^{\alpha_3}\right] \tilde \psi_n \\
&= \sum_\text{contractions} \text{(polarization factor)} \times \text{(trimmed wavefunction coefficient)}\,.
\end{align}
  Note that for scalars the trimmed wavefunction $\tilde\psi_n$ coincides with $\psi_n$.
    \item $\Delta^+=3$ de Sitter mode functions, e.g. massless scalars and gravitons: this enforces that the trimmed wavefunction $\tilde \psi_n$ is a rational function with overall momentum scaling $k^3$.
    \item The amplitude limit: this enforces the residue of the highest $k_T$-pole to be fixed in terms of a corresponding amplitude \cite{Maldacena:2011nz,Raju:2012zr} (see \cite{COT} for a derivation and an explicit formula). As we will see, we do need to use this rule since the MLT enforces this local amplitude limit automatically.
    \item Bose symmetry: this enforces invariance under permutations of the momenta of identical fields.
    \item Locality and the Bunch-Davies vacuum: this enforces that the only allowed poles for contact diagrams are in the total energy, $1/k_T^p$ with $p=1+\sum_A (\Delta_A-4)$ where the sum is over all vertices (only one for a $3$-point function) and $\Delta_A$ is their mass dimension (three plus the number of derivatives for cubic bosonic interactions). This is necessary but not sufficient for locality.
\end{itemize}
These Bootstrap Rules allows us to write a relatively simple \textit{bootstrap Ansatz}. In particular, the part of the cubic wavefunction coefficient of three scalars that survives when we compute the correlator must take the form
\begin{align} \label{ThreePointAnsatz}
\psi_3^{(p)}(k_{1},k_{2},k_{3})= \frac{1}{k_{T}^p} \sum_{n=0}^{\lfloor \frac{p+3}{3} \rfloor} \sum_{m=0}^{\lfloor \frac{p+3-3n}{2} \rfloor} C_{mn} k_{T}^{3+p-2m-3n} e_2^m e_3^n,
\end{align}
where $e_{2}$ and $e_{3}$ are the elementary symmetric polynomials in Section \ref{Conventions}, $\lfloor \dots \rfloor$ is the floor function, and $C_{mn}$ are constant coefficients that are real by unitarity \cite{COT} and correspond to linear combinations of coupling constants in the bulk. Indeed, the COT tells us that \cite{COT}
\begin{align}
\psi_{3}(k)+\psi_{3}(-k)^{\ast}=0\,,
\end{align} 
and in \eqref{ThreePointAnsatz} scale invariance has ensured that $\psi_{3}(k) = - \psi_{3}(-k)$ and so we need $\psi_{3}(k) = \psi_{3}(k)^{\ast}$. Given that these couplings are real, these contributions to the wavefunction also contribute to the correlator which is given by the real part of $\psi_{3}$. The total number $N_{\text{ansatz}}$ of free coefficients in the bootstrap Ansatz \eqref{ThreePointAnsatz} is equal to the number of non-negative integer solutions to $3+p \geq 2m+3n$ which is given by \cite{Mahmoudvand}
\begin{align}
N_{\text{ansatz}}(p) = \sum_{q=0}^{\lfloor \frac{p+3}{2} \rfloor} \left(1+ \lfloor \frac{p+3-2 q}{3} \rfloor \right). 
\end{align}
For $p=0$, $\psi_{3}$ can also contain an IR-divergent logarithm which we will consider separately. To compute all possible $\psi_3$ we now fix the $C_{mn}$ using the MLT. \\

\noindent Now, the MLT \eqref{MLT3Point} imposes the following recursion relations on the $C_{mn}$:
\begin{align} \label{MLTConstraints}
&\sum_{m=0}^{\lfloor \frac{p+3}{2}\rfloor}\,m\,C_{m0\,}\, \me_1^{4-2m} \, \me_2^{m-1} +\sum_{m=0}^{\lfloor \frac{p+3}{2}\rfloor}\,(3-2m)\,C_{m0}\, \me_1^{2-2m} \, \me_2^m +\sum_{m=0}^{\lfloor \frac{p}{2}\rfloor}\,C_{m1}\,\me_1^{-2m}  \, \me_2^{m+1}\,=0\,,
\end{align}
where we have used the non-vanishing partial derivatives with respect to $k_{3}$
\begin{align}
\dfrac{\partial e_1}{\partial k_{3}}=1\,,\qquad \dfrac{\partial e_2}{\partial k_{3}}=\me_1\,,\qquad \dfrac{\partial e_3}{\partial k_{3}}=\me_2\,,
\end{align}
and have defined the elementary symmetric polynomials for the two remaining variables,
\begin{align}\label{defmathbb}
\mathbb{e}_1=k_1+k_2\,,\qquad \mathbb{e}_2=k_1\,k_2\,.
\end{align}
It is simple to see that the second term in \eqref{MLTConstraints} contains all the powers of $\mathbb{e}_1$ that are contained in the first and third terms, and so the total number of constraints is equal to the number of terms in the sum $\sum_{m=0}^{\lfloor \frac{p+3}{2}\rfloor}\,(3-2m)\,C_{m0}\, \me_1^{2-2m} \, \me_2^m$ which is simply   
\begin{align}
N_{\text{constraints}}(p)=1+\lfloor \frac{p+3}{2}\rfloor\,.
\end{align}
After imposing all constraints, the final number of IR-finite $3$-point functions arising from manifestly local theories is therefore
\begin{align} \label{Ntotal}
N_{\text{total}}(p) = N_{\text{ansatz}}(p) - N_{\text{constraints}}(p) = \sum_{q=0}^{\lfloor \frac{p+3}{2}\rfloor}\, \lfloor \dfrac{p+3-2q}{3}\rfloor. 
\end{align}
Let's now look at the constraints in more detail. By looking at the terms with the largest power of $\me_2$ in \eqref{MLTConstraints} we conclude that
\begin{align}
\text{odd $p$}&: \quad  C_{\frac{p+3}{2} 0} = 0, \label{LeadingConstraint1} \\
\text{even $p$}&: \quad C_{\frac{p}{2}1} = (p-1) C_{\frac{p+2}{2}0}. \label{LeadingConstraint2}
\end{align}
For odd $p$, \eqref{LeadingConstraint1} tells us that the residue of the leading $k_{T}$ pole cannot be independent of $e_{3}$. This constraint can be understood according to the following argument \cite{BBBB}. The residue of the highest $k_T$ pole is a cubic scattering amplitude $\mathcal{A}_{3}$. For this amplitude to be manifestly local i.e. not contain any inverse powers of external energies, the numerator of $\psi_{3}$ must contain at least one power of $e_3$. Indeed, a tree-level $3$-point amplitude for a single scalar in a boost-breaking theory is a symmetric polynomial in the energies of the external particles \cite{PSS}. A complete basis is provided by the two symmetric polynomials $e_{2}$ and $e_{3}$ since for scattering amplitudes energy is conserved: $k_{T}=0$. A general amplitude for a manifestly local theory therefore takes the schematic form $\mathcal{A}_{3} \sim e_{2}^{\alpha} e_{3}^{\beta}$ where $2\alpha+3\beta = p$. For massless fields we have \cite{COT}
\begin{align}\label{amplim}
\lim_{k_{T} \rightarrow 0} \text{Re}(\psi_{3}) \sim e_{3}\frac{\text{Re}(i^p \mathcal{A}_{3})}{k_{T}^p},
\end{align}
and so at least one power of $e_{3}$ should appear on the leading $k_{T}$ pole of the $3$-point function, as ensured by the MLT. For even $p$, scale invariance guarantees that the leading $k_{T}$ pole contains at least one factor of $e_{3}$ and so the corresponding amplitude is guaranteed to be manifestly local. However, there is a sub-leading $k_{T}$ pole of degree $p-1$ whose residue is independent of $e_{3}$. If the coefficient of this term was unconstrained then one could cancel the leading $k_{T}$ pole such that this sub-leading pole became leading which would in turn yield an amplitude that could not come from a manifestly local theory. The MLT deals with this and indeed for even $p$, \eqref{LeadingConstraint2} fixes the coefficient of the $k_{T}$ pole of degree $p-1$ in terms of the coefficient of the leading $k_{T}$ pole. The remaining constraints, for both odd and even $p$, are similar in nature and constrain the coefficients of the $e_{3}$ independent terms. Let's now look at some examples to illustrate the power of the MLT.  

\paragraph{$\mathbf{p=0}$} To begin with, consider the bootstrap Ansatz for $p=0$
\begin{align}
    \psi_3^{(0)}=C_{00} k_T^3+C_{10} k_T e_2 +C_{01} e_3+\log(-\eta_{0} k_T) \left[ \tilde C_{00} k_T^3+\tilde C_{10} e_2 k_T+\tilde C_{01} e_3 \right]\,,
\end{align}
where we allowed for a $\log(-k_T \eta_{0})$, representing the liming case $p\to 0$. The constraints from the MLT give
\begin{align}
    C_{10} &= -3 C_{00} - \tilde C_{00},& \tilde C_{10}&= -3 \tilde C_{00},& C_{01} &= 3 C_{00} + 4 \tilde C_{00},& \tilde C_{01} &= 3 \tilde C_{00}\,,
\end{align}
and hence at this order there are only two allowed shapes
\begin{align}
    \psi_3^{(0)}&=C_{00}(k_{T}^3 - 3 k_{T}e_{2} + 3 e_{3}) + \tilde C_{00} \left[ 4 e_{3} - e_{2} k_T + (3 e_3 - 3 e_2 k_T + k_T^3) \log(-k_T \eta_{0} ) \right]\\
    &=\tilde C_{00}  \left[ 4 e_{3} - e_{2} k_T + (3 e_3 - 3 e_2 k_T + k_T^3) \log(-k_T \eta_{0} / \mu) \right]\,.
\end{align}
A few comments are in order. The first term, which is proportional to $C_{00}$, is the well-known ``local" non-gaussianity \cite{LNG} that arises by taking the free theory for $\phi({\bf x})$ and performing the local field redefinition $\phi({\bf{x}}) \rightarrow \phi({\bf{x}}) + \phi^2({\bf{x}})$ to leading order. This does not alter the $S$-matrix and indeed this term is finite on the total energy pole, $k_T \to 0$. Note that this is the unique field redefinition that generates a manifestly local, and scale invariant $3$-point function from the free theory. The reason is that for any field redefinition $\Delta \phi$, scale invariance requires us to have as many derivatives as inverse derivatives\footnote{This is because in real space our massless scalar is a singlet under a scale transformation; only in momentum space does it transform.}, but manifest locality forbids any inverse spatial Laplacians, and so the only possibility is a zero-derivative polynomial redefinition. For future reference we define 
\begin{align} \label{localNG}
\psi_{3}^{\text{local}} = k_{T}^3 - 3 k_{T}e_{2} + 3 e_{3}.
\end{align}
The second term, which is proportional to $\tilde C_{00}$, corresponds to the polynomial interaction\footnote{It was noticed in \cite{Seery:2008qj} that the result for this interaction in (25) of \cite{Zaldarriaga:2003my} missed a few terms. For a corrected and pedagogical derivation see Sec 3.3 of these \href{https://www.dropbox.com/s/nwau7po9w11w6kx/field_theory_in_Cosmology.pdf?dl=0}{lecture notes}.} $\phi^3$. Indeed this term is singular on the total energy pole and in fact it has a corresponding amplitude\footnote{In $p=0$ case the amplitude limit in \eqref{amplim} is modified, as derived in \cite{COT}.}, which is just a constant. In slow-roll inflation this term is present and is second order in the slow-roll parameters; however, in the limit $\eps \to 0$ this is the only surviving term and it enjoys an approximated conformal invariance \cite{Pajer:2016ieg}. When using soft theorems instead of the MLT to fix the free parameters in the bootstrap Ansatz, one of the above constraints is missed and one needs to resort to conformal invariance \cite{BBBB}. Here we show that the MLT has no trouble dealing with this interaction or with the log. Finally, we note that the real and rational part of the wavefunction automatically satisfies the unitarity constraint imposed by the COT. Conversely, the $\log$ term does not satisfy the COT by itself. Indeed, the COT demands that log term always appears in the combination $\log(-k_T \eta_{0}) + i \pi /2$ (see \cite{COT} for a detailed discussion). One might expect the log term to come with a term of the form $\gamma_{E} \times \text{Poly}_{3}$ where $\gamma_{E}$ is Euler's constant. From the bulk calculation one can verify that this polynomial is precisely the local non-gaussianity in \eqref{localNG} and so it is correctly captured by our derivation.


\paragraph{$\mathbf{p=1}$} The new Ansatz for $p=1$ is expanded to
\begin{align}
  \text{Ansatz: }  \psi_3^{(1)}=\frac{1}{k_T} \left[ C_{00} k_T^4+C_{10} k_T^2 e_2 +C_{01} k_Te_3 + C_{20} e_2^2\right] \,,
\end{align}
where we neglected to write the log because the MLT obliges it to only arises at $p=0$, and we have already discussed that case above. The MLT yields the following constraints
\begin{align}
C_{20} = 0, \qquad 3C_{00} = C_{01} = -C_{10},
\end{align}
and so there are no new $\psi_3$ allowed at this level. This is to be expected because there exist no boost-breaking cubic amplitudes for three scalars. The only term allowed by dimensional analysis and Bose symmetry would be $k_T$, but this is the total energy and vanishes for amplitudes. Equivalently, the $k_T$ factor cancels the $k_T $ pole hence reducing us to the $p=0$ case. Remarkably, the cubic scalar wavefunction generated by gravitational interactions has precisely $p=1$ \cite{Maldacena:2002vr}. Here we cannot see it because it is associated with a non-manifestly local amplitude. This is to be expected since this interaction arises from solving the GR constraints for the lapse and the shift, which requires inverting the Laplace operator. For an extended discussion of this non-manifestly local contribution see \cite{BBBB}.


\paragraph{$\mathbf{p=2}$} Now consider $p=2$, with the Ansatz from \eqref{ThreePointAnsatz}. The MLT yields the following constraints
\begin{align}
C_{20} = C_{11}, \quad C_{10} =- 3C_{00} , \quad C_{01} = -2C_{11} + 3C_{00},
\end{align}
so the only new $3$-point function is 
\begin{align}
\psi_{3}^{(2)} = \frac{C_{11}}{4} \psi_{3}^{\text{DBI}} + \text{lower $k_T$-singularity},
\end{align}
where we have defined
\begin{align} \label{DBI}
\psi_{3}^{\text{DBI}}  = -k_{T}^3 + 3k_{T}e_{2} - 11 e_{3} + \frac{4 e_{2}^2}{k_{T}}+ \frac{4 e_{2}e_{3}}{k_{T}^2}.
\end{align}
This is the $3$-point function of the DBI limit of the EFT of inflation \cite{GJS,DBIInflation,EFTofI}. As explained in \cite{GJS}, despite the EFT of inflation operators having three-derivatives, the leading $k_{T}$ pole for the DBI limit is degree 2 due to its vanishing amplitude in the flat-space limit, which in turn is due to the non-linearly realised $ISO(1,4)$ symmetry in that limit. 

\paragraph{$\mathbf{p=3}$} Now consider $p=3$. The MLT yields the following set of constraints
\begin{align}
C_{30} = 0, \quad C_{20} = C_{11}, \quad C_{10} = -3C_{00}, \quad C_{01} = 3C_{00} -2C_{11}.
\end{align}
Hence, the new $3$-point function with a non-vanishing $k_T^{-3}$ pole is simply
\begin{align}
\psi_{3}^{\text{EFT1}} &= \frac{e_{3}^2}{k_{T}^3} + \text{lower $k_T$-singularity}. \label{PsiThreeEFT1} 
\end{align}
This is the $3$-point function arising from the boost-breaking $\phi'^{3}$ (EFT1) term in the EFT of inflation \cite{EFTofI}. There is a second three-derivative self-interaction for this goldstone mode, namely $\phi'(\nabla \phi)^2$ (EFT2). This yields a wavefunction coefficient that is a linear combination of our $p=2$ and $p=3$ MLT solutions. Indeed, $\psi_{3}^{\text{DBI}} = 12 \psi_{3}^{\text{EFT1}} - \psi_{3}^{\text{EFT2}}$, which is the unique combination in the EFT of inflation for which the leading $k_{T}$ pole is degree 2 rather than 3.

\paragraph{$\mathbf{p=4}$} As a final example consider $p=4$. The MLT constraints are
\begin{align}
C_{21} = 3C_{30}, \quad C_{10} = -3C_{00}, \quad C_{01} = 3C_{00}-2C_{20}, \quad C_{20}= C_{11}+3C_{30}, 
\end{align}
and after imposing these constraints we can write
\begin{align}
\psi_{3}^{(4)} =C_{30}\psi_{3}^{\phi \phi''^2} + \text{lower $k_T$-singularity},
\end{align}
where we have defined
\begin{align}
\psi_{3}^{\phi \phi''^2} = \frac{1}{k_{T}^4}(-3k_{T}^2e_{2}e_{3}+k_{T}e_{2}^3+3e_{2}^2e_{3}).
\end{align}
This is the wavefunction coefficient arising from a $\phi \phi''^2$ operator in the bulk. There are indeed other four-derivative operators one can write down at cubic order in $\phi$, but they are all degenerate with $\phi \phi''^2$, up to the presence of lower derivative operators, after integration by parts and use of the scalar's equation of motion. This is made completely manifest in our bootstrap approach since the MLT only allows for a single wavefunction coefficient with a leading $1/k_{T}^4$ pole.\\

\noindent We have therefore seen that a sub-set of the Bootstrap Rules of \cite{BBBB} combined with the Manifestly Local Test (MLT) provides a conceptually transparent and computationally very efficient way to derive bispectra. Not only does the MLT ensure that the leading $k_{T}$ poles yield manifestly local amplitudes, it also fixes the \textit{full shapes} of manifestly local $3$-point functions. For example, the highly non-trivial structure of \eqref{DBI} is completely fixed by the MLT and any deviations from these tunings for the sub-leading $k_{T}$ poles would represent a deviation from manifest locality and/or unitary time evolution in the bulk. Our results therefore contain the full $3$-point functions for the EFT of inflation up to any order in derivatives, which captures some constraints that were missed in \cite{BBBB}. Furthermore, as shown in appendix \ref{AppendixC}, the number of real wavefunction coefficients is equal to the number of amplitudes plus one,
\begin{align}
    N_{\text{total}}(p) = N_{\text{amplitudes}}(p) + 1= 1 + \sum_{q=0}^{\lfloor \frac{p+3}{2}\rfloor}\, \lfloor \dfrac{p+3-2q}{3}\rfloor\,,
\end{align}
to any order $p$ in derivatives (note that here we have included the logarithmic term in the counting). The one extra $3$-point function can be traced back to the fact that the $S$-matrix is invariant under perturbative field redefinitions whereas $\psi_3$ is not. In particular, the much studied local non-Gaussianity arises from the only scale invariant and manifestly local field redefinition at this order: $\phi({\bf x}) \rightarrow \phi({\bf x}) + \phi^2({\bf x})$. Each of the remaining $3$-point functions are tied to an amplitude that in turn can be derived from a manifestly local operator in Minkowski space. This implies that each of the $3$-point functions we have derived from the MLT come from manifestly local operators in dS space.

\paragraph{The imaginary part \label{image}} So far we concentrated on the real part of the wavefunction since this is what contributes to the correlator\footnote{This is true for parity even interactions.  For parity odd interactions the correlator picks up the imaginary part of the wavefunction coefficient.  However, parity odd interactions have more derivatives and do not generate the imaginary IR-divergent terms that we bootstrap below.}.  However, our methods can also constrain the imaginary part of the wavefunction,  where there can be inverse powers of $\eta_{0}$.  First of all, notice that by scale invariance we have
\begin{align}
\psi_{3}(\lambda k,\eta_{0} \lambda^{-1}) = \lambda^{3}\psi_{3}(k,\eta)\,.
\end{align}
Furthermore, the only $k$-dependent denominators one can have are powers of $k_{T}$, due to the choice of the Bunch-Davies vacuum, and the residues of these poles must contain an amplitude which cannot depend on time. Therefore the most generic Bootstrap Ansatz in the late-time limit $  \eta_{0}\to 0 $ is
\begin{align} \label{3pointGeneral}
\psi_{3}=\frac{\text{Poly}_{3+p}}{k_{T}^{p}}+\log(-\eta_{0}k_{T})\text{Poly}_{3}+\frac{\text{Poly}_{2}}{\eta_{0}}+\frac{\text{Poly}_{1}}{\eta_{0}^{2}}+\frac{\text{Poly}_{0}}{\eta_{0}^{3}}\,,
\end{align}
where $\text{Poly}_{n} $ are polynomials in momenta of degree $  n $. We have already bootstrapped the first two terms in this Ansatz so here we concentrate on the latter three where, a priori, these polynomials can have complex coefficients. Now the COT dictates \cite{COT}
\begin{align} \label{COTContact}
\psi_{3}(k)+\psi_{3}(-k)^{\ast}=0\,,
\end{align}
and so $  \text{Poly}_{2} $ and $  \text{Poly}_{0} $ must be pure imaginary since they are invariant under $k \rightarrow -k$, while the $  \text{Poly}_{1} $ must be real. We will now constrain these polynomials using the MLT. \\

\noindent The numerator of the most divergent term, $  \eta_{0}^{-3} $, must be a number by scale invariance, and so it is not constrained further by the MLT. We will denote it as $a$. It is easy to see that $\text{Poly}_{1} $ must vanish. Indeed, there is no linear combination of the energies that can satisfy the MLT, since the latter forbids all linear terms. Finally, the least trivial case is that of $\eta_{0}^{-1}$ where by writing down a general ansatz for $\text{Poly}_{2}$ and imposing the MLT, we find
\begin{align}
\text{Poly}_{2}= i b \left( k_{1}^{2}+ k_{2}^{2}+ k_{3}^{2} \right) = i b(k_{T}^2 - e_{2}),
\end{align}
where $b$ is a real number. Here we have imposed Bose symmetry given that in this section our focus is on the three-point function of identical scalars. However, if we had considered the three-point function for three distinct fields that all satisfy \eqref{MLT3Point}, $\text{Poly}_{1}$ would still have to be set to zero, whereas for $\text{Poly}_{2}$, $k_{1}^2$, $k_{2}^2$ and $k_{3}^2$ could have arbitrary coefficients. So a combination of the MLT and the COT tells us that for three identical scalars
\begin{align}\label{bandc}
\lim_{\eta_{0}\to 0}\psi_{3}\supset i b \frac{\left(k_{T}^2 - e_{2} \right)}{\eta_{0}} + i \frac{a}{\eta_{0}^{3}}\,.
\end{align}
This result means that the late-time, time-dependent part of the full $\psi_{3}$ (other than the log term) is actually purely imaginary. This ensures that the correlator does not see any of these late-time divergences since the correlator only depends on the real part of $\psi_{3}$. Notice that, when converted to real space, both contributions above are contact terms that vanish at separated points. This is to be expected from experience with AdS/CFT where these contact terms are non-universal and are indeed related to non-universal divergences that depend on how the IR limit is regulated.\\

\noindent In fact, this result can quite easily be extended to any $n$-point functions. Indeed, with three spatial dimensions, all $n$-point wavefunction coefficients satisfy 
\begin{align}
\psi_{n}(\lambda k,\eta_{0} \lambda^{-1}) = \lambda^{3}\psi_{n}(k,\eta)\,,
\end{align}
and so any poles at $\eta_{0} = 0$ take the form shown in \eqref{3pointGeneral} with the polynomials a function of the $n$ external energies but still with the shown degrees. Now the COT $\psi_{n}(k)+\psi_{3}(-k)^{\ast}=0$ is valid for any $n$-point function \cite{COT} and so it remains true that only the $1 / \eta_{0}^2$ term can have real coefficients and therefore have the chance of appearing in the correlator. However, if all external fields are subject to the MLT, which is the case if they have massless mode functions, then the MLT ensures that no such $\text{Poly}_{1}$ is allowed and therefore no $\eta_{0} = 0$ poles appear in the correlator.

\subsection{A massless scalar coupled to gravity}
\label{grav1}
We now consider the case where our massless scalar $\phi$ couples to gravity. We consider minimal coupling to a massless graviton, and bootstrap the $\psi_{\phi \phi \gamma}$ for $p=2$, corresponding to minimal coupling to gravity as in General Relativity. In contrast, note that we cannot bootstrap $\psi_{\phi \phi \phi}$ from graviton interactions because this arises after integrating out the non-dynamical lapse and shift in the ADM formalism \cite{Maldacena:2002vr}, which leads to non-manifestly local interactions involving inverse Laplacians. A detailed discussion of this scalar bispectrum appeared in \cite{BBBB}, together with other bispectra involving the graviton. In a future publication we will provide a general result for any bispectra of spinning particles to all orders in derivatives \cite{JPSS}. \\

\noindent The bootstrap rules of \cite{BBBB}, which we reviewed earlier for scalars, apply to gravitons too and so we only need to make a few small tweaks to our previous Ansatz. There are two main differences compared with the scalar case. First, by little group scaling, $\psi_{\phi \phi \gamma}$ must contain a polarisation factor $\epsilon^{h}_{ij}({\bf{k}}_{3})$ where $h = \pm 2$ is the graviton's helicity. Now to form a little group invariant we need to contract these two indices with $k_{1}^i$ or $k_{2}^i$ given that the polarisation tensor is transverse and traceless. By momentum conservation, each of the three choices are degenerate so we write $\epsilon_{ij}({\bf{k}}_{3})k^{i}_{1}k^{j}_{2}$ as the appropriate little group invariant without loss of generality. Second, there is no reason for this $3$-point function to be symmetric under the exchange of one of the scalar's momentum and the graviton's momentum since Bose symmetry only applies to identical fields. This is an improvement over the treatment in \cite{BBBB} where such symmetry needed an ad hoc assumption. We therefore only assume symmetry under the interchange of $k_{1}$ and $k_{2}$. So the variables we will use to write the boostrap Ansatz are $\mathbb{e}_1$, $\mathbb{e}_2$ (defined in \eqref{defmathbb}) and $k_3$. Taking these into account and including a possible logarithm, our general Ansatz for the real part of $\psi_{3}$ up to $p=2$ is 
\begin{align} \label{ThreePointAnsatzGraviton}
\psi_{\phi \phi \gamma}^{(2)}(k_{1},k_{2},k_{3})= \frac{\epsilon_{ij}({\bf{k}}_{3}) k^{i}_{1}k_{2}^{j}}{k_{T}^2}[D_{00}k_{3}^3 + D_{10}k_{3}^2\mathbb{e}_1 + D_{01}k_{3}\mathbb{e}_2 + D_{20}k_{3}\mathbb{e}_1^2+D_{11}\mathbb{e}_2 \mathbb{e}_1 \nonumber \\ + D_{30}\mathbb{e}_1^3 + k_{T}^2\log(-k_{T}\eta_{0})(\tilde{D}_{00}k_{3}+\tilde{D}_{10}\mathbb{e}_1)],
\end{align}
where in the following we will drop the $\phi \phi \gamma$ subscript. We now have to apply the Manifestly Local Test (MLT) to one of the scalar energies and the graviton's energy. Let's first consider the scalar. Without loss of generality we consider $k_{2}$ and so the MLT is 
\begin{align} 
\dfrac{\partial }{\partial k_{2}}\psi_3^{(2)}(k_1,k_2,k_{3})\Big|_{k_{2}=0}=0\,,
\end{align}
which yields the constraints
\begin{align} \label{ScalarConstraints}
D_{30} = -D_{11}, \qquad
2D_{11}=D_{01}, \qquad
D_{01} = D_{10}-2D_{20}, \qquad
D_{10}=2D_{00}, \qquad \tilde{D}_{10} = 0.
\end{align}
Now for the graviton we have
\begin{align} 
\dfrac{\partial }{\partial k_{3}}\psi_3^{(2)}(k_1,k_2,k_{3})\Big|_{k_{3}=0}=0\,,
\end{align}
which, after imposing \eqref{ScalarConstraints}, yields
\begin{align}
D_{20}=-2D_{00}, \qquad \tilde{D}_{00} = 0.
 \end{align}
After imposing all of these constraints we are left with a unique $p=2$ $3$-point function which turns out to be \textit{fully symmetric} and can be written as 
 \begin{align} \label{ThreePointFinalGraviton}
\psi_3^{(2)}= \frac{D_{00}\epsilon_{ij}({\bf{k}}_{3}) k^{i}_{1}k_{2}^{j}}{k_{T}^2}(k_{T}^3- k_{T}e_{2}- e_{3}),
\end{align}
which is the $3$-point function arising from the familiar minimal coupling $\gamma^{ij}\partial_{i}\phi \partial_{j}\phi$. \\

\noindent A few comments are in order. First, notice that the MLT forces the trimmed wavefunction of $\psi_{\phi\phi\gamma}$, namely what's left after stripping off the polarization factor $\epsilon_{ij} k^{i}_{1}k_{2}^{j}$, to be fully symmetric under permutation of all three momenta, which is not a consequence of Bose symmetry. From the bulk point of view, this property arises from the fact that the graviton has the same mode functions as the massless scalar and so the integral over time of the bulk-to-boundary propagators is fully permutation invariant. It is remarkable that the MLT enforces this property as well. This happens because we have imposed the MLT for the graviton as well as for the scalars, which brings in the information that their mode functions are the same. Indeed this will not be the case in next example involving two conformally coupled scalars and a graviton. Second, the MLT teaches us that the lowest derivative interaction has two derivatives and correspond to minimal coupling in General Relativity. This is true even though we allow for boost breaking interactions. This is a particular case of a more general set of results obtained in \cite{Creminelli:2014wna,Bordin:2017hal,Bordin:2020eui}, where it was also concluded that the first boost breaking $\psi_{\phi \phi \gamma}$ generated from coupling to the inflaton foliation of time comes at three-derivatives. In that work the authors work at the level of the Lagrangian and use field redefinitions to remove redundant couplings in the EFT of inflation. Our on-shell approach should be able to reproduce in an efficient manner all their more general results, including the graviton non-Gaussianities. We plan to discuss this in \cite{JPSS}. Finally, we note that this $3$-point function was derived in \cite{Mata:2012bx} using de Sitter symmetries. Here we see that the MLT efficiently forces the result even when no assumptions about de Sitter boosts have been made. \\

\noindent In this case one can also bootstrap the imaginary part of the wavefunction. Given that the polarisation factor already scales as $k^2$, the only term we haven't yet discussed has a $1  / \eta_{0}$ late-time singularity. The COT ensures that its coefficient is a pure imaginary number which of course passes the MLT. Indeed, as explained in \cite{Gordon}, the polarisation factor appearing in this three-point function should be kept fixed on the LHS of \eqref{COTContact} and so the only way to satisfy this unitarity constraint is to have an imaginary coefficient. We therefore have
\begin{align} \label{ImaginaryGraviton}
\psi_{3} \supset \epsilon_{ij}({\bf{k}}_{3}) k^{i}_{1}k_{2}^{j} \frac{i}{ \eta_{0}},
\end{align}
as the only possible late-time divergent contribution to the wavefunction.


\subsection{A conformally coupled scalar coupled to gravity}
\label{grav2}
We now turn to interactions between a conformally coupled scalar and gravity. We consider $\psi_{\varphi \varphi \gamma}$ which, in contrast to $\psi_{\varphi \varphi \varphi}$, can indeed have a non-vanishing real part. The bootstrap rules of \cite{BBBB} were not directly applied to conformally coupled scalars but many of the arguments presented there for massless fields also apply to conformally coupled fields. Indeed, symmetries of the theory and the choice of a Bunch-Davies initial state still dictate that a general IR-finite $3$-point function is given by a polynomial in the three external energies, with the appropriate symmetry as dictated by Bose statistics, divided by a power of $k_{T}$ (multiplied by the appropriate little group invariants if we have spinning fields). However, the scaling with momenta differs for conformally coupled fields. The overall scaling of an $n$-point function due to scale invariance is (see e.g. \cite{COT})
\begin{align}
\psi_{n} \sim k^{3(1-n) + \sum_{a=1}^{n}\Delta_{a}^{+}},
\end{align}
and so for two conformally coupled scalars ($\Delta^{+} = 2$) and one graviton ($\Delta^{+} = 3$), we have $\psi_{3} \sim k$. \\

\noindent Given the above discussion, let's now bootstrap $\psi_{\varphi \varphi \gamma}$ for $p=2$. We leave a more general analysis to \cite{JPSS}. Our general Ansatz is
\begin{align} \label{ThreePointAnsatzGravitonCC}
\psi_3^{(2)}(k_{1},k_{2},k_{3})= \frac{\epsilon_{ij}({\bf{k}}_{3}) k^{i}_{1}k_{2}^{j}}{k_{T}^2 \eta_{0}^2}(E_{00}k_{3} + E_{10}\mathbb{e}_1),
\end{align}
where again we have dropped the $\varphi \varphi \gamma$ subscript and have add two factors of $\eta_{0}$ in the denominator as required by dimensional analysis. In contrast to the massless case, scale invariance and little group scaling do not allow for a logarithm in this Ansatz. Now, as explained in Section \ref{MLTforMassive}, the MLT for a conformally coupled scalar is automatically satisfied by the form of our Ansatz and so we only need to apply the MLT to the graviton's energy. We have 
\begin{align} 
\dfrac{\partial }{\partial k_{3}}\psi_3^{(2)}(k_1,k_2,k_{3})\Big|_{k_{3}=0}=0\,,
\end{align}
which fixes $E_{00} = 2E_{10}$, and so the unique manifestly local $3$-point function is
\begin{align} \label{ThreePointGravitonCC}
\psi_3^{(2)}(k_{1},k_{2},k_{3})= \frac{E_{10}\epsilon_{ij}({\bf{k}}_{3}) k^{i}_{1}k_{2}^{j}}{k_{T}^2 \eta_{0}^2}(2k_{3} + \mathbb{e}_1),
\end{align}
which is the wavefunction coefficient arising from the minimal coupling operator $\gamma^{ij}\partial_{i}\varphi \partial_{j} \varphi$ in the bulk. In comparison to massless scalars, we therefore see that there cannot be leading order $k_{T}$ poles of degree $< 2$. Also, in contrast to the case of $\psi_{\phi\phi\gamma}$, the MLT does not force the $3$-point function to be fully symmetric in the external energies, which is to be expected since the mode function of a conformally coupled scalar is different from that of the massless graviton. The only imaginary term we can add to this wavefunction is also \eqref{ImaginaryGraviton}. In Section \ref{Results} we will use this $3$-point function to bootstrap the $4$-point function for conformally coupled scalars due to graviton exchange.


\section{Bootstrapping 4-point functions using partial energy shifts} \label{PartialEnergyShifts}
In the previous sections we exploited the Manifestly Local Test (MLT) to bootstrap contact diagrams for various $3$-point interactions. This begs the question: \textit{how do we bootstrap exchange diagrams from their constituent contact subdiagrams?} In this section we introduce a systematic way for computing \textit{rational} $4$-point exchange diagrams from their constituent $3$-point functions, and we expect our formalism to generalise to higher-point functions too. We present a three-step procedure. In the Step I we leverage the analyticity of $4$-point functions by use of the Cauchy's integral theorem, which relates the desired $4$-point function to the residues and boundary term associated with a meromorphic function $\tilde \psi_4(z)$ of a single complex variable $z$. The function $\tilde \psi_4(z)$ is obtained by an appropriate shift by $z$ of the partial energies in the arguments of $\psi_4$, as we discuss in the next section.
All the residues of $\tilde{\psi}_4(z)/z$ on its $z$ poles, both leading and subleading, are \textit{completely} fixed by the Cosmological Optical Theorem (COT). The sum over the residues will generically \textit{not} satisfy the COT by itself and so in Step II a suitable boundary term will be added to ensure that our bootstrapped $4$-point function arises from unitary time evolution. Finally, in Step III we use the $6$-point MLT to fix the remaining parts of the boundary term. After going through these three steps, the $4$-point function satisfies both the COT and the MLT. However, we still have the liberty to add any function of the kinematics that itself satisfies $(i)$ the COT in \eqref{HomogeneousCOT} and $(ii)$ the MLT in \eqref{MLT}. We expect that when combined with a generalisation of the bootstrap rules of \cite{BBBB}, these conditions give us an on-shell \textit{definition} of contact $4$-point functions. Here we assume a rational ansatz for the wavefunction coefficients. This is a valid assumption at tree-level for massless scalar and tensor fields in the absence of logarithmic IR-divergences. In Section \ref{Results}, we work through a number of examples where we explicitly find the local quartic operators that account for the difference between our bootstrap procedure and bulk computations. 


\subsection{Step I: partial energy recursion relations}\label{sec:step1}
Locality implies that wavefunction coefficients can only diverge when: $(i)$ the sum of energies entering a subdiagram vanishes, these are partial energy poles, or $(ii)$ when the sum of all external energies vanishes which is the total energy pole. Unitarity, in the form of the cosmological single-cut rules \cite{COT}, relates the (leading and subleading) singular behaviour near each of the partial energy singularities to the \textit{sub-diagrams} that emerge after cutting an appropriate internal line. For concreteness, 
consider an $s$-channel $4$-point diagram of a single massless field, represented by $\psi_4(k_1,k_2,k_3,k_4,s)$. The allowed singularities are at
\begin{align}
    E_L=k_1+k_2+s=0\,,\qquad E_R=k_3+k_4+s=0\,, \qquad k_T=k_1+k_2+k_3+k_4=0\,.
\end{align}
We take $\psi_{4}$ to be symmetric in the external pairs $(k_{1},k_{2})$ and $(k_{3},k_{4})$, but generalisations should be straightforward. To make the following expressions algebraically simpler, we make the following change of variables in the arguments of the $4$-point function and its $3$-point subdiagrams:
\begin{align}
    &\psi_4: \qquad (k_1,k_2,k_3,k_4,s)\to (E_L,E_R,k_1 k_2, k_3 k_4,s)\,,\\ 
    &\psi_3^L: \qquad (k_1,k_2,s)\to (E_L,k_1k_2,s)\,,\\ 
    &\psi_3^R: \qquad (k_3,k_4,s)\to (E_R,k_3k_4,s)\,.
\end{align}
Notice that with this new set of variables, the total energy is not an independent quantity but is given by $k_T=E_L+E_R-2s$. It is straightforward to verify that $\psi_4$ and $\psi_3$ retain their rational format upon performing this change of variables. Now, near the $E_L = 0$ pole, $\psi_4$ admits the Laurent series
\begin{align}\label{Laurant}
    \psi_4=\sum\limits_{0<n\leq m}\dfrac{R_n(E_R,k_1\,k_2,k_3\,k_4,s)}{E_L^n}+{\cal O}(E_L^0)\,,
\end{align}
where $m$ is an integer that encodes the degree of the leading pole. Notice that we always symmetrise between the left and right vertices and so the same expansion holds near the $E_R=0$ pole, upon permuting $E_L$ with $E_R$ and $k_1k_2$ with $k_3 k_4$. We want to prove that unitarity fully fixes the coefficients of this expansion except for the last analytical part. Writing the Cosmological Optical Theorem with the new kinematical variables we have
\begin{align}
    \psi_4(E_L,E_R,k_1k_2,k_3k_4,s)+\psi^{\ast}_4(-E_L+2s,-E_R+2s,k_1k_2,k_3k_4,s)= \RHS 
\end{align}
where for future convenience we have denoted by $\RHS$ the right-hand side of the COT\footnote{As compared with \eqref{finalCOT}, here we used the contact COT to write $\psi_3^\ast(-E_L+2s,k_1k_2,s)=-\psi_3(E_L-2s,k_1k_2,-s)$, as in \cite{COT}.}:
\begin{align}
    \RHS= P(s)\left(\psi_3(E_L,k_1 k_2,s)-\psi_3(E_L-2s,k_1k_2,-s)\right) \\
    \qquad \times \left(\psi_3(E_R,k_3 k_4,s)-\psi_3(E_R-2s,k_3k_4,-s)\right).\nonumber
\end{align}
The key observation is that the second term on the left-hand side of this expression is analytic around $E_L=0$, and so can be dropped in the limits $E_L\to 0$ or $E_R\to 0$. This implies that the right-hand side side of the COT determines all of the leading and sub-leading partial energy poles $R_n$ of $\psi_{4}$. This is more information than what is provided by the factorization results recently employed in \cite{CosmoBootstrap3}, which fix the leading singularity. For reference, we summarise the singularities of the components of the COT in Table \ref{sing}. We present the singularities that involve $E_{L}$ but those for $E_{R}$ are again the same with appropriate change of arguments in $\psi_{3}$. The COT identifies the integer $m$ with the degree of the total energy pole in the $3$-point function i.e.
\begin{align}
    \lim_{E_L\to 0}\psi_3\propto \dfrac{1}{E_L^m}\,,\qquad m=\text{dimension of the vertex}-3\,.
\end{align}
Moreover, the COT gives the coefficients $R_n$ in terms of the partial derivatives of its right-hand side with respect to the partial energy $E_L$ as
\begin{align}
    R_n(E_R,k_1k_2,k_3k_4,s)&=\dfrac{1}{(m-n)!}\dfrac{\partial^{m-n}}{\partial E_L^{m-n}}\left[ E_L^m\, \RHS (E_L,E_R,k_1k_2,k_3k_4,s)\right]_{E_L=0}\,.
\end{align}
\begin{table}[] 
\begin{center}
\begin{tabular}{c c c c c}
  &  {$\psi_4(k_a,s)$}  & {$\psi_4(-k_a,s)$}  & {$\psi_3(k_1,k_2,s)$} & {$\psi_3(k_1,k_2,-s)$}   \\
\hline
(partial energy pole)\tikzmark{a11}$\,E_{L}=0$ & \tikzmark{a12}\checkmark  & \tikzmark{a13}\ding{55}  & \tikzmark{a14}\checkmark & \tikzmark{a15}\ding{55} \\

(collinear pole)\tikzmark{a21}$\,E_{L}=2s$  & \tikzmark{a22}\ding{55} & \tikzmark{a23}\checkmark & \tikzmark{a24}\ding{55} & \checkmark \\

(total energy pole)\tikzmark{a31}$\,E_L+E_R=2s$  & \tikzmark{a32}\checkmark  & \tikzmark{a33}\checkmark  & \tikzmark{a34}\ding{55} & \ding{55} 
\end{tabular}
\caption{Singularities of the elements apearing in the $4$-point exchange COT. The same applies to $E_R$ singularities as well with the substitution $\psi_3(k_1,k_2,s)\to \psi_3(k_3,k_4,s)$.}
\label{sing}
\end{center}
\end{table}
\noindent What about the analytical part of the expansion? It might appear that it is not constrained by unitarity at all, precluding us from bootstraping the full $4$-point function. This is, however, a rushed judgment as we have not yet used the full knowledge of the allowed poles. Recall that $\psi_4$ must be regular in the collinear limit, i.e. $E_R=2s$ (or $E_L=2s$), \textit{iff} we keep $E_L$ (or $E_{R}$) finite. However, the coefficients of its Laurent expansion will generically inherit such spurious poles from $\RHS$ which requires the \textit{non-singular} part of $\psi_4$ to come to the rescue and cancel these bad singularities. Let us see how this happens in a concrete example. Consider the $4$-point function of a massless scalar in flat space arising from the cubic interaction $\phi^3$. It is given by
\begin{align}
    \psi_4=\dfrac{1}{E_L\,E_R\,(E_L+E_R-2s)}=\dfrac{1}{E_L}\dfrac{1}{E_R(E_R-2s)}+\sum_{n\geq 1}\dfrac{(-1)E_L^{n-1}}{E_R\,(E_R-2s)^{n+1}}\,.
\end{align}
We see that by expanding the total energy pole around $E_L=0$ one generates an infinite number of terms analytic in $E_L$ that are singular at the collinear limit, and yet the full $4$-point is free of the latter singularity. There is still one more property that the Laurent expansion should satisfy: it should reproduce a similar expansion around $E_R=0$. Ensuring the cancellation of spurious poles and the correct Laurent expansion around each partial energy pole turns out to be very restrictive.  \\

\begin{figure}
    \centering
    \includegraphics[width=10cm]{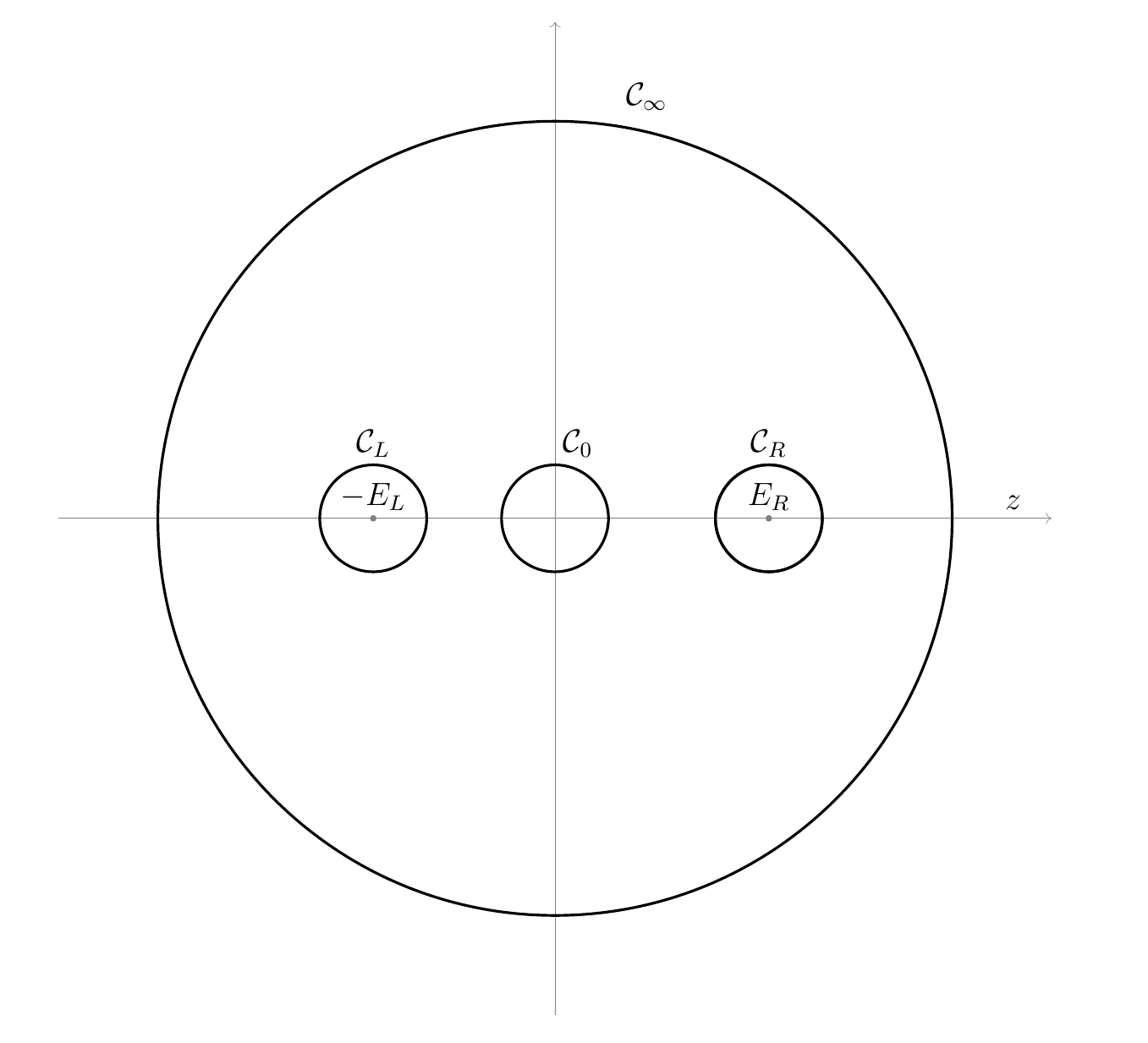}
    \caption{The countour integrals adopted for relating the value of $\tilde{\psi}(z)$ at the origin to its residues and an associated boundary term at infinity.}
    \label{contour}
\end{figure}

\noindent It is very natural to then seek an integrated approach in order to satisfy these properties all at once. The most pedestrian way forward is to insert the most generic Ansatz for $\psi_4$, namely
\begin{align}
\label{ansats}
    \psi_4=\dfrac{\text{Poly}_{2+4m}(E_L,E_R,k_1 k_2,k_3 k_4,s)}{E_L^m\,E_R^m\,(E_L+E_R-2s)^{2m-1}}\,,
\end{align}
into the COT and fix the free coefficients appearing in the polynomial in the numerator as much as possible (here the degree of the total energy pole is fixed by the power counting argument of \cite{BBBB}, and $\text{Poly}_l$ is a polynomial of energy dimension $l$ with its degree fixed by scale invariance for external fields with massless mode functions). The downside of this approach is the proliferation of parameters needed to write down such an Ansatz as we increase the degree of the singularity of the vertices. We instead take a different approach and bootstrap $\psi_4$ using the miracles of Cauchy's integral theorem. The trick is to shift the arguments of $\psi_4$ by a \textit{single complex variable} $z$ and subsequently arrive at a shifted four-point function $\tilde{\psi}_4(z)$ such that:
\begin{itemize}
    \item[$(a)$] $\tilde{\psi}_4(z=0)=\psi_4$,
 \item[$(b)$] $\tilde{\psi}_4(z)$ is an analytic function of $z$ except for isolated poles, 
    \item[$(c)$] the residues of $\tilde{\psi_4}(z)/z$ at $z\neq 0$ are fixed by the Cosmological Optical Theorem. 
\end{itemize}
A shift that satisfies all of these requirements is the following \textit{partial energy shift}\footnote{Energy shifts were introduced in \cite{Arkani-Hamed:2017fdk} to fix the residues of simple poles. Our partial-energy shifts, combined with the Cosmological Optical Theorem, enable us to fix the leading \textit{and all subleading} partial energy poles which is crucial for computing inflationary correlators.}
\begin{align}
    \psi_4(E_L,E_R,k_1k_2,k_3k_4,s)\to \tilde{\psi}_4(z)=\psi_4(E_L+z,E_R-z,k_1k_2,k_3k_4,s)\,.
\end{align}
Let us verify that $(a)-(c)$ are satisfied. Condition $(a)$ is trivial since the shift vanishes at $z=0$. Condition $(b)$ is satisfied since $\tilde{\psi}_4(z)$ inherits the analytical properties of $\psi_4$. Indeed, it is an analytic function in the complex plane of $z$ except for two isolated poles located at
\begin{align}
    \text{singularities of } \tilde{\psi}_4(z): \qquad z=-E_L \qquad \text{and}\,\quad z=E_R\,.
\end{align}
The singular part of the Laurent expansion of $\tilde{\psi}_4$ around these poles is dictated by the right-hand side $\RHS$ of the COT i.e. 
\begin{align}
\label{Laurentz}
    \tilde{\psi}_4(z)&=\sum_{0<n\leq m}\dfrac{A_n(E_R,E_L,k_1k_2,k_3k_4,s)}{(z+E_L)^n}+{\cal O}(z+E_L)\,,\\ \label{An}
    A_n &=\dfrac{1}{(m-n)!} \left[\partial^{m-n}_z (z+E_L)^m\, \RHS(E_L+z,E_R-z,k_1k_2,k_3k_4,s)\right]_{z=-E_L}\,.
\end{align}
Notice that, in principle, the coefficients $A_n$ can be expressed in terms of $R_n$. As a corollary to $(a)$ and $(b)$, we can use the residue theorem to write,
\begin{align}
    \psi_4(E_L,E_R,k_1k_2,k_3k_4,s)=\dfrac{1}{2\pi i}\oint\limits_{{\cal C}_0}dz\, \dfrac{\tilde{\psi}_4(z)}{z}\,,
\end{align}
where ${\cal C}_0$ is a contour that rotates around the origin (see Figure \ref{contour}). We adopt the clockwise direction for contour integration throughout. So condition $(c)$ is also satisfied since the Laurent expansion of $\tilde{\psi}_4$ around $z=-E_L$ and $z=E_R$ directly follows from that of $\psi_4$ at $E_L=0$ and $E_R=0$ respectively which are in turn fixed by the COT in terms of lower-point vertices, as we explained above. Using that Laurent expansion, we can straightforwardly compute the residues of $\tilde{\psi}_4/z$ at these locations. They are given by
\begin{align}
    \dfrac{1}{2\pi i}\oint\limits_{{\cal C}_L}dz\, \dfrac{\tilde{\psi}_4(z)}{z}=\text{Res}\left[\dfrac{\tilde{\psi}_4(z)}{z}\right]_{z=-E_L}&=-\sum_{0<n\leq m}\dfrac{A_n(E_L,E_R,k_1k_2,k_3k_4,s)}{E_L^n}\,, \label{AL}\\ 
     \dfrac{1}{2\pi i}\oint\limits_{{\cal C}_R}dz\, \dfrac{\tilde{\psi}_4(z)}{z}=\text{Res}\left[\dfrac{\tilde{\psi}_4(z)}{z}\right]_{z=E_R}&=-\sum_{0<n\leq m}\dfrac{A_n(E_R,E_L, k_3k_4,k_1k_2,s)}{E_R^n}\,. \label{AR}
\end{align}
Notice that our partial energy shift was carefully chosen such that the total energy is independent of $z$:
\begin{align}
k_T(z)=(E_L-z)+(E_R+z)-2s=k_T.    
\end{align}
Any shift that does not have this property will introduce additional singularities at $z_T$ (defined via $k_T(z_T)=0$). The residues of $\tilde{\psi}_4/z$ at such poles cannot be fixed by the COT since the total energy poles precisely cancel on each side of the equation, see e.g. Table \ref{sing}. \\

\noindent We now make use of the analyticity of $\tilde{\psi}_4(z)$ in order to deform the ${\cal C}_0$ contour and arrive at 
\begin{align} \label{Psi4Residue}
    \psi_4(E_L,E_R,k_1k_2,k_3k_4,s) &=-\text{Res}\left[\dfrac{\tilde{\psi}_4(z)}{z}\right]_{z=-E_L}-\text{Res}\left[\dfrac{\tilde{\psi}_4(z)}{z}\right]_{z=E_R}+B \\
    & = \psi_{\text{Res}} + B \,,
\end{align}
where $B$ is a boundary contribution at infinity i.e.
\begin{align}\label{bound}
    B=\dfrac{1}{2\pi i}\oint\limits_{{\cal C}_\infty}dz\, \dfrac{\tilde{\psi}_4(z)}{z}\,,
\end{align}
and 
\begin{align}
    \psi_{\text{Res}}&=\sum_{0<n\leq m}\dfrac{A_n(E_L,E_R,k_1k_2,k_3k_4,s)}{E_L^n}+\sum_{0<n\leq m}\dfrac{A_n(E_R,E_L, k_3k_4,k_1k_2,s)}{E_R^n}\,.\label{psireshere}
\end{align}
Notice that we could perform the integral in Cauchy's theorem in full generality, for any interaction at once. Using the above formula, the calculation of $\psi_{\text{Res}}$ is reduced to the elementary process of taking derivatives of linear combinations of lower-point functions, as dictated by \eqref{An}. As compared to the Feynman rules for representing $\psi_n$ as integrals in time in the bulk of de Sitter, \textit{\eqref{psireshere} provides an alternative and equivalent representation of (the singular part of) the wavefunction coefficients, where time has completely disappeared}.\\

\noindent As we illustrate in a number of examples below, the boundary term is generically non-vanishing and cannot be fixed by the COT alone because both shifted terms on the left-hand side of the COT diverge as $z$ goes to infinity. However, we know the following facts:
\begin{itemize}
    \item $B$ cannot have any partial energy poles\footnote{This is very similar to BCFW for amplitudes where with a particular momentum shift, the boundary term can in principle have a pole as one out of the three Mandelstam variables is taken to zero. This is because for a given shift there is always a sum of two momenta that is independent of the deformation parameter.} and the Laurent expansion of $\psi_4(E_L+z,E_R-z,...)$ around $z=\infty$ is analytic in $E_L$ and $E_R$. This follows from noticing that for any finite constant $c$ we have $B(E_R,E_L)=B(E_R+c,E_L-c)$, and so it cannot have isolated $E_L$ or $E_R$ poles.
     \item All the partial energy singularities of $\psi_4$ are in $\psi_{\text{Res}}$ (since none are in $B$). Furthermore the Laurent expansion of $\psi_{\text{Res}}$ around each partial energy singularity only has negative powers of that partial energy. This is because shifting $\psi_4$ in \eqref{Psi4Residue} for a second time and computing the contour integral at infinity should be consistent with Equation \eqref{bound}, which cannot happen unless the shifted residue sector vanishes at $z\to \infty$.
     \item The partial energy shift forces $\psi_{\text{Res}}$ to have a \textit{total energy pole}. At first glance, this seems odd as the right-hand side $\RHS$ of the COT---the very origin of the residue sector--- does not admit such a pole. The reason is that $\RHS$ has a pole in the collinear limit ($E_R-2s=0$), so after shifting the kinematics and evaluating $\RHS$ and its derivatives at $z_L=-E_L$, the collinear pole transmutes into a total energy singularity for $A_n$'s, simply because $k_T=E_L+E_R-2s=(E_L+z_L)+(E_R-z_L)-2s$. 
\end{itemize}
In summary, in Step I we determine the part $\psi_{\text{Res}}$ of $\psi_{4}$ that has any partial energy singularities using \eqref{psireshere}, \eqref{An} and the right-hand side  $\RHS$ of the Cosmological Optical Theorem for a given $\psi_{3}$. This determines $\psi_4$ up to the boundary term $B$ in \eqref{Psi4Residue} that has no partial energy singularities. The goal of the next two steps is to fix $B$. We identify three contributions to the boundary term
\begin{align}
    B=\BCOT+\BMLT+\Bcontact\,.
\end{align}
The first contribution $\BCOT$ ensures that $\psi_4$ satisfies the COT and is computed in Step II. The second contribution $\BMLT$ ensures that $\psi_4$ satisfies the MLT and is computed in Step III. Finally, we conjecture that the last contribution $\Bcontact$ always corresponds to the $\psi_4$ induced by a contact diagram corresponding to a (quartic) local operator. This cannot be fixed unless further input is given about the model under consideration, in comparison to the boundary term for scalar amplitudes in the BCFW formalism \cite{DSJS}.


\subsection{Step II: back to the Cosmological Optical Theorem}\label{sec:step2}

In deriving $\psi_{\text{Res}}$ in \eqref{Laurentz}, we have only used the structure of the right-hand side of the Cosmological Optical Theorem (COT). Therefore, in general $\psi_{\text{Res}}$ does \textit{not} satisfies the COT. As we illustrate in a number of examples in Section \ref{Results}, generically one needs to include a necessary boundary term $\BCOT$ to ensure that $\psi_4$ is consistent with unitarity. To see this, consider the right-hand side of the COT for a give cubic wavefunction $\psi_3$. We claim that the most general form of its Taylor expansion around $s=0$ is
\begin{align}
    \RHS&=\alpha_0\,s^3+\sum_{l\geq 1}s^{l+3}\,f_l(E_L,E_R,k_1k_2,k_3k_4)\,,\label{alpha}
\end{align}
where 
\begin{align}
   \alpha_0=\dfrac{1}{3!}\dfrac{\partial}{\partial s}\RHS (E_L,E_R,s,k_1k_2,k_3k_4)\Big|_{s=0}\,.
\end{align}
The reason behind this expansion is that (\textit{i}) $\RHS$ must scale as $k^3$ by scale invariance, (\textit{ii}) it only contains odd positive powers of $s$ (no $s$ singularities are allowed by manifest locality) and (\textit{iii}) a linear term in $s$ is precisely obliged to vanish by the MLT (we remind the reader that we are working with external massless fields). Therefore, to have the right scaling all $f_l$'s must have denominators. Since all $k_T^{(4)}$ poles cancel in the COT, the $f_l$'s must be singular at $E_L=0$ and $E_R=0$. This is in contrast with the boundary term $B$ in \eqref{Psi4Residue}, which we argued cannot have partial energy singularities. It follows that the COT splits into two distinct conditions given by
\begin{align}
    \psi_{\text{Res}}(k_a,s)+\psi_{\text{Res}}(-k_a,s)&=\sum_{l\geq 1}s^{l+3}\,f_l(E_L,E_R,k_1k_2,k_3k_4)\,, \\
     B(k_a,s)+B(-k_a,s)&=\alpha_0\,s^3.
\end{align}
By construction $\psi_{\text{Res}}$ should contain all the partial energy singularities, so the first equality has to hold. The second equality has the particular solution 
\begin{align}\label{BCOT}
    \BCOT(k_a,s)=\dfrac{1}{2}\alpha_0\,s^3\,,
\end{align}
and leaves the remaining terms to satisfy the homogeneous equation (the contact COT of \cite{COT})
\begin{align} \label{HomogeneousCOT}
     \left(B(k_a,s)-\BCOT(k_a,s)\right)+\left(  B(-k_a,s)- \BCOT(-k_a,s)\right )=0\,.
\end{align}
Given that $ B$ can only have a total energy pole, we can write
\begin{align}
    \Delta B\equiv B(k_a,s)-\BCOT(k_a,s)=\dfrac{\text{Poly}_{2m+2}(k_a,s)}{k_T^{2m-1}}\,,
\end{align}
where we have used scale invariance to fix the degree of the polynomial in the numerator. Plugging this Ansatz into \eqref{HomogeneousCOT} yields
\begin{align}
    \text{Poly}_{2m+2}(k_a,s)-\text{Poly}_{2m+2}(-k_a,s)=0\,.
\end{align}
This implies that the polynomial cannot have odd powers of $s$. Therefore, $\Delta B$ is an analytic function of $s^2=k_1^2+k_2^2+2\bfk_1.\bfk_2$, and we can rewrite \eqref{HomogeneousCOT} as
\begin{align}
\label{contcot}
    \Delta B(k_a;\dots)+\Delta B(-k_a;\dots)=0\,,
\end{align}
which is precisely the COT for a contact term and we have added the $\ldots$ to allow for rotation invariant contractions of the momenta. This equation is compatible with the hypothesis that $\Delta B$ is a contact term, but in the next section we show that this is not always the case. Instead, there is a part of $\Delta B$ that must necessarily be attributed to the exchange $4$-point function $\psi_4$. \\

\noindent In summary, in Step II we reconstruct the part $\BCOT$ of $B$ by demanding that the COT is satisfied. The result is given in \eqref{BCOT} with $\alpha_{0}$ defined in \eqref{alpha} in terms of the cubic couplings.


\subsection{Step III: constraining the boundary term with the Manifestly Local Test} \label{BoundaryTerm}
The previous step has guaranteed that our $4$-point function satisfies the Cosmological Optical Theorem (COT) and so is consistent with unitary time evolution in the bulk. This fixed $\BCOT$ in \eqref{BCOT}. However, we should also make sure that higher-point diagrams that include our $4$-point function as a sub-diagram are consistent with manifest locality and unitarity. In particular, the $4$-point diagram we are bootstraping must obey the Manifestly Local Test (MLT) (see Section \ref{sec:MLT})
\begin{align}
\dfrac{\partial}{\partial k_a}\psi_4(\{k \},s,\{\bfk\})\Big|_{k_a=0}=0\,,
\end{align}
where we are reverting back to the original kinematical parameters $(\{k \},s)$ in which the MLT takes a more compact form. The MLT ensures that the $6$-point function shown in Figure \ref{6pt} has the correct singularities for a manifestly local theory of massless scalars.  Now, it is not guaranteed that $\psi_{\text{Res}} + \BCOT$ satisfies the MLT. When it does not, we must add an additional contribution $\BMLT$ to the boundary term $B$. In cases where  
\begin{align}
    \dfrac{\partial }{\partial k_a} \psi_{\text{Res}}(\{k \},s,\{\bfk\})\Big|_{k_a=0}\neq 0\,,
\end{align}
$\BMLT$ is found by solving
\begin{align} \label{BMLTdefinition}
    \dfrac{\partial }{\partial k_a}\BMLT \Big|_{k_a=0}= -\dfrac{\partial }{\partial k_a}\psi_{\text{Res}}\Big|_{k_a=0}\,,
\end{align}
where we remind the reader that when a derivative is taken with respect to an external energy, all other variables are kept fixed (and so $B_{\text{COT}}$ does not contribute). Let us emphasise that this is a very non-trivial constraint. Since $B_{\text{MLT}}$ is only allowed to have $k_{T}$ poles, if we differentiate with respect to say $k_{1}$, the left-hand side of \eqref{BMLTdefinition} can only have a pole at $k_{2}+k_{3}+k_{4} = 0$. This must also be the case on the right-hand side and given that $\psi_{\text{Res}}$ can have $k_{T}$, $E_{L}$ and $E_{R}$ poles, many cancellations must occur. As always, this equation must hold for the external energy of any field with a massless mode function and so can in principle yield a system of constraints. For conformally coupled fields (or any field with $\Delta^{+}=2$) the MLT is again automatically satisfied. In Section \ref{Results} we illustrate the power of the $6$-point MLT in a number of informative examples. \\

\noindent With all the compulsory elements added to the boundary term, we are still free to add any correction $\Delta \psi_4(E_L,E_R, k_1k_2,k_3k_4,s)$ which must: 
\begin{itemize}
    \item only have $k_{T}$ poles,
    \item satisfy the homogeneous Cosmological Optical Theorem \eqref{HomogeneousCOT},
    \item satisfy the $4$-point Manifestly Local Test \eqref{MLTGeneral}.
\end{itemize}
If $\Delta \psi_4$ contains only real couplings and has the correct momentum scaling as dictated by scale invariance, it will always satisfy the homogeneous COT since this is equivalent to the COT for contact terms. As for the MLT, as we discussed around equation \eqref{contcot}, $\Delta\psi_4$ can only depend on even powers of $s$ which can in turn be written in terms of inner products of the external momenta. So the MLT actually takes the form for a contact term i.e. 
\begin{align}
     \dfrac{\partial }{\partial k_a}\Delta \psi_4(\{k \},\{\bfk\})\Big|_{k_a=0}=0\,.
\end{align}
In Section \ref{BootstrapThreePoint} we saw that the Bootstrap Rules of \cite{BBBB} along with the MLT allows one to bootstrap \textit{all} $3$-point functions arising from manifestly local theories. With an adapted form of the Bootstrap Rules of \cite{BBBB}, the $4$-point contact MLT, and the fact that $\Delta \psi_4$ is only permitted to have $k_T$ poles, we expect that this is enough to provide an \textit{on-shell definition} of a contact $4$-point function. This will be discussed in \cite{trispectrum}. In Section \ref{Results} we consider a number of examples where we compute the difference between our bootstrap result for an exchange $4$-point function, derived following our three-step procedure, and the result of the bulk computation and we show that the difference is given by contact diagrams of quartic interactions. Note that here we are bootstrapping the $s$-channel of an exchange process, so when we compare our result to the bulk computation we sum over permutations to also include the $t$ and $u$ channels. This ensures that any potential difference is a sum of contact terms with the correct symmetry.  \\

\noindent In summary, Step III of our bootstrap procedure requires one to solve for $ B_{\text{MLT}}$ by demanding that full the $4$-point function satisfies the MLT. 

\subsection{Comparison to BCFW momentum shifts}
The partial energy shifts we have introduced in this section are reminiscent of the BCFW momentum shifts \cite{BCFW} extensively used in the $S$-matrix bootstrap programme, yet there are important differences. Below we compare the two methods:
\begin{itemize}
    \item BCFW shifts \cite{BCFW} are defined as complex deformations of a sub-set of the external spinors appearing in the spinor helicity formalism used to compactly write $4$-point scattering amplitudes of massless particles (see e.g. \cite{TASI, Elvang:2013cua,Benincasa:2013} for reviews and \cite{DSJS} for a discussion in the context of boost-breaking amplitudes). Two out of the four external momenta are deformed in such a way that the new momenta remain on-shell and satisfy momentum conservation. In contrast, our partial-energy shifts act on a sum of energies while keeping the external momenta held fixed. This allows us to isolate the partial-energy poles of exchange diagrams without introducing branch-cuts in the deformation parameter.  
    
    \item BCFW shifts isolate two out of the three allowed poles of full $4$-point amplitudes. The allowed poles occur when one of the Mandelstam variables goes to zero, and the residues are fixed by consistent factorisation (see also \cite{PSS, Schuster:2008nh,McGady:2013sga,Arkani-Hamed:2017jhn}). One of the Mandelstam variables is independent of the deformation parameter and so is analogous to $k_{T}$ in the above discussion. The two deformed Mandelstam variables are analogous to $E_{L}$ and $E_{R}$.
    
    \item The fact that BCFW shifts isolate two of the Mandelstam variables at the same time, means that these shifts do not rely on individual Feynman diagrams. Rather, the shifts act on the full $4$-point amplitude. This has made the shifts particularly powerful for deriving recursion relations that fix the full tree-level $S$-matrix in Yang-Mills and General Relativity in terms of their $3$-point amplitudes \cite{BCFW,BCF,Benincasa:2007qj}, and is the basis of the four-particle test of Benincasa and Cachazo \cite{Benincasa:2007xk}. In contrast, our partial-energy shifts act diagram-by-diagram. It would be very interesting to generalise our partial-energy shifts such that they relate the poles of different channels. This will be particularly important for bootstrapping wavefunction coefficients with external spinning fields.
\end{itemize}


\section{Explicit examples} \label{Results}
In this section, we use the three-step procedure outlined in the previous section to bootstrap various tree-level $4$-point exchange diagrams. For each case we summarise the main features in Table \ref{BootstrapResults}. We state whether boundary terms are required to satisfy the Cosmological Optical Theorem (non-zero $\alpha_{0}$) or the Manifestly Local Test (non-zero $B_{\text{MLT}}$), and state whether our bootstrap result is equal to the bulk result, or differs by a sum of contact $4$-point interactions $\Delta \mathcal{L}_{\text{int}}$. 
\begin{table}[h!] 
\begin{center}
\begin{tabular}{c c c c c}
Cubic vertex  &  {$\alpha_{0}$}  & {$ B_{\text{MLT}}$}  & Bootstrap - Bulk  \\
\hline
$\phi^3$ in Minkowski  & \tikzmark{a13}\ding{55} & \tikzmark{a13}\ding{55}  & 0  \\

$\gamma^{ij} \partial_{i}\phi \partial_{j}\phi$  & \tikzmark{a22}\ding{55} & \tikzmark{a23}\checkmark  in \eqref{DeltaBGravitonExchange} & 0  \\

$\gamma^{ij} \partial_{i}\varphi \partial_{j}\varphi$  & \tikzmark{a22}\ding{55}  & \tikzmark{a22}\ding{55}  & 0   \\

$\phi \phi'^2$   & \tikzmark{a32}\checkmark  in \eqref{Alpha0}    & \tikzmark{a33}\checkmark  in \eqref{DeltaBMLT}   & $\Delta \mathcal{L}_{\text{int}}$  in \eqref{ContactDifference}     \\

$\phi'^3$  & \tikzmark{a22}\ding{55}  & \tikzmark{a22}\ding{55}  & $\Delta \mathcal{L}_{\text{int}}$  in \eqref{EFT1Contact}  \\

$\phi' (\nabla \phi)^2$  & \tikzmark{a32}\checkmark  in \eqref{AlphaEFT2}  & \tikzmark{a33}\checkmark  in \eqref{DeltaBEFT2}  & $\Delta \mathcal{L}_{\text{int}}$  in \eqref{EFT2Contact} 
\end{tabular}
\caption{Here we summarise the main features of various $4$-point exchange functions induced by two copies of the cubic vertex given in the first column and computed with our three-step procedure. Unless stated otherwise, the background geometry is de Sitter.}
\label{BootstrapResults}
\end{center}
\end{table}


\subsection{A flat-space warm-up}\label{sec:6p1}

All of the results we derived so far apply \textit{mutatis mutandis} to flat spacetime where the Manifestly Local Test (MLT) is the same as for conformally coupled fields, see \eqref{MLTConformal}. Since in Minkowksi we find only \textit{simple} partial and total energy poles\footnote{In de Sitter, the order $p$ of the $k_T$ pole is $p=1+\sum_A \Delta_A -4$ where $\Delta_A$ are the mass dimensions of all the vertices in the diagram \cite{BBBB}. In the bulk representation, the factors of $1/k_T$ arise from all the inverse metric factors that contract derivatives bringing factors of $ a^{-1} \sim \eta$ which are then schematically integrated against $e^{i k_T \eta}$. In contrast, in Minkowski the integrals are always exponential functions $e^{i k_T t}$ leading to simple poles irrespectively of the number of derivatives}, even for derivative interactions, this is a great place to start to demonstrate our techniques in explicit examples. Let's consider the following simple cubic polynomial interaction
\begin{align}
H_{\text{int}}=\lambda\frac{ \phi^{3}}{3!}\,.
\end{align}
The cubic wavefunction can be easily bootstrapped adapting the Bootstrap Rules of \cite{BBBB} to Minkowksi. Since $\psi_3$ must scale as $k^{-1}$ for a dimension three interaction and it must have a simple $1/k_T$ pole, we must have $\psi_3 \propto k_T^{-1}$. The overall coefficient follows from matching the $k_T$ residue to the corresponding amplitude, $A_3 = \lambda$. This precisely agrees with the bulk calculation
\begin{align}\label{psi3}
\psi_{3}=i \times \frac{\lambda}{3!} \times 3! \times \int d\eta \,e^{ik_T \eta}=\frac{\lambda}{k_{T}}\,.
\end{align}
It's easy to check that this expression solves the contact Cosmological Optical Theorem (COT) in \eqref{cotcon}. Following our discussion in Section \ref{sec:step1}, we can write the tree-level, $s$-channel $4$-point exchange generated by the above interaction as
\begin{align}\label{residflat}
\psi_{4}(E_{L},E_{R},s)=\sum_{0<m\leq n} \left(\frac{A_{m}(E_{L},E_{R})}{E_{L}^{m}}+ \frac{A_{m}(E_{R},E_{L})}{E_{R}^{m}}\right)+B\,,
\end{align}
where 
\begin{align}
\lim_{z\to -E_{L}}\psi_{4}(E_{L}+z,E_{R}-z)&=\sum_{0<m\leq n}\frac{A_{m}(E_{L},E_{R})}{(z+E_{L})^{m}}+\text{finite}\,,\\
\lim_{z\to E_{R}}\psi_{4}(E_{L}+z,E_{R}-z)&=\sum_{0<m\leq n}\frac{A_{m}(E_{R},E_{L})}{(E_{R}-z)^{m}}+\text{finite}\,,
\end{align}
and $B$ is a boundary term that is non-singular when either of the partial energies vanishes, $E_L=0$ or $E_R=0$. To compute the residues $  A_{m} $ we use the COT 
\begin{align}\label{flatcot}
    \disc \left[ i \psi_4 \right] = i P(s) \disc_{\bfs} \left[ i\psi_3(k_1,k_2,s) \right]\disc_{\bfs} \left[ i\psi_3(k_3,k_4,s) \right]\,,
\end{align}
where $P(s)=1/2s$ is the Minkowski power spectrum for a massless scalar field.
As outlined in Section \ref{sec:step1}, the second term in the definition of $\disc$ on the left-hand side is not singular in the $  E_{L}\to 0 $ limit. This follows from noticing that the $s$-channel $\psi_4$ has only partial and total energy poles $E_L$, $E_R$ and $k_T$. Upon analytic continuation of the external momenta and energies, these singularities are moved to 
\begin{align}
   \psi_4^\ast(-k_a,s) &\sim  \frac{1}{(- k_1 -k_2 +s)}\frac{1}{(- k_3 -k_4 +s)} \frac{1}{-k_T} + \text{finite} \\
   &\sim - \frac{1}{(2s-E_L)(2s-E_R)k_T}\,,
\end{align}
which is finite as $E_L \to 0$ for $s\neq 0$. So
\begin{align}
\lim_{E_L\to 0} \disc \left[ i \psi_{4}(k_{a},s)\right] = \lim_{E_L\to 0}   i \psi_{4}(k_{a},s) +\text{finite}\,.
\end{align}
The right-hand side of the COT hence gives us the residues we are after
\begin{align}
\disc \left[ i \psi_{3}(k_{1},k_{2},s) \right] &=\frac{i\lambda}{k_{1}+k_{2}+s}+\frac{i\lambda}{-k_{1}-k_{2}+s}=i\lambda\left[ \frac{1}{E_{L}}-\frac{1}{E_{L}-2s} \right]\,,\\
\disc \left[ i \psi_{3}(k_{3},k_{4},s)\right] &=\frac{i\lambda}{k_{3}+k_{4}+s}+\frac{i\lambda}{-k_{3}-k_{4}+s}=i\lambda\left[ \frac{1}{E_{R}}-\frac{1}{E_{R}-2s} \right]\,.
\end{align}
Using these expressions and the COT in \eqref{flatcot}, we find that 
\begin{align}
    \lim_{E_L \to 0} i \psi_4 &= \frac{i}{2s} \frac{-2is\lambda}{E_R(E_R-2s)}\frac{-2is\lambda}{E_R(E_R-2s)}\,.
\end{align}
Hence, upon shifting $E_L \to E_L +z$ and $E_R \to E_R -z$ the residue of the pole at $z=-E_L$ is found using \eqref{An}
\begin{align}\label{A1flat}
A_{1}(E_{L},E_{R})&=\lim_{z\to -E_L} (z+E_L)\psi_4(E_L+z,E_R-z,k_T)\\ 
&= \frac{\lambda^{2}}{(E_{L}+E_{R})k_{T}}\,,
\end{align}
which is the same as the residue at $z=E_R$. Notice the non-trivial appearance of $  k_{T}=k_T^{(4)}=k_{1}+k_{2}+k_{3}+k_{4} $ from a combination of cubic wavefunctions $\psi_3$ that individually know nothing about $k_{T}$. Plugging the only non-vanishing residue we found \eqref{A1flat} into \eqref{residflat} we conclude
\begin{align}\label{res}
\psi_{4}&=\frac{A_{1}(E_{L},E_{R})}{E_{L}}+\frac{A_{1}(E_{R},E_{L})}{E_{R}}+B\\
&=-\frac{\lambda^{2}}{k_{T}E_{L}E_{R}}+B\,.
\end{align}
What can we say about the boundary term? Following Steps II and III of our procedure, outlined in Sections \ref{sec:step2} and \ref{BoundaryTerm}, we plug this expression for $\psi_4$ back into the COT \eqref{flatcot} and find that it is satisfied with $B=0$. The MLT \eqref{MLTConformal} is also satisfied for $B=0$ since in flat-space the MLT only requires the $4$-point function to be regular as one external energy is taken soft. Therefore \eqref{res} with $B=0$ is our final result for $\psi_{4}$. This matches the expression derived by performing the two time integrals in the bulk representation. 


\subsection{Graviton exchange in de Sitter}

We now turn to bootstrapping scalar $4$-point functions arising due to graviton exchange in dS space\footnote{We will concentrate on a single diagram due to the exchange of the transverse, traceless graviton to illustrate how our methods can fix the structure of the poles. In the full gauge theory there may also be additional diagrams that we are omitting here.}. This trispectrum was first computed in \cite{Seery:2008ax} using traditional methods. We first consider a massless scalar then a conformally coupled scalar. For a massless scalar the relevant $3$-point function, as bootstrapped in Section \ref{BootstrapThreePoint}, is 
 \begin{align}
\psi_3^{\phi \phi \gamma}=-\lambda_{\phi} \frac{\epsilon_{ij}({\bf{k}}_{3}) k^{i}_{1}k_{2}^{j}}{k_{T}^2}( k_{T}^3- k_{T}e_{2}- e_{3}),
\end{align}
which matches the result of \cite{Maldacena:2002vr} from the bulk operator
\begin{align}
\mathcal{L}_{\text{int}} = \frac{\lambda_{\phi} }{2}a^2(\eta) \gamma^{ij}\partial_{i}\phi \partial_{j}\phi.
\end{align}
We now follow Step I of Section \ref{PartialEnergyShifts} and compute the right-hand side of the COT followed by extracting the residues $A_{n}= A_{n}(E_{L},E_{R},k_{1}k_{2},k_{3}k_{4},s)$. After summing over the helicities of the exchanged graviton, the right-hand side of the COT contains the overall polarisation factor
\begin{align}
\sum_{\lambda = \pm 2} \epsilon^{\lambda}_{ij}({\bf s})\epsilon^{\lambda}_{lm}({\bf -s})k^{i}_{1}k^{j}_{2}k^{l}_{3}k^{m}_{4},
\end{align}
that does not affect the residues. Now since the $k_{T}$ pole of the $3$-point function is degree 2, we have $A_{n} = 0$ for $n \geq 3$ while the remaining residues can be straightforwardly computed according to \eqref{An},
\begin{align}
A_{2} &= \frac{2 \lambda_{\phi}^2 k_{1}k_{2}s}{k_{T}^2(E_{L}+E_{R})^2}[2k_{3}k_{4} + (E_{L}+E_{R})k_{T}], \\
A_{1}&= \frac{\lambda_{\phi}^2}{k_{T}^3(E_{L}+E_{R})^3}\sum_{n=0}^4 a_{n}s^{n}, 
\end{align}
where we have defined
\begin{align}
a_{0} &= 2k_{1}k_{2}(E_{L}+E_{R})^2[(E_{L}+E_{R})^2 + 2k_{3}k_{4}], \\
a_{1} &= -4k_{1}k_{2}(E_{L}+E_{R})[(E_{L}+E_{R})^2 - 2k_{3}k_{4}], \\
a_{2} &= -2(E_{L}+E_{R})^4 -4 (E_{L}+E_{R})^2(k_{1}k_{2}+k_{3}k_{4})-16k_{1}k_{2}k_{3}k_{4}, \\
a_{3} &= 8(E_{L}+E_{R})[(E_{L}+E_{R})^2 + k_{1}k_{2}+k_{3}k_{4}], \\
a_{4} &= -8(E_{L}+E_{R})^2.
\end{align}
Plugging these expressions into \eqref{psireshere} yields
\begin{align} \label{GravExchangeStep1}
\psi_{4} = 2 \lambda_{\phi}^2 \sum_{\lambda = \pm 2} \epsilon^{\lambda}_{ij}({\bf s})\epsilon^{\lambda}_{lm}({\bf -s})k^{i}_{1}k^{j}_{2}k^{l}_{3}k^{m}_{4}f_{\phi}(k_{a},s) + B,
\end{align}
where 
\begin{align}
f_{\phi}(k_{a},s) = &-\frac{s^2}{k_{T}E_{L}E_{R}}+\frac{s k_{1}k_{2}}{k_{T}E_{L}^2E_{R}}+\frac{s k_{3}k_{4}}{k_{T}E_{R}^2E_{L}}+\frac{2s k_{1}k_{2}k_{3}k_{4}}{k_{T}^2 E_{L}^2 E_{R}^2} \nonumber \\ &-\frac{s(k_{1}k_{2}+k_{3}k_{4})}{k_{T}^2 E_{L}E_{R}}
+ \frac{k_{1}k_{2}}{k_{T}^2E_{L}}+\frac{k_{3}k_{4}}{k_{T}^2E_{R}} + \frac{2k_{1}k_{2}k_{3}k_{4}}{k_{T}^3E_{L}E_{R}}.
\end{align}
Again we notice the welcome appearance of $1/k_{T}$ factors even though we have not demanded anything about the residues of $k_{T}$ poles. Following Step II, we now plug this result back into the COT and find that it is satisfied for any $B$ satisfying \eqref{HomogeneousCOT} i.e. the boundary term is not required to contain an $s^3$ term and we can choose $\alpha_0=0$. Finally, for Step III we apply the MLT to the full \eqref{GravExchange} to constrain the boundary term further. Given the symmetries of this $4$-point function, we only need to apply the MLT to one of the external energies, say $k_{1}$. We find
\begin{align}
\frac{\partial \psi_{4}}{\partial k_{1}}\Big|_{k_{1}=0} = ~~ \frac{2 \lambda_{\phi}^2\sum_{\lambda} \epsilon^{\lambda}_{ij}({\bf s})\epsilon^{\lambda}_{lm}(-{\bf s})k^{i}_{1}k^{j}_{2}k^{l}_{3}k^{m}_{4}}{(k_{2}+k_{3}+k_{4})^2} +~~\frac{\partial B_{\text{MLT}}}{\partial k_{1}}\Big|_{k_{1}=0},
\end{align}
and so the only way to satisfy the MLT \eqref{MLT} is to set
\begin{align}
 B_{\text{MLT}} = \frac{2 \lambda_{\phi}^2\sum_{\lambda } \epsilon^{\lambda}_{ij}({\bf s})\epsilon^{\lambda}_{lm}(-{\bf s})k^{i}_{1}k^{j}_{2}k^{l}_{3}k^{m}_{4}}{k_{T}}. \label{DeltaBGravitonExchange}
\end{align}
Recall that this boundary term is only defined up to the presence of terms that themselves satisfy the homogeneous COT and the MLT, which we would naturally define as contact terms. Our final answer for $\psi_{4}$, with the inclusion of this boundary term, is exactly what one finds by performing the bulk computation, as was done in \cite{COT} (the corresponding trispectrum was first derived in \cite{Seery:2008ax}). We emphasise that in this example, our method has fixed \textit{all} the leading and sub-leading total energy and partial energy poles.\\

\noindent We can do the same for a conformally coupled scalar. The relevant $3$-point function is 
\begin{align} 
\psi_3^{\varphi \varphi \gamma}= - \lambda_{\varphi} \frac{\epsilon_{ij}({\bf{k}}_{3}) k^{i}_{1}k_{2}^{j}}{k_{T}^2}(2k_{3} + \mathbb{e}_1),
\end{align}
which comes from the bulk operator
\begin{align}
\mathcal{L}_{\text{int}} = \frac{\lambda_{\varphi} }{2}a^2(\eta) \gamma^{ij}\partial_{i}\varphi \partial_{j}\varphi.
\end{align}
After Step I we find that the $4$-point function due to graviton exchange is 
\begin{align} \label{GravExchange}
\psi_{4} = 2 \lambda^2_{\varphi} \sum_{\lambda = \pm 2} \epsilon^{\lambda}_{ij}({\bf s})\epsilon^{\lambda}_{lm}(-{\bf s})k^{i}_{1}k^{j}_{2}k^{l}_{3}k^{m}_{4}f_{\varphi}(k_{a},s) + B,
\end{align}
where 
\begin{align}
f_{\varphi}(k_{a},s) = \frac{2}{k_{T}^3E_{L}E_{R}}+\frac{1}{k_{T}^2E_{L}E_{R}^2}+\frac{1}{k_{T}^2E_{L}^2E_{R}}-\frac{1}{k_{T}E_{L}^2E_{R}^2}.
\end{align}
We remind the reader that although the external fields are conformally coupled, the graviton is massless and so $P(s) \sim 1/s^3$. It is simple to see that this expression satisfies the COT without the need for an $s^3$ term in $B$, just as in the massless case. Finally, since all external fields in this $4$-point function are conformally coupled, the MLT \eqref{MLTcc} simply requires the $4$-point function to be finite as one of the external energies is taken soft, and this is trivially satisfied with $B=0$. We therefore conclude that, up to contact terms, we can fix $B=0$ giving us the final form of the $4$-point function which matches the bulk expression and that derived by factorisation in \cite{CosmoBootstrap3}. Note that the structure of this final $4$-point function differs from that of massless scalars in that it does not contain a regular term at $E_{L,R} = 0$. We see that this is a consequence of the MLT, which only dictates the presence of a boundary term in the massless case.

\subsection{$\phi \phi'^2$ self-interaction in de Sitter}
Before moving to the effective field theory of inflation, let's consider the simplest IR finite self-interaction of a massless scalar in de Sitter, which arises at two-derivatives. We take the interaction to be 
\begin{align}
\mathcal{L}_{\text{int}} = \frac{g}{2} a^{2}(\eta) \phi \phi'^2.
\end{align}
The $3$-point function is given by
\begin{align}
\psi_{3}^{\phi \phi'^2} =- \frac{g}{k_{T}^2}(e_{3} e_{2} + k_{T}e_{2}^2 - 2 k_{T}^2 e_{3}).
\end{align}
Now by following Step I of our bootstrap procedure we can compute the residues $A_{n}$ and then $\psi_{\text{Res}}^{\phi \phi'^2}$. The expression is somewhat complicated and so we present it in Appendix \ref{AppendixResidue}. By taking this expression and plugging it into the Cosmological Optical Theorem we find that a boundary is indeed required by unitarity and we need to set
\begin{align} 
\alpha_{0}^{\phi \phi'^2} = 2 g^{2}. \label{Alpha0}
\end{align}
Turning to Step III, we now plug our expression for $\psi_{\text{Res}}^{\phi \phi'2}$ into the MLT. We find that $ B_{\text{\tiny MLT}}^{\phi \phi'^2}$ is non-zero and is given by
\begin{align}
 B_{\text{\tiny MLT}}^{\phi \phi'^2} = - \frac{g^{2}s^{2}}{k_{T}}(k_{T}^2-s^2). \label{DeltaBMLT}
\end{align}
Our final expression is therefore 
\begin{align}
\psi_{4}^{\phi \phi'^2} = \psi_{\text{Res}}^{\phi \phi'^2} +  g^2 s^3 +  B_{\text{\tiny MLT}}^{\phi \phi'^2}.
\end{align}
Lets now compare this bootstrap result to the expression one finds from the bulk computation. Following the bulk prescription reviewed in Appendix \ref{AppendixA} we compute the $4$-point function and find that it differs from our bootstrap result. After summing over permutations i.e. adding also the $t$ and $u$ channels, the difference can be accounted for by the contact diagrams of the following local operators
\begin{align}
\Delta \mathcal{L}_{\text{int}}^{\phi \phi'^2} = \frac{5g^2}{4} a^2(\eta) \phi^2 \phi'^2 -  \frac{g^2}{2} a^2(\eta) \phi^2[\phi'^2 - (\nabla \phi)^2]. \label{ContactDifference}
\end{align}
Note that the final two terms in $\Delta \mathcal{L}_{\text{int}}^{\phi \phi'^2}$ arise from taking the free theory and performing the field redefintion
\begin{align}
\phi({\bf{x}}) \rightarrow \phi({\bf{x}}) - \frac{g^2}{6}\phi^3({\bf{x}}). 
\end{align}
\subsection{Effective field theory of inflation}
We now turn to the self-interactions of the shift-symmetric Goldstone mode in the effective field theory of inflation \cite{EFTofI}. At cubic order the two self-interactions are
\begin{align}
\mathcal{L}_{\text{int}} = \frac{g_{1}}{3 !} a(\eta) \phi'^3+\frac{g_{2}}{2} a(\eta) \phi' (\nabla \phi)^2,
\end{align}
and we will refer to these to operators as EFT1 and EFT2 respectively. Let's begin by bootstrapping the exchange diagram due to two copies of EFT1. The $3$-point function is
\begin{align} \label{ThreePointEFT1}
\psi_{3}^{\text{\tiny EFT1}} = -\frac{2 g_{1} e_{3}^2}{k_{T}^3}.    
\end{align}
Following Step I of Section \ref{sec:step1}, we use \eqref{ThreePointEFT1} to compute the right-hand side of the COT, and then extract the non-zero residues
\begin{align}
A_{3} = -\frac{4 g_{1}^2(k_{1}k_{2}k_{3}k_{4}s)^2}{k_{T}^3(E_{L}+E_{R})^3}&[3(E_{L}+E_{R})^2 - 6(E_{L}+E_{R})s + 4s^2], \\
A_{2} =-\frac{48 g_{1}^2(k_{1}k_{2}k_{3}k_{4}s)^2}{k_{T}^4(E_{L}+E_{R})^4}&[(E_{L}+E_{R})^3 - 3(E_{L}+E_{R})^2s+4(E_{L}+E_{R})s^2-2s^3], \\
A_{1} =-\frac{24 g_{1}^2(k_{1}k_{2}k_{3}k_{4}s)^2}{k_{T}^5(E_{L}+E_{R})^5}&[5(E_{L}+E_{R})^4 - 20(E_{L}+E_{R})^3s \nonumber \\&+40(E_{L}+E_{R})^2s^2-40(E_{L}+E_{R})s^3+16s^4],
\end{align}
from which it follows that
\begin{align}
\psi_{4}^{\text{\tiny EFT1}} = -4g_{1}^2(k_{1}k_{2}k_{3}k_{4}s)^{2} \left[\frac{6}{k_{T}^5 E_{L}E_{R}} + \frac{3}{k_{T}^4 E_{L}E_{R}} \left(\frac{1}{E_{L}} + \frac{1}{E_{R}} \right)+ \frac{1}{k_{T}^3 E_{L}E_{R}} \left(\frac{1}{E_{L}} + \frac{1}{E_{R}} \right)^2 \nonumber \right. \\
\left. +\frac{1}{k_{T}^2 E_{L}^2E_{R}^2} \left(\frac{1}{E_{L}}  + \frac{1}{E_{R}} \right)+\frac{1}{k_{T}E_{L}^3 E_{R}^3}  \right] + B.
\end{align}
One can check that this expression satisfies the COT for any $B$ and so we do not need to add an $s^3$ term. This expression also satisfies the Manifestly Local Test (MLT) with $B=0$ thanks to the overall factor of $(k_{1}k_{2}k_{3}k_{4})^2$ that ensures that the first derivative with respect to any $k_{a}$ vanishes at $k_{a}=0$. We therefore set $B=0$. Let's now compare this result to the bulk result which we compute in Appendix \ref{AppendixA}. We find that, after summing over channels, the two expressions are not equivalent, but the difference can be accounted for by the local operator\footnote{Note that all the scale factors drop out in this interation: there are four positive powers of $a(\eta)$ from the measure, and four negative powers due to the four derivatives.}
\begin{align}
\Delta \mathcal{L}^{\text{\tiny EFT1}}_{\text{int}} = -\frac{g_{1}^2}{4!} \phi'^4. \label{EFT1Contact}
\end{align}

\noindent Finally, consider the $4$-point function due to two copies of the EFT2 vertex. The relevant $3$-point function is 
\begin{align} \label{EFT2ThreePoint}
\psi_{3}^{\text{\tiny EFT2}} &= - \frac{g_{2}}{2k_{T}^3} (k_{T}^6 - 3k_{T}^4 e_{2} + 11k_{T}^3 e_{3} - 4 k_{T}^2 e_{2}^2 - 4k_{T}e_{2}e_{3}+12 e_{3}^2).
\end{align}
Due to the complexity of this $3$-point function, the residues $A_{n}$ take complicated forms and the resulting $\psi_{\text{Res}}^{\text{EFT2}}$ is a long expression, which we provide in Appendix \ref{AppendixResidue}. With this expression in hand we can then move to Steps II and III to constrain the boundary term. In contrast to the $\phi'^3$ self-interaction, here the COT and MLT require a non-zero boundary term. Taking $\psi_{\text{Res}}^{\text{EFT2}}$ and plugging it into the COT, with \eqref{EFT2ThreePoint} used to compute the right-hand side, we find that
\begin{align}
\alpha^{\text{\tiny EFT2}}_{0} = \dfrac{25}{2} g_{2}^2. \label{AlphaEFT2}
\end{align}
Furthermore, the sum of these two components, from Step I and Step II, does not satisfy the MLT. Indeed we are required to also add 
\begin{align}
B_{\text{\tiny MLT}}^{\text{\tiny EFT2}} = -\frac{12g_{2}^{2}(k_{1}^2k_{2}^2+k_{3}^2k_{4}^2)s^2}{k_{T}^3} + \frac{4 g_{2}^2s^4}{k_{T}} - 5g_2^2  k_{T} s^2. \label{DeltaBEFT2}
\end{align}
Our final $4$-point function is therefore
\begin{align}
\psi_{4}^{\text{\tiny EFT2}} = \psi_{\text{Res}}^{\text{ \tiny EFT2}} + \frac{25}{4}g_{2}^2 s^3 + B_{\text{\tiny MLT}}^{\text{\tiny EFT2}}.
\end{align}
Let's now compare our result to the one arising from the bulk calculation, after we have summed over channels. Again the two expressions do not agree, but the difference can be accounted for by the following linear combination of local operators (we remind the reader that we are working in units with $H=1$)
\begin{align}
\label{EFT2Contact}
\frac{1}{g_{2}^2}\Delta \mathcal{L}^{\text{\tiny EFT2}}_{\text{int}}& = - \frac{5}{2} \phi'^4 +2 \phi'^2 (\nabla \phi)^2  + 9 a(\eta) \phi \phi'^3 -17 a^2(\eta) \phi^2 \phi'^2 +\frac{17}{2} a^2(\eta) \phi^2[\phi'^2 - (\nabla \phi)^2]\,,
\end{align}
where again the final two terms arise from the field redefintion $\phi({\bf{x}}) \rightarrow \phi({\bf{x}}) + \frac{17}{6}g_{2}^2\phi^3({\bf{x}})$. The details of this expression are not so important, the main point is that in all examples we have studied our bootstrap result recovers the bulk calculation up to a boundary term that is a contact diagram from local operators.


\section{Summary and future directions}

In this paper, we have introduced two new bootstrap tools for efficiently computing wavefunction coefficients/cosmological correlators. First, in Section \ref{sec:MLT}, we introduced a \textit{Manifestly Local Test (MLT)} that must be passed by wavefunction coefficients arising from manifestly local interactions of fields with de Sitter mode functions. Our test, given in \eqref{MLT} for the mode functions of a massless scalar or graviton, applies to contact and exchange $n$-point functions alike. We extended the MLT to massive fields too, with particular attention paid to conformally coupled scalars, and expect our results to provide a useful tool in the context of cosmological collider physics \cite{Arkani-Hamed:2015bza,Lee:2016vti}. We have shown in Section \ref{BootstrapThreePoint} that when combined with a sub-set of the Bootstrap Rules of \cite{BBBB}, the MLT allows us to bootstrap all $3$-point functions for a self-interacting massless scalar, improving over the results of \cite{BBBB}, and for minimal couplings between a graviton and a massless or conformally coupled scalar. In the latter two cases, we provided an on-shell proof that the leading interactions in the EFT expansion have two-derivatives and correspond to the familiar minimal couplings between scalars and gravitons. We also used our techniques to bootstrap contributions with $1 / \eta_{0}$ late-time divergences and showed that a combination of the MLT and the COT ensures that these are always imaginary and so do not contribute to the correlators.  \\

\noindent In Section \ref{PartialEnergyShifts} we then turned to bootstrapping exchange diagrams and introduced \textit{partial energy recursion relations}, which allow for efficient computation of $4$-point exchange diagrams given a pair of $3$-point sub-diagrams. When used in conjunction with the Cosmological Optical Theorem (COT) \cite{COT} (see also upcoming work \cite{Gordon,sCOTt}), these recursion relations fix the residues of all leading and sub-leading partial energy poles and therefore fix the $4$-point exchange diagram up to the presence of a boundary term with only total energy poles. The boundary term is then fixed by the MLT and the COT up to contact contributions from quartic interactions, which can always be chosen at will. For a number of examples, including scalar $4$-point functions due to graviton exchange and the cubic self-interactions in the EFT of inflation \cite{EFTofI}, we have shown that the resulting $4$-point function is equivalent to the one derived from bulk computations, up to contact contributions. We emphasise that throughout our analysis we did not assume invariance under de Sitter boosts, and therefore our bootstrap methods for computing exchange $4$-point functions can be applied to the very generic and potentially large interactions in the Effective Field Theory of Inflation.   \\

\noindent There are a number of avenues for future research:
\begin{itemize}
    \item So far our analysis has been restricted to massless and conformally coupled external scalars which is the simplest arena for testing new bootstrap tools. We expect our recursion relations to also be useful when bootstrap spinning correlators but we expect that a generalisation that does not apply diagram-by-diagram might also be required. We plan to investigate this in future work.
    \item It would be interesting to investigate the extent to which our bootstrap approach can remove redundancies in the EFT of inflation. This has been considered in \cite{Creminelli:2014wna,Bordin:2017hal,Bordin:2020eui} at the level of the Lagrangian where field redefinitions were used to remove redundant interactions. Our on-shell approach should make this process particularly simple and transparent. 
    \item In this paper we have restricted ourselves to meromorphic functions but it would be very interesting to adapt our approach to situations where the analytical structure of $4$- and higher-point functions is more complicated. For example, branch cuts appear in $4$-point functions due to the exchange of massive fields and such non-analytic behaviour is the famous avatar of cosmological particle production in de Sitter \cite{Arkani-Hamed:2015bza}. One would like to bootstrap these highly non-trivial structures directly from locality and unitarity, complementing the approach taken in \cite{Arkani-Hamed:2018bjr}. Other types of non scale invariant branch cuts stem from secular growth in de Sitter space, where our partial energy recursion relations do not apply as formulated here. We leave the possibility of bootstrapping IR-divergent contributions to the correlators of light fields in de Sitter for future research. 
    \item Positivity bounds have become a powerful tool in the $S$-matrix programme given their power to constrain low-energy effective field theory couplings from some mild assumptions about the UV, see e.g. \cite{Positivity1,Bellazzini:2020cot,deRham:2017zjm,Melville:2019tdc,Arkani-Hamed:2020blm,Melville:2019wyy,Caron-Huot:2021rmr,Bellazzini:2016xrt,Tolley:2020gtv} and references therein. Very recently these bounds have been applied to boost-breaking amplitudes as relevant for the sub-horizon limit of the effective field theory of inflation \cite{Grall:2020tqc,Positivity2}. It would be very interesting to derive such bounds for cosmological correlators/wavefunction coefficients, away from the total-energy pole, given that such constraints can in principle constrain the size of couplings which is not possible with the current bootstrap methods presented here and elsewhere.
\end{itemize}


\section*{Acknowledgements}
We would like to thank Daniel Baumann, James Bonifacio, Giovanni Cabass, Harry Goodhew, Mang Hei Gordon Lee, Tanguy Grall, Aaron Hillman, Austin Joyce, Hayden Lee, Joao Melo, Scott Melville, Guilherme L. Pimentel, Sebastien Renaux-Petel, Jakub Supel and Dong-Gang Wang for useful discussions. SJ would like to dedicate this work to his  yet anonymous daughter who is due to arrive in this world in late March 2021. E.P. and D.S. have been supported in part by the research program VIDI with Project No. 680-47-535, which is (partly) financed by the Netherlands Organisation for Scientific Research (NWO). SJ is supported by the European Research Council under the European Union’s Horizon 2020 research and innovation programme (grant agreement No 758792, project GEODESI). 


\appendix


\section{Bulk calculations of the wavefunction of the universe} \label{AppendixA}
In this appendix we review the Feynman rules for computing perturbative contributions to the wavefunction coefficients. For simplicity, we focus on tree-level diagrams of a single scalar field theory a on de Sitter background with at most one time derivative per field\footnote{Any Lagrangian with more than one time derivative can be brought back to single-time-derivative form by recursively using the equation of motion.}.\\

\noindent Consider a tree diagram with $n$ external lines going to the boundary, $V$ vertices and $I=V-1$ internal lines. Attribute momenta $\bfk_a$ to the external legs and momenta $\bfp_m$ to the internal so that momentum (but not energy) is conserved at each vertex. The expression for the wavefunction coefficient $\psi_n(\lbrace k\rbrace, \lbrace p\rbrace, \lbrace\bfk\rbrace)$ can be computed by the following recipe,
\begin{itemize}
    \item Insert an overall factor of $(-i)$. 
    \item Insert the appropriate vertex accounting for the appropriate permutations, e.g. for $\dfrac{g}{n!}\phi^n$ insert $g$ (no factor of $i$).  
      \item To each vertex assign a time integral 
       \begin{equation}
       \int_{-\infty(1-i\epsilon)}^0 a^{4+n-D}(\eta_\alpha)d\eta_\alpha\,\,\qquad  \alpha=1,\dots V
       \end{equation}
       where $D$ is the mass dimension and $n$ the valency (number of legs) of the vertex.
    \item For each external line with energy $k_a=|\bfk_a|$, insert the bulk-to-boundary propagator $K(k_a,\eta_\alpha)$
   \item For each internal line with energy $p_m=|\bfp_m|$, insert the bulk-to-bulk propagator $G(p_m,\eta_\alpha,\eta_{\beta})$
\end{itemize}

\noindent Using these Feynman rules, below we provide the explicit expression for some of the diagrams we encountered in Section \ref{Results}. 
\subsection*{Contact terms}
\begin{align}
     \psi_{\phi'^3}(k_1,k_2,k_3)&=-i \int d\eta\,a(\eta)\,K'(k_1,\eta)\,K'(k_2,\eta)\,K'(k_3,\eta)\,,\\ 
    \psi_{\phi\phi'^2}(k_1,k_2,k_3)&=-i \int d\eta\,a^2(\eta)\, K'(k_1,\eta)\,K'(k_2,\eta)\,K(k_3,\eta)+\text{2 permutations}\,,\\ 
    \psi_{\phi'(\partial_i \phi)^2}(k_1,k_2,k_3)&=i(\bfk_2.\bfk_3)\int d\eta\,a(\eta)\,K'(k_1,\eta)K(k_2,\eta)K(k_3,\eta)+\text{2 permutations.}
\end{align}
\subsection*{Exchange Diagrams}
After integrating by parts to remove the time derivatives from the bulk-to-bulk propagator we arrive at\footnote{Notice that for vertices that have more than one internal line entering them, it is not always possible to strip off $\partial_\eta$ from all the bulk-to-bulk propagators.}, 
\begin{align}
\nonumber
 &\psi^s_{\phi'^3\times \phi'^3}=-i \int\,d\eta'\,d\eta\,G(s,\eta,\eta')\partial_\eta \left(a(\eta) K'(k_1,\eta)K'(k_2,\eta)\right)\,\partial_{\eta'} \left(a(\eta') K'(k_3,\eta')K'(k_4,\eta')\right)\,.\\ \nonumber
   & \psi^s_{\phi\phi'^2\times \phi\phi'^2}=-i \int\,d\eta\,d\eta'\,G(s,\eta,\eta')\\ \nonumber
&\left[-\partial_\eta\,(a^2(\eta)K'(k_1,\eta)K(k_2,\eta))-\partial_\eta\,(a^2(\eta)K'(k_2,\eta)K(k_1,\eta))+a^2(\eta)K'(k_1,\eta)K'(k_2,\eta)) \right],\\ \nonumber
&\times \left[-\partial_{\eta'}\,(a^2(\eta')K'(k_3,\eta')K(k_4,\eta'))-\partial_{\eta'}\,(a^2(\eta')K'(k_4,\eta')K(k_3,\eta'))+a^2(\eta')K'(k_3,\eta')K'(k_4,\eta')) \right].\\ \nonumber
&\psi^s_{\phi'(\partial_i\phi)^2\times \phi'(\partial_i\phi)^2}=-i\,\int\,d\eta\,d\eta'\,G(s,\eta,\eta')\\ \nonumber
&\qquad\qquad\qquad\times\left[\bfk_1.\bfk_2\,\partial_\eta\,\left(a(\eta)K(k_1,\eta)K(k_2,\eta)\right)\right.\\ \nonumber
&\qquad\qquad\qquad\quad \left. +a(\eta)(\bfk_1+\bfk_2).\bfk_2\,K'(k_1,\eta)K(k_2,\eta)+a(\eta)(\bfk_1+\bfk_2).\bfk_1\,K'(k_2,\eta)\,K(k_1,\eta)\right]\\ \nonumber
&\qquad\qquad\qquad\times\left[\bfk_3.\bfk_4\,\partial_{\eta'}\,\left(a(\eta')K(k_3,\eta')K(k_4,\eta')\right)\right.\\ \nonumber
&\qquad\qquad\qquad\quad \left. +a(\eta')(\bfk_3+\bfk_4).\bfk_3\,K'(k_3,\eta')K(k_4,\eta')+a(\eta')(\bfk_3+\bfk_4).\bfk_3\,K'(k_4,\eta')\,K(k_3,\eta')\right]\,,
\end{align}
where $\psi^s$ stands for the $s$-channel contribution to the $4$-point function. 


\section{Counting wavefunction coefficients and amplitudes} \label{AppendixC}

In this appendix we show that for a single scalar field to any order in derivatives, the number of independent manifestly local cubic amplitudes $A_3$ matches the number of real independent cubic wavefunctions $\psi_3$ minus one. The extra 3-point function corresponds to the only manifestly local field redefinition at this order. \\

\noindent If we include the unique $p=0$ logarithmic term $\psi_{3}^{\text{log}}$ we have, c.f. \eqref{Ntotal}, 
\begin{align}
N_{\text{total}}(p) &= 1 + \sum_{q=0}^{\lfloor \frac{p+3}{2}\rfloor}\, \lfloor \dfrac{p+3-2q}{3}\rfloor.
\end{align}
It follows that as we increase the leading $k_{T}$ pole from degree $p$ to $p+1$, the number of new $3$-point functions is 
\begin{align}
N_{\text{total}}(p+1)-N_{\text{total}}(p) &= \sum_{q=0}^{\lfloor \frac{p+4}{2}\rfloor}\, \lfloor \dfrac{p+4-2q}{3}\rfloor  - \sum_{q=0}^{\lfloor \frac{p+3}{2}\rfloor}\, \lfloor \dfrac{p+3-2q}{3}\rfloor,
\end{align}
which is easier to analyse for even and odd $p$ separately. For even $p$ we have 
\begin{align}
N_{\text{total}}(1)-N_{\text{total}}(0) &= 0, \label{NewCorrelators0} \\
N_{\text{total}}(p+1)-N_{\text{total}}(p) &= \sum_{q=0}^{\frac{p}{2}-1} I(p+4-2q), \quad  p \geq 2,  \label{NewCorrelatorsP}
\end{align}
where 
\begin{align}
I(x) &= 1 ~~ \text{if} ~~ x ~ | ~ 3 \\
& = 0 ~~ \text{otherwise}.
\end{align}
Here $x ~ | ~ 3$ stands for $x$ ``divides" $3$, e.g. $I(2) = I(4) = 0$ and $I(3) = I(6) = 1$. Let's now consider the number of amplitudes with exactly $p+1$ derivatives, $N_{\text{amplitudes}}(p+1)$, which is simply the number of integer solutions to $2 \alpha+3 \beta=p+1$. We have $N_{\text{amplitudes}}(1) = 0$, and so as we increase the number of derivatives from $p=0$ to $p=1$ there are no new amplitudes and no new $3$-point functions c.f. \eqref{NewCorrelators0}. For even $p \geq 2$ we have
\begin{align} \label{NewAmplitudes}
N_{\text{amplitudes}}(p+1) = \sum_{q=0}^{\frac{p}{2}-1}I(p+1-2q) = \sum_{q=0}^{\frac{p}{2}-1}I(p+4-2q),
\end{align}
where we have used $I(x+3) = I(x)$. So for even $p \geq 2$, the number of new $3$-point functions as we go from $p$ to $p+1$ derivatives \eqref{NewCorrelatorsP} equals the number of new amplitudes \eqref{NewAmplitudes}. Similarly, for odd $p$ we have
\begin{align} \label{NewCorrelators1}
N_{\text{total}}(p+1)-N_{\text{total}}(p) = \sum_{q=0}^{\frac{p+3}{2}} I(p+4-2q) = \sum_{q=0}^{\frac{p+1}{2}} I(p+4-2q),
\end{align}
while the number of amplitudes is given by
\begin{align} \label{NewAmplitudes1}
N_{\text{amplitudes}}(p+1) = \sum_{q=0}^{\frac{p+1}{2}}I(p+1-2q) = \sum_{q=0}^{\frac{p+1}{2}}I(p+4-2q).
\end{align}
So for odd $p$ we again see that the number of new $3$-point functions  as we increase from $p$ to $p+1$ derivatives \eqref{NewCorrelators1} is equal to the number of new amplitudes \eqref{NewAmplitudes1}. To compute the final $N_{\text{total}}(p) - N_{\text{amplitudes}}(p)$ we therefore only need to compare the number of $3$-point functions to the number of amplitudes for $p=0$. There are two $3$-point functions: $\psi_{3}^{\text{local}}$ and $\psi_{3}^{\text{log}}$ for $p=0$, while there is only a single $p=0$ amplitude which is simply a constant \cite{PSS}. We therefore conclude that
\begin{align}
N_{\text{total}}(p) = N_{\text{amplitudes}}(p) + 1.
\end{align}

\section{Expressions for $\psi_{\text{Res}}$} \label{AppendixResidue}
In this appendix we collect some of the longer expressions for $\psi_{\text{Res}}$. For an exchange diagram due to two copies of the $\phi \phi'^2$ vertex, we have \vspace{0.2cm} \\
\tiny
$\psi_{\text{Res}}^{\phi \phi'^2} = -\frac{g^2}{64 E_{L}^2 E_{R}^2 k_{T}^3}[\frac{1}{2} E_{L} E_{R} k_{T}^8-2 E_{L}^2 E_{R} k_{T}^7+2 E_{R} k_{1} k_{2} k_{T}^7-E_{L}^2 E_{R}^2 k_{T}^6-E_{L}^3 E_{R} k_{T}^6-2 E_{R}^2 k_{1} k_{2} k_{T}^6-12 E_{L} E_{R} k_{1} k_{2} k_{T}^6+2 k_{1} k_{2} k_{3} k_{4} k_{T}^6+12 E_{L}^3 E_{R}^2 k_{T}^5-8 E_{R} k_{1}^2 k_{2}^2 k_{T}^5+4 E_{L}^4 E_{R} k_{T}^5-4 E_{R}^3 k_{1} k_{2} k_{T}^5+6 E_{L} E_{R}^2 k_{1} k_{2} k_{T}^5+30 E_{L}^2 E_{R} k_{1} k_{2} k_{T}^5-20 E_{L} k_{1} k_{2} k_{3} k_{4} k_{T}^5-3 E_{L}^3 E_{R}^3 k_{T}^4-4 E_{L}^4 E_{R}^2 k_{T}^4-8 E_{R}^2 k_{1}^2 k_{2}^2 k_{T}^4+32 E_{L} E_{R} k_{1}^2 k_{2}^2 k_{T}^4-E_{L}^5 E_{R} k_{T}^4+4 E_{R}^4 k_{1} k_{2} k_{T}^4+16 E_{L} E_{R}^3 k_{1} k_{2} k_{T}^4-28 E_{L}^2 E_{R}^2 k_{1} k_{2} k_{T}^4-40 E_{L}^3 E_{R} k_{1} k_{2} k_{T}^4-16 k_{1}^2 k_{2}^2 k_{3} k_{4} k_{T}^4+40 E_{L}^2 k_{1} k_{2} k_{3} k_{4} k_{T}^4+36 E_{L} E_{R} k_{1} k_{2} k_{3} k_{4} k_{T}^4-20 E_{L}^4 E_{R}^3 k_{T}^3-10 E_{L}^5 E_{R}^2 k_{T}^3+8 E_{R}^3 k_{1}^2 k_{2}^2 k_{T}^3+40 E_{L} E_{R}^2 k_{1}^2 k_{2}^2 k_{T}^3-48 E_{L}^2 E_{R} k_{1}^2 k_{2}^2 k_{T}^3-2 E_{L}^6 E_{R} k_{T}^3+2 E_{R}^5 k_{1} k_{2} k_{T}^3-4 E_{L} E_{R}^4 k_{1} k_{2} k_{T}^3+16 E_{L}^2 E_{R}^3 k_{1} k_{2} k_{T}^3+52 E_{L}^3 E_{R}^2 k_{1} k_{2} k_{T}^3+30 E_{L}^4 E_{R} k_{1} k_{2} k_{T}^3+48 E_{L} k_{1}^2 k_{2}^2 k_{3} k_{4} k_{T}^3+48 E_{R} k_{1}^2 k_{2}^2 k_{3} k_{4} k_{T}^3-40 E_{L}^3 k_{1} k_{2} k_{3} k_{4} k_{T}^3-88 E_{L}^2 E_{R} k_{1} k_{2} k_{3} k_{4} k_{T}^3+10 E_{L}^4 E_{R}^4 k_{T}^2+15 E_{L}^5 E_{R}^3 k_{T}^2+6 E_{L}^6 E_{R}^2 k_{T}^2+8 E_{R}^4 k_{1}^2 k_{2}^2 k_{T}^2+16 E_{L} E_{R}^3 k_{1}^2 k_{2}^2 k_{T}^2+40 E_{L}^2 E_{R}^2 k_{1}^2 k_{2}^2 k_{T}^2+32 E_{L}^3 E_{R} k_{1}^2 k_{2}^2 k_{T}^2+32 k_{1}^2 k_{2}^2 k_{3}^2 k_{4}^2 k_{T}^2+E_{L}^7 E_{R} k_{T}^2-2 E_{R}^6 k_{1} k_{2} k_{T}^2-4 E_{L} E_{R}^5 k_{1} k_{2} k_{T}^2-12 E_{L}^2 E_{R}^4 k_{1} k_{2} k_{T}^2-32 E_{L}^3 E_{R}^3 k_{1} k_{2} k_{T}^2-34 E_{L}^4 E_{R}^2 k_{1} k_{2} k_{T}^2-12 E_{L}^5 E_{R} k_{1} k_{2} k_{T}^2-48 E_{L}^2 k_{1}^2 k_{2}^2 k_{3} k_{4} k_{T}^2-48 E_{R}^2 k_{1}^2 k_{2}^2 k_{3} k_{4} k_{T}^2-64 E_{L} E_{R} k_{1}^2 k_{2}^2 k_{3} k_{4} k_{T}^2+20 E_{L}^4 k_{1} k_{2} k_{3} k_{4} k_{T}^2+12 E_{L}^2 E_{R}^2 k_{1} k_{2} k_{3} k_{4} k_{T}^2+32 E_{L}^3 E_{R} k_{1} k_{2} k_{3} k_{4} k_{T}^2+8 E_{L} E_{R}^4 k_{1}^2 k_{2}^2 k_{T}+8 E_{L}^2 E_{R}^3 k_{1}^2 k_{2}^2 k_{T}-8 E_{L}^3 E_{R}^2 k_{1}^2 k_{2}^2 k_{T}-8 E_{L}^4 E_{R} k_{1}^2 k_{2}^2 k_{T}-64 E_{L} k_{1}^2 k_{2}^2 k_{3}^2 k_{4}^2 k_{T}-2 E_{L} E_{R}^6 k_{1} k_{2} k_{T}-6 E_{L}^2 E_{R}^5 k_{1} k_{2} k_{T}-4 E_{L}^3 E_{R}^4 k_{1} k_{2} k_{T}+4 E_{L}^4 E_{R}^3 k_{1} k_{2} k_{T}+6 E_{L}^5 E_{R}^2 k_{1} k_{2} k_{T}+2 E_{L}^6 E_{R} k_{1} k_{2} k_{T}+16 E_{L}^3 k_{1}^2 k_{2}^2 k_{3} k_{4} k_{T}+16 E_{R}^3 k_{1}^2 k_{2}^2 k_{3} k_{4} k_{T}-16 E_{L} E_{R}^2 k_{1}^2 k_{2}^2 k_{3} k_{4} k_{T}-16 E_{L}^2 E_{R} k_{1}^2 k_{2}^2 k_{3} k_{4} k_{T}-4 E_{L}^5 k_{1} k_{2} k_{3} k_{4} k_{T}+56 E_{L}^3 E_{R}^2 k_{1} k_{2} k_{3} k_{4} k_{T}+12 E_{L}^4 E_{R} k_{1} k_{2} k_{3} k_{4} k_{T}-64 E_{L} E_{R} k_{1}^2 k_{2}^2 k_{3}^2 k_{4}^2+32 E_{L} E_{R}^3 k_{1}^2 k_{2}^2 k_{3} k_{4}+64 E_{L}^2 E_{R}^2 k_{1}^2 k_{2}^2 k_{3} k_{4}+32 E_{L}^3 E_{R} k_{1}^2 k_{2}^2 k_{3} k_{4}-24 E_{L}^3 E_{R}^3 k_{1} k_{2} k_{3} k_{4}-32 E_{L}^4 E_{R}^2 k_{1} k_{2} k_{3} k_{4}-8 E_{L}^5 E_{R} k_{1} k_{2} k_{3} k_{4} + (E_{L} \rightarrow E_{R}, k_{1}k_{2} \rightarrow k_{3}k_{4})]$.  \vspace{0.2cm} \\
\normalsize
For the exchange diagram for two copies of the EFT2 vertex, $\phi' (\nabla \phi)^2$, we have \vspace{0.2cm}\\
\tiny
$\psi_{\text{Res}}^{\text{EFT2}} = -\frac{g_{2}^2}{8 k_{T}^5 E_{L}^3 E_{R}^3} [\frac{1}{2} E_{L}^2 E_{R}^2 k_{T}^{10}-E_{L}^3 E_{R}^2 k_{T}^9+2 E_{L} E_{R}^2 k_{1} k_{2} k_{T}^9-5 E_{L}^3 E_{R}^3 k_{T}^8-5 E_{L}^4 E_{R}^2 k_{T}^8-12 E_{R}^2 k_{1}^2 k_{2}^2 k_{T}^8-12 E_{L}^2 E_{R}^2 k_{1} k_{2} k_{T}^8+2 E_{L} E_{R} k_{1} k_{2} k_{3} k_{4} k_{T}^8+30 E_{L}^4 E_{R}^3 k_{T}^7+10 E_{L}^5 E_{R}^2 k_{T}^7-12 E_{R}^3 k_{1}^2 k_{2}^2 k_{T}^7+28 E_{L} E_{R}^2 k_{1}^2 k_{2}^2 k_{T}^7-10 E_{L} E_{R}^4 k_{1} k_{2} k_{T}^7-4 E_{L}^2 E_{R}^3 k_{1} k_{2} k_{T}^7+30 E_{L}^3 E_{R}^2 k_{1} k_{2} k_{T}^7-24 E_{R} k_{1}^2 k_{2}^2 k_{3} k_{4} k_{T}^7-20 E_{L}^2 E_{R} k_{1} k_{2} k_{3} k_{4} k_{T}^7-15 E_{L}^4 E_{R}^4 k_{T}^6-20 E_{L}^5 E_{R}^3 k_{T}^6-5 E_{L}^6 E_{R}^2 k_{T}^6+48 E_{R}^4 k_{1}^2 k_{2}^2 k_{T}^6+8 E_{L} E_{R}^3 k_{1}^2 k_{2}^2 k_{T}^6-4 E_{L}^2 E_{R}^2 k_{1}^2 k_{2}^2 k_{T}^6+72 k_{1}^2 k_{2}^2 k_{3}^2 k_{4}^2 k_{T}^6+10 E_{L} E_{R}^5 k_{1} k_{2} k_{T}^6+36 E_{L}^2 E_{R}^4 k_{1} k_{2} k_{T}^6-14 E_{L}^3 E_{R}^3 k_{1} k_{2} k_{T}^6-40 E_{L}^4 E_{R}^2 k_{1} k_{2} k_{T}^6+96 E_{R}^2 k_{1}^2 k_{2}^2 k_{3} k_{4} k_{T}^6+32 E_{L} E_{R} k_{1}^2 k_{2}^2 k_{3} k_{4} k_{T}^6+36 E_{L}^2 E_{R}^2 k_{1} k_{2} k_{3} k_{4} k_{T}^6+40 E_{L}^3 E_{R} k_{1} k_{2} k_{3} k_{4} k_{T}^6-10 E_{L}^5 E_{R}^4 k_{T}^5-5 E_{L}^6 E_{R}^3 k_{T}^5-E_{L}^7 E_{R}^2 k_{T}^5-12 E_{R}^5 k_{1}^2 k_{2}^2 k_{T}^5-32 E_{L} E_{R}^4 k_{1}^2 k_{2}^2 k_{T}^5+52 E_{L}^2 E_{R}^3 k_{1}^2 k_{2}^2 k_{T}^5-24 E_{L}^3 E_{R}^2 k_{1}^2 k_{2}^2 k_{T}^5-48 E_{L} k_{1}^2 k_{2}^2 k_{3}^2 k_{4}^2 k_{T}^5-14 E_{L}^2 E_{R}^5 k_{1} k_{2} k_{T}^5+2 E_{L}^3 E_{R}^4 k_{1} k_{2} k_{T}^5+46 E_{L}^4 E_{R}^3 k_{1} k_{2} k_{T}^5+30 E_{L}^5 E_{R}^2 k_{1} k_{2} k_{T}^5-144 E_{R}^3 k_{1}^2 k_{2}^2 k_{3} k_{4} k_{T}^5-48 E_{L} E_{R}^2 k_{1}^2 k_{2}^2 k_{3} k_{4} k_{T}^5+24 E_{L}^2 E_{R} k_{1}^2 k_{2}^2 k_{3} k_{4} k_{T}^5-88 E_{L}^3 E_{R}^2 k_{1} k_{2} k_{3} k_{4} k_{T}^5-40 E_{L}^4 E_{R} k_{1} k_{2} k_{3} k_{4} k_{T}^5+10 E_{L}^5 E_{R}^5 k_{T}^4+15 E_{L}^6 E_{R}^4 k_{T}^4+6 E_{L}^7 E_{R}^3 k_{T}^4+E_{L}^8 E_{R}^2 k_{T}^4-12 E_{R}^6 k_{1}^2 k_{2}^2 k_{T}^4+8 E_{L} E_{R}^5 k_{1}^2 k_{2}^2 k_{T}^4+16 E_{L}^2 E_{R}^4 k_{1}^2 k_{2}^2 k_{T}^4-8 E_{L}^3 E_{R}^3 k_{1}^2 k_{2}^2 k_{T}^4-4 E_{L}^4 E_{R}^2 k_{1}^2 k_{2}^2 k_{T}^4-192 E_{L}^2 k_{1}^2 k_{2}^2 k_{3}^2 k_{4}^2 k_{T}^4+32 E_{L} E_{R} k_{1}^2 k_{2}^2 k_{3}^2 k_{4}^2 k_{T}^4-2 E_{L} E_{R}^7 k_{1} k_{2} k_{T}^4-4 E_{L}^2 E_{R}^6 k_{1} k_{2} k_{T}^4-12 E_{L}^3 E_{R}^5 k_{1} k_{2} k_{T}^4-32 E_{L}^4 E_{R}^4 k_{1} k_{2} k_{T}^4-34 E_{L}^5 E_{R}^3 k_{1} k_{2} k_{T}^4-12 E_{L}^6 E_{R}^2 k_{1} k_{2} k_{T}^4+96 E_{R}^4 k_{1}^2 k_{2}^2 k_{3} k_{4} k_{T}^4-48 E_{L} E_{R}^3 k_{1}^2 k_{2}^2 k_{3} k_{4} k_{T}^4-64 E_{L}^2 E_{R}^2 k_{1}^2 k_{2}^2 k_{3} k_{4} k_{T}^4-24 E_{L}^3 E_{R} k_{1}^2 k_{2}^2 k_{3} k_{4} k_{T}^4+12 E_{L}^3 E_{R}^3 k_{1} k_{2} k_{3} k_{4} k_{T}^4+32 E_{L}^4 E_{R}^2 k_{1} k_{2} k_{3} k_{4} k_{T}^4+20 E_{L}^5 E_{R} k_{1} k_{2} k_{3} k_{4} k_{T}^4-12 E_{L} E_{R}^6 k_{1}^2 k_{2}^2 k_{T}^3+8 E_{L}^2 E_{R}^5 k_{1}^2 k_{2}^2 k_{T}^3+80 E_{L}^3 E_{R}^4 k_{1}^2 k_{2}^2 k_{T}^3+88 E_{L}^4 E_{R}^3 k_{1}^2 k_{2}^2 k_{T}^3+28 E_{L}^5 E_{R}^2 k_{1}^2 k_{2}^2 k_{T}^3-48 E_{L}^3 k_{1}^2 k_{2}^2 k_{3}^2 k_{4}^2 k_{T}^3-256 E_{L}^2 E_{R} k_{1}^2 k_{2}^2 k_{3}^2 k_{4}^2 k_{T}^3-2 E_{L}^2 E_{R}^7 k_{1} k_{2} k_{T}^3-6 E_{L}^3 E_{R}^6 k_{1} k_{2} k_{T}^3-4 E_{L}^4 E_{R}^5 k_{1} k_{2} k_{T}^3+4 E_{L}^5 E_{R}^4 k_{1} k_{2} k_{T}^3+6 E_{L}^6 E_{R}^3 k_{1} k_{2} k_{T}^3+2 E_{L}^7 E_{R}^2 k_{1} k_{2} k_{T}^3-24 E_{R}^5 k_{1}^2 k_{2}^2 k_{3} k_{4} k_{T}^3+112 E_{L} E_{R}^4 k_{1}^2 k_{2}^2 k_{3} k_{4} k_{T}^3-16 E_{L}^2 E_{R}^3 k_{1}^2 k_{2}^2 k_{3} k_{4} k_{T}^3-40 E_{L}^3 E_{R}^2 k_{1}^2 k_{2}^2 k_{3} k_{4} k_{T}^3-32 E_{L}^4 E_{R} k_{1}^2 k_{2}^2 k_{3} k_{4} k_{T}^3+56 E_{L}^4 E_{R}^3 k_{1} k_{2} k_{3} k_{4} k_{T}^3+12 E_{L}^5 E_{R}^2 k_{1} k_{2} k_{3} k_{4} k_{T}^3-4 E_{L}^6 E_{R} k_{1} k_{2} k_{3} k_{4} k_{T}^3-12 E_{L}^2 E_{R}^6 k_{1}^2 k_{2}^2 k_{T}^2-48 E_{L}^3 E_{R}^5 k_{1}^2 k_{2}^2 k_{T}^2-72 E_{L}^4 E_{R}^4 k_{1}^2 k_{2}^2 k_{T}^2-48 E_{L}^5 E_{R}^3 k_{1}^2 k_{2}^2 k_{T}^2-12 E_{L}^6 E_{R}^2 k_{1}^2 k_{2}^2 k_{T}^2+144 E_{L}^4 k_{1}^2 k_{2}^2 k_{3}^2 k_{4}^2 k_{T}^2-256 E_{L}^2 E_{R}^2 k_{1}^2 k_{2}^2 k_{3}^2 k_{4}^2 k_{T}^2-192 E_{L}^3 E_{R} k_{1}^2 k_{2}^2 k_{3}^2 k_{4}^2 k_{T}^2-48 E_{L} E_{R}^5 k_{1}^2 k_{2}^2 k_{3} k_{4} k_{T}^2+128 E_{L}^2 E_{R}^4 k_{1}^2 k_{2}^2 k_{3} k_{4} k_{T}^2+136 E_{L}^3 E_{R}^3 k_{1}^2 k_{2}^2 k_{3} k_{4} k_{T}^2-16 E_{L}^4 E_{R}^2 k_{1}^2 k_{2}^2 k_{3} k_{4} k_{T}^2+24 E_{L}^5 E_{R} k_{1}^2 k_{2}^2 k_{3} k_{4} k_{T}^2-24 E_{L}^4 E_{R}^4 k_{1} k_{2} k_{3} k_{4} k_{T}^2-32 E_{L}^5 E_{R}^3 k_{1} k_{2} k_{3} k_{4} k_{T}^2-8 E_{L}^6 E_{R}^2 k_{1} k_{2} k_{3} k_{4} k_{T}^2-432 E_{L}^3 E_{R}^2 k_{1}^2 k_{2}^2 k_{3}^2 k_{4}^2 k_{T}+432 E_{L}^4 E_{R} k_{1}^2 k_{2}^2 k_{3}^2 k_{4}^2 k_{T}-72 E_{L}^4 E_{R}^3 k_{1} k_{2} k_{3}^2 k_{4}^2 k_{T}-72 E_{L}^5 E_{R}^2 k_{1} k_{2} k_{3}^2 k_{4}^2 k_{T}+72 E_{L}^4 E_{R}^3 k_{1}^2 k_{2}^2 k_{3} k_{4} k_{T}+72 E_{L}^5 E_{R}^2 k_{1}^2 k_{2}^2 k_{3} k_{4} k_{T}+864 E_{L}^3 E_{R}^3 k_{1}^2 k_{2}^2 k_{3}^2 k_{4}^2+864 E_{L}^4 E_{R}^2 k_{1}^2 k_{2}^2 k_{3}^2 k_{4}^2 + (E_{L} \rightarrow E_{R}, k_{1}k_{2} \rightarrow k_{3}k_{4})]$.
\normalsize

\bibliographystyle{JHEP}
\bibliography{refs}

\end{document}